\documentclass[11pt]{article}

\usepackage[margin=1in]{geometry}
\usepackage[utf8]{inputenc}
\usepackage{amsmath,amssymb,amsfonts,amsthm}
\usepackage[font=small,labelfont=bf]{caption}
\usepackage{bold-extra}
\usepackage{graphicx}

\usepackage{changepage}
\newenvironment{tight}{\begin{adjustwidth}{1cm}{}}{\end{adjustwidth}}

\usepackage{listings}

\lstdefinelanguage[90]{Fortran}{
  language=Fortran,
  morekeywords={write,implicit,none,program,use,class,character,allocatable,contains,module,procedure,public,private,type,dimension,interface,integer,real,logical,subroutine,function,call,if,then,else,end,do,allocate,nullify,null},
  sensitive=true,
  morecomment=[l]!,
  morestring=[b]",
}

\usepackage[dvipsnames,svgnames]{xcolor}
\usepackage{url}
\usepackage{hyperref}
\hypersetup{
  colorlinks   = true, 
  urlcolor     = blue, 
  linkcolor    = blue, 
  citecolor   = blue 
}
\usepackage[numbers,sort&compress]{natbib}
\let\OLDthebibliography\thebibliography
\renewcommand\thebibliography[1]{
  \OLDthebibliography{#1}
  \setlength{\parskip}{4pt}
  \setlength{\itemsep}{0pt plus 0.3ex}
}

\allowdisplaybreaks
\usepackage{multicol}
\usepackage{enumitem}

\usepackage[capitalise]{cleveref}
\usepackage{array}

\newcommand{\TB}[1]{{\color{black}#1}}

\newcommand{\evortran}{\texttt{evortran}}

\begin{document}

\thispagestyle{empty}

\def\thefootnote{\fnsymbol{footnote}}

\begin{flushright}
  \footnotesize
  IFT-UAM/CSIC-25-76
\end{flushright}

\vspace*{0.4cm}

\begin{center}

{\Large
  \textbf{
  evortran: a modern Fortran package for
  genetic algorithms with\\[0.4em]applications from LHC
  data fitting to LISA signal reconstruction
  }
}

\vspace{1.8em}

Thomas Biek\"otter$^1$\footnote{thomas.biekoetter@desy.de}

\vspace*{0.8em}

$^1$\textit{
  Instituto de F\'isica Te\'orica UAM/CSIC,
  Calle Nicolás Cabrera 13-15,\\
  Cantoblanco, 28049, Madrid, Spain
}

\vspace*{0.2cm}

\begin{abstract}
\texttt{evortran} is a modern Fortran library designed for
high-performance genetic algorithms and evolutionary
optimization. \texttt{evortran} can be used to tackle
a wide range of problems in high-energy physics
and beyond, such as derivative-free parameter optimization,
complex search taks, parameter scans and fitting experimental
data under the presence of instrumental noise.
The library is built as an \texttt{fpm} package
with flexibility and efficiency in mind, while
also offering a simple installation process, user interface
and integration into existing Fortran (or Python) programs.
\texttt{evortran} offers a variety of selection, crossover,
mutation and elitism strategies, with which users can tailor
an evolutionary algorithm to their specific needs.
\texttt{evortran} supports different abstraction levels:
from operating directly on individuals  and populations,
to running full evolutionary cycles, and even enabling
migration between independently evolving populations to
enhance convergence and maintain diversity.
In this paper, we present the functionality of the
\texttt{evortran} library, demonstrate its capabilities with
example benchmark applications,
and compare its performance with existing genetic
algorithm frameworks.
As physics-motivated applications, we use \texttt{evortran}
to confront extended Higgs sectors with LHC data and to
reconstruct gravitational wave spectra
and the underlying physical parameters from LISA mock data,
demonstrating its effectiveness in realistic, data-driven scenarios.
\end{abstract}

\vspace*{4em}

\end{center}

\renewcommand{\thefootnote}{\arabic{footnote}}
\setcounter{footnote}{0}

\newpage

{
  \hypersetup{linkcolor=black}
  \tableofcontents
}

\section{Introduction}
\label{sec:introduction}

Genetic algorithms~(GAs) are a class of
optimization techniques inspired by
natural selection and evolution.
Unlike gradient-based optimization methods,
which rely on derivative information to
navigate the solution space, GAs explore
the solution space through stochastic processes
that mimic the principles of biological
evolution. A so-called \textit{population}
of candidate solutions, each represented
as an \textit{individual},
is iteratively
evolved over successive generations.
Each individual is represented by a set of
parameters referred to as \textit{genes}, and the
\textit{fitness} of an individual is
a measure of how well it solves the optimization
problem at hand. The fitness is obtained
by computing the so-called fitness function
that depends on the genes of the
individual.\footnote{The fitness function in
GAs does not necessarily depend solely on the
genes of a single individual.
It can also incorporate external parameters that
change over time to simulate environmental changes,
or depend on the genes of a subset or the entire
population to define a relative fitness value.
This allows for more complex selection mechanisms,
such as competitive or cooperative fitness evaluation,
but also increases the number of calls of the
fitness function.
\evortran\ supports such generalized fitness functions,
but for brevity, all examples in this paper will
assume the most common case, where the fitness function
maps the genes of each individual directly to a
fitness value without further dependencies.}
In order to evolve one or a
set of populations,
a GA applies genetic operations that act
on the individuals:
\begin{enumerate}[itemsep=0pt,topsep=4pt]
  \item \textbf{Selection:} The selection step
    involves choosing individuals from the
    population for reproduction based on their fitness,
    with fitter individuals having a higher
    probability of being selected to
    reproduce and pass on their genes.
  \item \textbf{Crossover:} The crossover (also
    called mating) step involves combining
    the genes of two or more parent individuals
    to create one or more offspring individuals,
    with the goal of producing new solutions that
    inherit desirable traits from the parent
    individuals.
  \item \textbf{Mutation:} The mutation step introduces
    random changes to the genes of a subset of
    the offspring individuals, ensuring diversity
    within the population and helping to prevent
    premature convergence.
  \item \textbf{Elitism:} The elitism step involves
    preserving a certain number of the fittest
    individuals from one generation to the next,
    ensuring that the best solutions are kept
    and not lost due to the stochastic nature
    of the previous selection, crossover and
    mutation steps.
\end{enumerate}
In addition to the total number of individuals
contained in the populations,
a specific GA is defined by the precise operations
performed at each of the four steps and the
probabilities assigned to each of these
operations. Different optimization problems often
require vastly different choices for
these operations and probabilities.
It is therefore essential to have a flexible
framework that allows for easy customization
and adaptation to different optimization tasks.

The stochastic-driven approach allows GAs to explore
complex, high-dimensional, non-convex, discontinuous and noisy
solution spaces where traditional methods may struggle
due to premature convergence to local minima or
undefined gradients.
GAs have shown great potential in several areas of
high-energy physics and cosmology.
For instance, in string theory GAs have been used to explore
the vast landscape of viable string vacua with phenomenologically
consistent properties~\cite{Damian:2013dq,
Damian:2013dwa,Blaback:2013qza,Abel:2014xta,Ruehle:2017mzq,
Bull:2018uow,Cole:2019enn,AbdusSalam:2020ywo,CaboBizet:2020cse,
Loges:2021hvn,Berglund:2023ztk}.
In particle physics phenomenology, GAs have been
applied to scan the parameter spaces of
beyond-the-Standard-Model~(BSM) theories in order to identify
regions of parameter space that are compatible with theoretical and
experimental constraints~\cite{Allanach:2004my,Akrami:2009hp,
Camargo-Molina:2017klw,Abel:2018ekz,
Biekotter:2021ovi,Biekotter:2023jld,
Biekotter:2023oen,
Wessen:2024pzm}.
GAs have also played an important role in numerical fitting
tasks, such as in the determination of parton distribution
functions~\cite{NNPDF:2014otw,
Carrazza:2015aoa,Carrazza:2017udv,Moutarde:2019tqa};
in the analyse, reconstruction and classification
of (detector-level) events at
particle accelerators~\cite{Abdullin:2006nu,
Tani:2020dyi,Nowak:2021xmp,
Bonilla:2021ize};
and in the extraction of
neutrino oscillation~\cite{Ohlsson:2001ck,LAGUNA-LBNO:2014cdn,
DUNE:2015lol,Alekou:2022emd,2023PhRvD.108j2002R},
astrophysical~\cite{Luo:2019qbk,Vargas:2021bgb},
and cosmological parameters~\cite{Crowder:2006wh,
Bogdanos:2009ib,Arjona:2020doi,
Arjona:2020axn,Kamerkar:2022dfu,
Alestas:2022gcg,Medel-Esquivel:2023nov}.
These examples are not intended to be exhaustive,
but rather to illustrate the broad applicability of
GAs across both experimental and theoretical
fields of physics.

GAs often require significant computational
costs in terms of evaluations of the fitness function
due to their stochastic nature, and they may 
converge slower to the optimal solution compared
to derivative-based techniques, such as gradient descent,
especially when a high degree of precision is required.
This highlights the need for fast and scalable
implementations of GAs to make them suitable for
large-scale high-dimensional optimization problems.
Several well-established libraries exist already that
provide implementations of GAs across different
programming languages. In Python, popular open-source
packages include
\texttt{DEAP}~\cite{DEAP_JMLR2012},
a flexible evolutionary computation framework, and
\texttt{PyGAD}~\cite{gad2021}
that supports training deep learning models
created with \texttt{Keras}.
GA framworks in C  and C++ which provide
efficient and customizable implementations
are, for instance,
\texttt{CMAES}~\cite{6790628}
\texttt{GAUL}~\cite{gaul}, 
\texttt{EO}~\cite{Keijzer2001} and
\texttt{OpenGA}~\cite{mohammadi2017openga}.
Furthermore, \texttt{PGAPack} is a general-purpose
GA framework written in C 
and using the
Message Passing Interface~(MPI) for parallelization~\cite{pgapack}.
In the Fortran ecosystem, a well-known optimization
library based on GAs is \texttt{Pikaia}~\cite{pikaia1},
which was originally developed for astrophysical
applications and recently has become available
as a modern Fortran package converted
to free-form source and with a new object-oriented
user interface~\cite{pikaiamodern}.

While several GA libraries with varying degree of
flexibility exist, \evortran\ was developed to
provide a modern Fortran-based alternative that
balances performance and ease of use.
Like \texttt{Pikaia}, it is available as a
Fortran package manager~(\texttt{fpm})~\cite{ehlert:hal-03355768,
curcic2021modernfortrantoolingthriving}
package, making the compilation, installation
and integration into other Fortran programs seamless.
Compared to C/C++ libraries, \evortran\ offers
a more user-friendly interface, while still
making use of the increased computational performance
of a low-level programming language.
In addition, its main optimization routines can be
accessed from Python via lightweight wrappers,
allowing seamless integration into Python-based workflows
and outperforming native Python implementations.
One of the key strengths of GAs is their inherently
modular structure, which allows for straightforward and
efficient parallelization, operating on individuals
in parallel across one or more populations.
To take advantage of this property, \evortran\ includes
support for parallel execution using \texttt{OpenMP},
which significantly reduces the runtime on multi-core systems.
This makes \evortran\ and attractive choice for
computationally intensive optimization problems
in scientific computing.
Moreover, \evortran\ is designed with a modular
structure that allows new GA operations for
selection, crossover, mutation and selection to
be easily added, such that the library can naturally
grow and evolve over time.
Finally, \evortran\ stands out due to its flexible
design which allows users to work at different
abstraction levels: (1) the user can operate on
individuals directly, (2) the user can evolve
whole populations, (3) and the user can evolve a
set of populations in parallel with the possibility
of periodically migrating individuals in
between the populations.

The migration approach is particularly
useful for optimization problems where the goals
is not only to find the globally optimal
solution but also to identify multiple diverse solutions
with sufficiently good fitness
(see also Ref.~\cite{Wessen:2024pzm}).
This feature is relevant for a variety of optimization tasks
encountered in high-energy physics.
For example, in studies of the phenomenology of
BSM theories, one often performs
extensive parameter space scans involving models with
a large number of free parameters, while at the same time
different sets of theoretical and experimental requirements
constrain the allowed parameter space
(see, e.g.~Refs.~\cite{Allanach:2004my,Akrami:2009hp,Abel:2018ekz,
Biekotter:2021ovi,
Biekotter:2023jld}).
These scans are constrained by experimental measurements,
which come with their own uncertainties.
In such contexts, the goal of the parameter scan
is not merely to find a single best-fit point,
but rather to identify all distinct regions of
parameter space that satisfy the imposed constraints
within the uncertainties.
The migration-based approach of \texttt{evortran} is especially
well-suited to this task, as independently evolving populations
can converge to different viable regions of the
parameter space.
This enables a more comprehensive exploration and
characterization of phenomenologically acceptable solutions
which might be missed in simple random scans or
when using gradient-based
optimization methods.

The outline of the paper is as follows:
in \cref{sec:structure}, we describe the design
and core functionality of the \evortran\ library
in detail. \cref{sec:instructs} provides user instructions,
including installation steps and usage guidelines.
In \cref{sec:apps}, we present example applications
that illustrate the capabilities of the library,
focusing on well-known benchmark functions for
optimization problems in \cref{sec:benchfuncs} and on
realistic physics applications in \cref{sec:physapps}.
Finally, \cref{sec:conclus} contains our conclusions.
Readers who are already familiar with
GAs and are primarily interested in how to install
and use \evortran\ may wish to skip directly to
\cref{sec:instructs}.

\section{Library design and functionality}
\label{sec:structure}

The design of \evortran\ is centered around flexibility,
modularity, and performance, enabling users to employ
GAs 
using varying degrees of computational resources
depending on the
complexity and requirements of their optimization tasks.
This section outlines the core components and features
of the library, introducing the derived types and
utilities that structure the implementation.
As already mentioned in \cref{sec:introduction},
a key concept underlying \evortran\ is its multi-layered
interface, which allows users to work at different
abstraction levels: (1) work directly on arrays of individuals,
(2) handle entire populations of individuals using
population-level methods, and (3) apply evolution
strategies acting on one or more populations
in parallel through a migration framework,
with optional inter-population exchange of individuals.
This design makes \evortran\ suitable for both
fine-grained control and high-level optimization applications.
The subsections that follow below are organized accordingly.
We begin by introducing the derived types for the
individual and population data structures
in \cref{sec:individuals} and \cref{sec:populations},
respectively, which serve as the most fundamental
building blocks of the library.
We then present the available GA operations
in \cref{sec:operations}, followed by the
two key utilities that implement full population
evolution in \cref{sec:evolution} and migration-based
evolution in \cref{sec:migration}.
Finally, we summarize the core numerical tools
and auxiliary utilities that are included in
\evortran\ in \cref{sec:cores}.

\subsection{Individuals}
\label{sec:individuals}

The most fundamental building block of \evortran\
is the \texttt{individual} derived type,
which represents a single candidate solution within
a GA. \evortran\ contains two types of individuals,
one for integer valued genes and one for genes that
can take continuous values as floating-point numbers.

\subsubsection{Integer individuals}
\label{sec:intinds}

We start by describing the version of individuals
with integer-valued genes, hereafter referred to
as \textit{integer individuals}.
The declaration of the \texttt{individual} type for integer
individuals is as follows:

\begin{center}
\begin{tight}
\begin{lstlisting}[
  label={lst:individual_type},
  linewidth=\textwidth]
type, public :: individual
  integer :: length
  integer :: base_pairs
  integer, dimension(:), allocatable :: genes
  procedure(func_abstract), pointer, private :: func => null()
  real(wp), private :: fitness = 0.0e0_wp
  logical, private :: fitness_calculated = .false.
contains
  procedure, public :: calc_fitness
  procedure, public :: get_fitness
  procedure, public :: reset_fitness
end type individual

interface individual
  procedure create_individual
end interface individual
\end{lstlisting}
\end{tight}
\end{center}
The components of the derived type are as follows:
\begin{itemize}[label=-, topsep=2pt, itemsep=1pt, parsep=0pt]
\item \texttt{length}: The number of genes in the individual.
It is set during initialization and must be greater than 1.
\item \texttt{base\_pairs}: The number of distinct values
that each gene can take. The default is 2,
corresponding to binary genomes where each gene
is either 0 or 1.
\item \texttt{genes}: An allocatable integer array of
size \texttt{length} that stores the genome of the individual.
\item \texttt{func}: A procedure pointer that points to the
fitness function associated with this individual.
The fitness function is set during initialization.
\item \texttt{fitness}: A real number storing the most
recently computed fitness value.
\item \texttt{fitness\_calculated}: A logical flag indicating
whether the current fitness value corresponds to the
current genome. This avoids redundant calls of the
fitness function.
\end{itemize}
To create a new integer individual, users can use
the overloaded interface \texttt{individual}, which wraps the
internal \texttt{create\_individual} function.
This provides a clean and intuitive way to
construct an individual object by specifying the
number of genes, a fitness function, the number of
base pairs, and optionally a seed value for the
genes at initialization.
For example, a typical initialization might
look like:\footnote{For brevity, we do not show the
statements required to initialize the PRNG here and
in the following code snippets contained in this section.
These should be called at the beginning of any
program that uses the \evortran\ library.
For details, see the discussion in \cref{sec:prng}.}
\begin{center}
\begin{tight}
\begin{lstlisting}[
  label={lst:individual_type},
  linewidth=\textwidth]
use evortran__individuals_integer, only : individual

type(individual) :: ind

ind = individual(length=10, fit_func=fit_func, base_pairs=2)
\end{lstlisting}
\end{tight}
\end{center}
This constructs an individual with 10 binary genes
and associates it with the fitness function \texttt{fit\_func}.
Since the optional argument \texttt{seed} is not given,
the gene values are assigned randomly within
the allowed range.
The fitness function must be defined by the user
(either in the main program or in an external module)
following the abstract interface defined as:
\begin{center}
\begin{tight}
\begin{lstlisting}[
  label={lst:individual_type},
  linewidth=\textwidth]
subroutine func_abstract(ind, f)
  class(individual), intent(in) :: ind
  real(wp), intent(out) :: f
end subroutine func_abstract
\end{lstlisting}
\end{tight}
\end{center}
Thus, the fitness function is a subroutine that
takes an individual object as input and computes
the fitness value that is returned as the
second argument.
In contrast to many existing GA frameworks that maximize
the fitness function, \evortran\ is designed to
minimize it, and the fitness function should be
defined accordingly.
The type \texttt{real(wp)} uses the working precision
\texttt{wp} defined in the module \texttt{evortran\_\_util\_kinds}.
By default, \texttt{wp} corresponds to double precision,
but it can be changed to quad precision at compile
time by providing the preprocessor flag \texttt{-DQUAD}.
Further details on compilation and installation are
provided in \cref{sec:instructs}.
The declarations of the procedure associated to the derived type
of integer individuals are as follows:
\begin{center}
\begin{tight}
\begin{lstlisting}[
  label={lst:individual_type},
  linewidth=\textwidth]
subroutine calc_fitness(this)
  class(individual), intent(inout) :: this
  real(wp) :: f
end subroutine calc_fitness

function get_fitness(this) result(f)
  class(individual), intent(inout) :: this
  real(wp) :: f
end function get_fitness

subroutine reset_fitness(this)
  class(individual), intent(inout) :: this
end subroutine reset_fitness
\end{lstlisting}
\end{tight}
\end{center}
These procedures carry out the following tasks:
\begin{itemize}[label=--, topsep=2pt, itemsep=1pt, parsep=0pt]
  \item \texttt{calc\_fitness}: Calculates the fitness with
    the current gene values by calling the fitness function if
    the fitness has not yet been computed.
  \item \texttt{get\_fitness}: Same as \texttt{calc\_fitness},
    but also returns the computed fitness value.
  \item \texttt{reset\_fitness}: Resets the fitness to zero and
    marks it as not yet computed, which is necessary after
    any genetic operation that modifies the genes
    of the individual.
\end{itemize}

\subsubsection{Float individuals}
\label{sec:floatinds}

In addition to supporting individuals with
discrete integer-valued genes, \evortran\ also
provides a derived type for individuals with
genes represented by continuous floating-point
numbers. These \textit{float individuals} are useful
for optimization problems in continuous parameter spaces.
Below is the definition of the derived type used to
represent float individuals:
\begin{center}
\begin{tight}
\begin{lstlisting}[
  label={lst:individual_type},
  linewidth=\textwidth]
type, public :: individual
  integer :: length
  real(wp) :: lower_lim = 0.0e0_wp
  real(wp) :: upper_lim = 1.0e0_wp
  real(wp), dimension(:), allocatable :: genes
  procedure(func_abstract), pointer, private :: func => null()
  real(wp), private :: fitness = 0.0e0_wp
  logical, private :: fitness_calculated = .false.
contains
  procedure, public :: calc_fitness
  procedure, public :: get_fitness
  procedure, public :: reset_fitness
end type individual

interface individual
  procedure create_individual
end interface individual
\end{lstlisting}
\end{tight}
\end{center}
Here, \texttt{wp} is the real kind working precision with
which \evortran\ operates, defined in the module
\texttt{evortran\_util\_kinds}. By default, \evortran\
uses double precision with 15 significant digits, but
the precision can be changed to quadruple-precision
with 30 significant digits at compile time, see
\cref{sec:installation}.
The components of the float individual type are:
\begin{itemize}[label=--, topsep=2pt, itemsep=1pt, parsep=0pt]
  \item \texttt{length}: The number of genes in the individual.
It is set during initialization and must be greater than 1.
  \item \texttt{lower\_lim}: the lower bound of the valid
    range for gene values (default is 0.0).
  \item \texttt{upper\_lim}: the upper bound of the valid
    range for gene values (default is 1.0).
  \item \texttt{genes}: a real-valued array of size
    \texttt{length} that stores the gene values.
  \item \texttt{func}: a procedure pointer to the fitness function.
  \item \texttt{fitness}: the cached fitness value.
  \item \texttt{fitness\_calculated}: logical flag indicating
    whether the fitness has already been computed.
\end{itemize}
Furthermore, the derived type for the float individuals
are associated to the same procedures as present
in the type for integer individuals, which are
described above.
New float individual objects can be created using the
generic interface \texttt{individual}, which internally
calls the \texttt{create\_individual} function, in the same
way as for integer individuals.
A float individual is initialized by providing the gene
length, the fitness function, and optionally the lower and
upper bounds for the gene values, as well as a random seed:
\begin{center}
\begin{tight}
\begin{lstlisting}[
  label={lst:individual_type},
  linewidth=\textwidth]
use evortran__individuals_float, only : individual

type(individual) :: ind

ind = individual(  &
  length=10, fit_func=fit_func,  &
  lower_lim=-1.0e0_wp, upper_lim=1.0e0_wp, seed=0.0e0_wp)
\end{lstlisting}
\end{tight}
\end{center}
This creates a float individual with 10 floating-point
number genes and associates it with the fitness
function \texttt{fit\_func}.
The fitness function must be defined by the user using
the same abstract interface as for
the integer individuals discussed above.
The optional arguments
\texttt{lower\_lim} and \texttt{upper\_lim} enforce
that allowed gene values should lie between -1
and 1, and the optional \texttt{seed}
argument is provided to initialize
all 10 gene values with the value 0.

\subsubsection{Operate on individuals}
\label{sec:oponinds}

Although \texttt{evortran} is primarily designed to carry out
optimization tasks at higher abstraction levels,
through evolving entire populations or even sets of populations,
it is also possible to operate directly on individual objects.
This can be useful for testing, debugging, or for users who
require fine-grained control over their GA.

The following example program demonstrates how to manually
create and operate on float individuals.
It shows the initialization of two individuals,
application of simulated binary crossover
(see \cref{sec:crossover}) to generate
two offspring individuals, and it uses the
shuffle mutation routine (see \cref{sec:mutation})
to modify the genes of one of the two offspring
individuals.
\begin{center}
\begin{tight}
\begin{lstlisting}[
  label={lst:individual_manual},
  linewidth=\textwidth]
program operate_on_individuals

  use evortran__util_kinds, only : wp
  use evortran__individuals_float, only : individual
  use evortran__crossovers_sbx, only : simulated_binary_crossing
  use evortran__mutations_shuffle, only : shuffle_mutate
  use evortran__prng_rand, only : initialize_rands

  implicit none

  type(individual) :: ind1
  type(individual) :: ind2
  type(individual) :: ind3
  type(individual) :: ind4

  call initialize_rands(mode='twister')

  ind1 = individual(6, func)
  ind2 = individual(6, func)

  call simulated_binary_crossing(  &
    ind1, ind2, ind3, ind4, eta_c=30e0_wp, p_c=0.5e0_wp)

  call shuffle_mutate(ind3)

contains

  subroutine func(ind, f)

    class(individual), intent(in) :: ind
    real(wp), intent(out) :: f

    f = sum(ind

  end subroutine func

end program operate_on_individuals
\end{lstlisting}
\end{tight}
\end{center}
Specifically, this example illustrates the following
sequence of operations:
\begin{itemize}[label=-, topsep=2pt, itemsep=1pt, parsep=0pt]
  \item The pseudo-random number generator is initialized
    using the Mersenne Twister algorithm
    (see also \cref{sec:prng})
    by calling \texttt{initialize\_rands}.
  \item Two float individuals \texttt{ind1} and
    \texttt{ind2} are created with 6 genes each.
    They are associated with the fitness function
    \texttt{func}, which in this example computes
    the sum of squares of the gene values.
  \item The procedure \texttt{simulated\_binary\_crossing}
    performs crossover between the parent individuals
    \texttt{ind1} and \texttt{ind2}, producing two
    offspring individuals \texttt{ind3} and \texttt{ind4}.
    The parameters \texttt{eta\_c} and \texttt{p\_c}
    are optional arguments that
    control the shape and probability of the crossover,
    see \cref{sec:crossover} for details.
  \item A shuffle mutation is applied to one of the
    offspring (\texttt{ind3}) using \texttt{shuffle\_mutate},
    see \cref{sec:mutation} for details.
\end{itemize}
While this example provides insight into the inner
workings of the library, it is generally not the
preferred way to use \texttt{evortran} for actual
optimization tasks.
These are better handled at the level of
populations, as discussed in the following.

\subsection{Populations}
\label{sec:populations}

Instead of working with individuals directly, it is usually more
convenient to work with a set of individuals and perform different
steps of a GA on this set as a whole. \evortran\ defines for this
purpose the derived type \texttt{population} both for integer
and float individuals.
This enables users to apply selection, crossover, mutation, and
elitism operations in a modular and efficient way.

\subsubsection{Integer populations}
\label{sec:intpop}

The declaration of the \texttt{population} derived type for
integer individuals is given below:
\begin{center}
\begin{tight}
\begin{lstlisting}[
  label={lst:integer_population},
  linewidth=\textwidth]
type, public :: population
  integer :: popsize
  integer :: gene_length
  integer :: base_pairs
  type(individual), public, dimension(:), allocatable :: inds
  type(individual), public, dimension(:), allocatable :: selection
  type(individual), public, dimension(:), allocatable :: elite
  type(individual), public, dimension(:), allocatable :: offspring
  integer, public, dimension(:), allocatable :: indices_sorted_fitness
contains
  procedure, public :: calc_fitnesses
  procedure, public :: select_individuals
  procedure, public :: select_elite
  procedure, public :: produce_offspring
  procedure, public :: make_population_from_offspring
  procedure, public :: get_fittest_individual
end type population

interface population
  procedure create_population
end interface population
\end{lstlisting}
\end{tight}
\end{center}
The components of the integer popolation type are:
\begin{itemize}[label=-, topsep=2pt, itemsep=1pt, parsep=0pt]
\item \texttt{popsize}: Number of individuals in the population.
\item \texttt{gene\_length}: Length of the gene array for each individual.
\item \texttt{base\_pairs}: Number of possible discrete values each
  gene can take (default is 2).
\item \texttt{inds}: The array of actual individual objects
  in the population.
\item \texttt{selection}: Array storing selected individuals.
\item \texttt{elite}: Array for elite individuals that are preserved
  between generations.
\item \texttt{offspring}: Array holding new individuals generated from
  crossover and mutation.
\item \texttt{indices\_sorted\_fitness}: Indices of individuals sorted
  by fitness.
\end{itemize}
A new integer population object can be initialized using the constructor interface:
\begin{center}
\begin{tight}
\begin{lstlisting}[
  label={lst:integer_population},
  linewidth=\textwidth]
use evortran__populations_integer, only: population

type(population) :: pop

pop = population(  &
  popsize, gene_length, fit_func,  &
  base_pairs, seed, inds)
\end{lstlisting}
\end{tight}
\end{center}
Here, \texttt{popsize} is the number of individuals,
\texttt{gene\_length} the length of the gene array of each
individual, \texttt{fit\_func} the fitness procedure,
\texttt{base\_pairs} (optional) the number of possible gene
values, and \texttt{seed} (optional) sets the initial values of
the genes of all individuals contained in the population.
If \texttt{seed} is not given, the gene values of the initial
populations are randomly generated following a uniorm distribution.
Finally, the optional argument \texttt{inds}
is an array of integer individuals that shall be contained
in the initial population.
The derived type for integer populations contains the
following type-bound procedures:
\begin{center}
\begin{tight}
\begin{lstlisting}[
  label={lst:integer_population},
  linewidth=\textwidth]
subroutine calc_fitnesses(this)
  class(population), intent(inout) :: this
end subroutine

subroutine select_individuals(  &
    this, num, mode, tourn_size, wheele_size)
  class(population), intent(inout) :: this
  integer, intent(in) :: num
  character(len=*), intent(in) :: mode
  integer, intent(in), optional :: tourn_size
  integer, intent(in), optional :: wheele_size
end subroutine

subroutine select_elite(this, num, mode)
  class(population), intent(inout) :: this
  integer, intent(in) :: num
  character(len=*), intent(in) :: mode
end subroutine

subroutine produce_offspring(  &
    this, num, mating, mating_prob, mutate, mutate_prob,  &
    mutate_gene_prob, include_elite, uniform_mating_ratio)
  class(population), intent(inout) :: this
  integer, intent(in) :: num
  character(len=*), intent(in) :: mating
  real(wp), intent(in), optional :: mating_prob
  character(len=*), intent(in), optional :: mutate
  real(wp), intent(in), optional :: mutate_prob
  real(wp), intent(in), optional :: mutate_gene_prob
  logical, intent(in), optional :: include_elite
  real(wp), intent(in), optional :: uniform_mating_ratio
end subroutine

function make_population_from_offspring(  &
    this, fit_func, add_elite) result(pop)
  class(population), intent(inout) :: this
  procedure(func\_abstract) :: fit\_func
  logical, intent(in), optional :: add\_elite
  type(population) :: pop
end function

function get_fittest_individual(this) result(ind)
  class(population), intent(in) :: this
  type(individual) :: ind
end function
\end{lstlisting}
\end{tight}
\end{center}
These procedures carry out the following tasks:
\begin{itemize}[label=--, topsep=2pt, itemsep=1pt, parsep=0pt]
\item \texttt{calc\_fitnesses}:
  Computes the fitness values of all individuals in the
  population by calling each individual's
  \texttt{calc\_fitness} routine.
\item \texttt{select\_individuals}:
  Selects \texttt{num} individuals from the population
  using the method specified by \texttt{mode}.
  Supported selection modes include tournament, rank,
  and roulette wheel selection, see
  \cref{tab:selection_methods}.
  The optional arguments
  \texttt{tourn\_size} and \texttt{wheele\_size}
  control the subset size in tournament and roulette wheel
  selection, respectively.
\item \texttt{select\_elite}:
  Selects an elite of \texttt{num} high-performing
  individuals that can be preserved and reintroduced after
  crossover and mutation to ensure that they
  are maintained in the next generation of the population.
  So far the only \texttt{mode} supported is
  called \texttt{best\_fitness} which includes the individuals
  with the best fitness values in the elite.
  These individuals are stored in the
  type-bound \texttt{elite} array.
\item \texttt{produce\_offspring}:
  Generates an array of offspring individuals
  with size \texttt{num} by applying the specified
  crossover, mutation and elitisim routines.
  These individuals are stored in the
  type-bound \texttt{offspring} array.
  The argument \texttt{mating} selects the crossover method,
  see \cref{tab:crossover_methods},
  while \texttt{mating\_prob} defines the crossover probability.
  The argument \texttt{mutate} selects the mutation methiod,
  see \cref{tab:mutation_methods}, while
  mutation is applied according to a probability
  of \texttt{mutate\_prob}.
  If the mutation method accepts a gene-wise mutation
  probability, this probability can be given
  with the argument \texttt{mutate\_gene\_prob}.
  If \texttt{include\_elite} is set to true,
  the elite individuals are also added to the offspring.
  The \texttt{uniform\_mating\_ratio} is used if uniform
  crossover is selected, see the discussion in \cref{sec:mutation}.
\item \texttt{make\_population\_from\_offspring}:
  Returns an instance of type \texttt{population}
  with the individuals given by the offspring individuals.
  The fitness function \texttt{fit\_func} is associated
  with the new population.
  If \texttt{add\_elite} is set to true,
  the previously selected elite individuals are also added
  to the new population.
\item \texttt{get\_fittest\_individual}:
  Returns the single individual with the highest fitness
  value in the population.
\end{itemize}

\subsubsection{Float populations}
\label{sec:floatpops}

Similar to integer individuals, \texttt{evortran} provides
a \texttt{population} derived type to manage and evolve
collections of float individuals.
These float individuals have real-valued genes and are
suited for optimization problems defined over continuous search
spaces. The population type offers the same high-level
routines for selection, crossover, mutation, and elitism as
for integer populations.
The definition of the \textit{float population} type is shown below:
\begin{center}
\begin{tight}
\begin{lstlisting}[
  label={lst:float_population},
  linewidth=\textwidth]
type, public :: population
  integer :: popsize
  integer :: gene_length
  real(wp) :: lower_lim = 0.0e0_wp
  real(wp) :: upper_lim = 1.0e0_wp
  type(individual), public, dimension(:), allocatable :: inds
  type(individual), public, dimension(:), allocatable :: selection
  type(individual), public, dimension(:), allocatable :: elite
  type(individual), public, dimension(:), allocatable :: offspring
  integer, public, dimension(:), allocatable :: indices_sorted_fitness
contains
  procedure, public :: calc_fitnesses
  procedure, public :: select_individuals
  procedure, public :: select_elite
  procedure, public :: produce_offspring
  procedure, public :: make_population_from_offspring
  procedure, public :: get_fittest_individual
  procedure, public :: get_fittest_individuals
end type population

interface population
  procedure create_population
end interface population
\end{lstlisting}
\end{tight}
\end{center}
In addition to the components contained also in
the integer population type, see the discussion above,
the float population type contains the following
additional components:
\begin{itemize}[label=--, topsep=2pt, itemsep=1pt, parsep=0pt]
\item \texttt{lower\_lim}: a real number defining the lower
bound of possible gene values (default 0.0).
\item \texttt{upper\_lim}: a real number defining the upper
bound of possible gene values (default 1.0).
\end{itemize}
These values are respected when generating initial genes, and when
applying crossover and mutation operators, where the stochastic
nature of these operations might otherwise lead to gene
values outside of the interval defined by
\texttt{lower\_lim} and \texttt{upper\_lim}.

A new float population object can be created
using the same constructor interface as for integer populations:
\begin{center}
\begin{tight}
\begin{lstlisting}[
  label={lst:float_population},
  linewidth=\textwidth]
use evortran__populations_float, only: population

type(population) :: pop

pop = population(  &
  popsize, gene_length, fit_func,  &
  lower_lim, upper_lim, seed)
\end{lstlisting}
\end{tight}
\end{center}
Here, \texttt{lower\_lim} and \text{upper\_lim} (optional)
define the possible range the genes.
All other arguments behave as in the case of integer populations.

Besides the procedures available for integer populations
(see previous subsection), the float population type defines
one additional procedure
called \texttt{get\_fittest\_individuals}.
This routine returns the \texttt{n} fittest individuals in
the population, sorted by fitness in ascending order.

\subsubsection{Operate on populations}
\label{sec:oponpops}

Similar to operating on individuals, as discussed
in \cref{sec:oponinds}, \texttt{evortran} allows to create
and manipulate directly instances of the \texttt{population}
types.
While operating directly on populations is a step up in
abstraction compared to manipulating individual objects,
it is still not the most convenient to use \texttt{evortran}
for real-world optimization tasks.
The library provides higher-level functions which
allow users to evolve populations over multiple generations
and epochs with minimal code, see the discussions
in \cref{sec:evolution} and \cref{sec:migration}.
However, working at the population level offers
insights into the inner workings of these
functions and can be helpful for advanced users who
need more fine-grained control.

The following example demonstrates how to manually evolve
a population of float individuals over one generation,
using built-in selection, crossover, and mutation operations:
\begin{center}
\begin{tight}
\begin{lstlisting}[
  label={lst:population_manual},
  linewidth=\textwidth]
program operate_on_populations

  use evortran__util_kinds, only : wp
  use evortran__populations_float, only: population
  use evortran__individuals_float, only: individual
  use evortran__prng_rand, only : initialize_rands

  implicit none

  type(population) :: pop
  type(population) :: new_pop
  type(individual) :: best

  call initialize_rands(mode='twister')

  pop = population(100, 10, func)

  call pop

  call pop
    100,  &
    mating='blend',  &
    include_elite=.true.,  &
    mutate='uniform')

  new_pop = pop
  best = new_pop

  write(*,*) "Genes of best-fit ind.:  ", best
  write(*,*) "Fitness of best-fit ind.:", best

contains

  pure subroutine func(ind, f)

    class(individual), intent(in) :: ind
    real(wp), intent(out) :: f

    f = sum(ind

  end subroutine func

end program operate_on_populations
\end{lstlisting}
\end{tight}
\end{center}
This example demonstrates a typical evolutionary step operating on a float population:
\begin{itemize}[label=--, topsep=2pt, itemsep=1pt, parsep=0pt]
  \item The random number generator is initialized with the
    Mersenne Twister method.
  \item A population \texttt{pop} of 100 float individuals is
    created, each with 10 genes, and associated with
    the custom fitness function \texttt{func}.
  \item A subset of individuals is selected for reproduction
    using the roulette selection method, which are
    stored in \texttt{pop\%selection}.
    The optional argument \texttt{wheele\_size}
    is propagated to the roulette wheel selection routine
    (see \cref{sec:selection}).
  \item The procedure \texttt{produce\_offspring} is called
    to generate 100 new offspring individuals, which are
    stored in \texttt{pop\%offspring}.
    The arguments set blend crossover and uniform mutation
    as operations to produce the offspring, and the
    optional argument \texttt{include\_elite} is set to true
    to enforce that the fittest individual in the population
    is taken over into the set of offspring individuals.
  \item By calling \texttt{make\_population\_from\_offspring},
    a new population \texttt{new\_pop} is constructed
    from the offspring, using the same fitness function.
  \item The fittest individual of the new population is
    extracted by calling \texttt{get\_fittest\_individual},
    Finally, the gene values and the fitness value of the
    best-fit individual is printed.
\end{itemize}
This example provides a closer look at the operations
\texttt{evortran} offers for manipulating populations manually.
It gives users full control over each step of the
evolutionary process and illustrates how populations are
typically evolved internally when higher-level functions
are used.

\subsection{Genetic algorithm operations}
\label{sec:operations}

A GA progresses through repeated application of four
key operations that mimic the principles of natural
selection and evolution: selection, crossover, mutation,
and elitism. The specific strategies and implementations
of these operations greatly influence the effectiveness
and efficiency of the GA. The \evortran\ library provides
flexible and modular routines for each of these operations,
allowing users to tailor the GA to their demands.
In the following subsections, we describe in detail the
selection, crossover, mutation, and elitism mechanisms
currently implemented in \evortran,
highlighting their underlying principles and
typical use cases. We also comment on the advantages
and disadvantages of the different methods.

\subsubsection{Selection routines}
\label{sec:selection}

The first step in the evolution of a population
from one generation to the other is the
selection step. During the selections step,
samples of the initial population are selected
that are allowed to take part in the subsequent
crossover step. The selection of the individuals
is guided by their fitness values, with the aim
of increasing the likelihood of
producing fitter offspring. To this end,
the selection procedure favors individuals with
good fitness values.
In \cref{tab:selection_methods} we show the selection
routines that are currently implemented in \evortran.
These routines are implemented generically
and can act on both integer and float
individuals.

\begin{table}[t]
  \centering
  \setlength{\tabcolsep}{12pt}
  \renewcommand{\arraystretch}{1.4}
  \begin{tabular}{>{\bfseries}lcl}
     & Kinds & Optional arguments
      [default values] \\
    \hline
    tournament & I/F & \texttt{tourn\_size} [2] \\
    rank & I/F & - \\
    roulette & I/F & \texttt{wheele\_size} [3]
  \end{tabular}
  \renewcommand{\arraystretch}{1.0}
  \caption{Selection methods implemented in \evortran.
    The first column shows the names of the three
    selection methods
    in \evortran.
    The respective functions are labeled in the second
    column as I, F, or I/F, indicating the type of
    genes they operate on: I for individuals with
    integer-valued genes, F for those with
    floating-point genes, and I/F for functions that
    can handle both types. The last column shows
    optional arguments of each routine and their default
    values.}
  \label{tab:selection_methods}
\end{table}

One of the most commonly used selection strategies is
\textit{tournament} selection. In this method,
a subset of individuals is randomly sampled from
the population, and the individual with the highest
fitness within this subset is selected to proceed to
the crossover step. This process is repeated as many
times as needed to construct the mating pool.
The tournament size determines the selection pressure.
Larger tournament sizes increase the chance of
selecting individuals with good fitness, while smaller ones
help preserve diversity. In \evortran, the default
tournament size is set to 2, which provides a good
balance between selection pressure and population
diversity for many applications.

Another selection strategy implemented in \evortran\
is \textit{rank} selection. In this approach,
individuals in the population are first sorted according
to their fitness values, and the selection is based on
their rank rather than their absolute fitness.
This method helps prevent the premature domination of
highly fit individuals and maintains a more balanced
selection pressure, particularly in cases where fitness
values vary widely. In \evortran, rank selection is
implemented in a straightforward way: the size of the
mating pool is specified, and the top-ranking
individuals, i.e.~those with the best fitness values,
are selected until the pool is full. While this ensures
that only the most promising individuals participate in
the crossover step, it also reduces diversity more
aggressively than other selection methods.
Compared to tournament selection, rank selection offers
more deterministic control over selection pressure
but comes with a computational cost.
First, it requires the evaluation of the fitness of
all individuals contained in the population.
Second, a sorting step is required to rank the
individuals, which can become expensive for
large population sizes.
In contrast, tournament selection is more scalable,
as it only operates on small subsets of the population and
avoids global sorting.

The third selection method currently implemented in
\evortran\ is a modified version of
\textit{roulette wheel} selection, designed to
balance randomness and selection pressure in a
computationally more efficient way than the usual
wheel selection. Unlike standard fitness-proportionate
selection over the entire population,
the \evortran\ implementation first randomly selects a
small subset of individuals called the wheel size $k$.
The default value of the wheel size in \evortran\
is $k = 3$. Let the subset consist of individuals
$ i = \{1, 2, \dots, k\}$. Among these, the best
fitness value is determined as
\begin{equation}
f_{\rm best} = \min \{ f_1, f_2, \dots, f_k \} \, ,
\end{equation}
where $f_i$ is the fitness values of the
individual $i$.\footnote{We remind the reader that
\evortran\ minimizes the fitness function.}
Then, for each individual in the subset, a weight
parameter
\begin{equation}
w_i = \exp \left[ \frac{- (f_i - f_{\rm best})^2}
  {f_{\rm best}^2} \right] \, ,
\quad \Rightarrow \quad 0  < w_i \leq 1 \, ,
\quad w_{\rm best} = 1 \, ,
\end{equation}
is computed. The selection probabilities for the
individuals are computed using these weights,
instead of using the fitness values $f_i$ themselves
as in usual roulette wheel selection, via
\begin{equation}
p_i = \frac{w_i}{\sum_{j=1}^k w_j} \, .
\end{equation}
Then a single individual from the subset is selected
based on these probabilities.
The whole operation is repeated until the mating
pool is full.
Compared to tournament selection and rank selection,
this approach can better preserve diversity, with the
diversity generally increasing with increasing
wheel size $k$, while still biasing
toward candidates with better fitness in the population.
Unlike tournament selection, where only
the best-fitted individual in the subset is selected,
here all individuals in the subset have a non-zero probability
$p_i$ of being chosen.
This selection strategy can therefore be viewed as a more
relaxed version of tournament selection (with
the tournament size equal to the wheel size).
Compared to rank selection,
it avoids the need to sort the entire population,
making it computationally less expensive
for large populations.
However, the influence of the wheel size and fitness
distribution can significantly affect its behavior,
requiring carefully choosing the wheel size
$k$ for good performance in specific applications.

\subsubsection{Crossover routines}
\label{sec:crossover}

The crossover step is the process during
which new offspring individuals are created
whose genes are determined by exchanging
and potentially modifying the
genes of two or more parent individuals.
\evortran\ currently offers several crossover
routines that are summarized in
\cref{tab:crossover_methods}.

\begin{table}[t]
  \centering
  \setlength{\tabcolsep}{12pt}
  \renewcommand{\arraystretch}{1.4}
  \begin{tabular}{>{\bfseries}lcccl}
     & Kinds & $N_{\rm par}$ & $N_{\rm off}$ & Optional arguments
      [default values] \\
    \hline
    one-point & I/F & 2 & 2 & - \\
    two-point & I/F & 2 & 2 & - \\
    uniform & I/F & 2 & 2 & \texttt{ratio} [0.5] \\
    blend & F & 2 & 2 & \texttt{alpha} [0.5] \\
    sbx & F & 2 & 2 & \texttt{eta\_c} [1.0],
      \texttt{p\_c} [0.5]
  \end{tabular}
  \renewcommand{\arraystretch}{1.0}
  \caption{Crossover methods implemented in \evortran.
    The first column shows the names of the four
    crossover methods
    in \evortran.
    The respective functions are labeled in the second
    column as I, F, or I/F, indicating the type of
    genes they operate on: I for individuals with
    integer-valued genes, F for those with
    floating-point genes, and I/F for functions that
    can handle both types.
    The third and fourth columns show
    $N_{\rm par}$ and $N_{\rm off}$ which
    are the number of parent individuals and offspring
    individuals, respectively. The last column shows
    optional arguments of each routine and their default
    values.}
  \label{tab:crossover_methods}
\end{table}

The \textit{one-point} crossover routine takes
two parent individuals as input and returns
two offspring individuals. The genes of the offspring
individuals are created by selecting a random
point along the genes of the parent individuals and
exchanging their genes beyond this point.
One-point crossing should be used if one wants to
maintain some structure in the gene pool since it
preserves large segments of the genes of the parent
individuals. However, if the optimization problems
is high-dimensional and complex, one-point crossing
may not introduce sufficient diversity, leading
to premature convergence.
Moreover, it disrupts the positional dependencies
of neighbouring genes, which may be disadvantageous for
optimization problems in which certain genes
must remain next to each other for meaningful
solutions.

The \textit{two-point} crossover routine works
in a very similar way as the one-point crossover
routine. The only difference is that two-point
crossing uses two (instead of one)
randomly chosen points along the genes,
and only the genes between these two points
are swapped in order to create the genes of
two offspring individuals.
Compared to one-point crossing, the two-point
crossing method provides more genetic mixing,
while still preserving large segments of genetic
material. However, it also disrupts genetic
structure by separating neighbouring genes.

The \textit{uniform} crossover routine takes
two parent individuals as input, and their genes are
inherited randomly by two offspring individuals.
Each gene of the offspring individuals is randomly
assigned to be taken over from either the first
or the second parent individual.
Typically, the probability to inherit a gene
from parent A or parent B is set to be equal,
such that both offspring individuals on average
acquire 50\% of their genes from one parent and
50\% from the other. This is also the default
setting in \evortran. By disrupting the genes
of the parent individuals at various positions
along the genes, uniform crossing leads to
excellent diversity in the gene pool.
This makes it well suited for highly complex
optimization tasks because it is less affected
by premature convergence compared to one-
and two-point crossing. However, the highly
disruptive behaviour of uniform crossing
with 50\% exchange probability does not maintain
larger blocks of genes, which may cause
slow convergence.
In such cases, it can be useful to lower
the degree of diversity. This can be achieved
in \evortran\ by changing
the ratio of the genes assigned from
either parent A or parent B
via the optional argument \texttt{ratio}.
The value given for this argument is the
probability for each gene of parent A
to be inherited by the first offspring
individual, and the second offspring
individual inherits the corresponding
gene from parent B. Accordingly, the probability for
the second offspring individual to inherit
a gene from parent A is one minus the
value given for \texttt{ratio}.

The \textit{blend} crossover routine
(also called BLX-$\alpha$) is a crossing procedure
that can only be applied to individuals with
genes consisting of continuous numbers and
not integers. This method generates two offspring
individuals by selecting new gene values
within an extended range between the gene values
of two parent individuals.
With $a_i$ being the genes from parent~A and
$b_i$ the genes from parent~B, the genes
$c_i$ and $d_i$
of the offspring individuals~C and~D
are given, respectively, by randomly
and uniformly selecting a number in the ranges
\begin{equation}
  c_i \in [ a_i - \alpha ( b_i - a_i ), b_i +
    \alpha ( b_i - a_i )] \quad \textrm{and} \quad
  d_i \in [ b_i - \alpha ( a_i - b_i ), a_i +
    \alpha ( a_i - b_i )] \, ,
\end{equation}
for all $i = 1,\dots,N_{g}$.
If the above operation leads to gene values $c_i$
and/or $d_i$ that fall outside of the allowed range
of the genes of the individuals, their values are
clipped to the nearest valid value, i.e.~either the
lower or the upper limit.
The parameter $\alpha$ controls the range
beyond which the genes of the offspring individuals
can extend beyond the ones of the parent
individuals. In many applications it is set
to $\alpha = 0.5$, and this is also the default
setting in \evortran. The presence of this
parameter allows tunable exploration of the
solution space, in contrast to the other
implemented crossover strategies.
However, in many cases a good choice for $\alpha$
can only be obtained on heuristic grounds
by ``trial and error''.
Blend crossing is fundamentally
different from the other
crossover methods due to its continuous nature,
creating gene values for the offspring individuals
that are similar but not identical to the genes
contained in the parent individuals.
This feature makes blend crossing often more
suitable for continuous optimization problems
in which small variations can give rise to
significant improvements in the fitness of
an individual.

Finally, \textit{simulated binary crossover} (usually
abbreviated as SBX) is a widely used operator in
GAs for individuals with floating-point genes.
Inspired by one-point crossover in binary-coded GAs,
SBX simulates a similar effect in continuous search
spaces. For each corresponding gene pair $(a_i,b_i)$
from two parent individuals, a random number
$0 \leq u_i \leq 1$ is drawn from a uniform distribution.
From the random variables $u_i$, spread factor
$\beta_{q,i}$ are computed using the distribution index
$\eta_c$ via
\begin{equation}
\beta_{q,i}(u) =
\begin{cases}
\left( 2u_i \right)^{1/(\eta_c + 1)}, & \text{if }
  u_i \leq 0.5 \\
\left( \frac{1}{2(1 - u_i)} \right)^{1/(\eta_c + 1)}, &
  \text{if } u_i > 0.5
\end{cases} \, .
\end{equation}
Then the corresponding pair of offspring genes
$c_i$ and $d_i$ are computed as
\begin{equation}
\begin{aligned}
c_i &= 0.5 \left[ (1 + \beta_{q,i}) a_i +
  (1 - \beta_{q,i}) b_i \right]\, , \\
d_i &= 0.5 \left[ (1 - \beta_{q,i}) a_i +
  (1 + \beta_{q,i}) b_i \right] \, .
\end{aligned}
\end{equation}
This operation is applied independently to each gene
pair ($i = 1, \dots, N_g$) with a certain gene-wise
crossover probability $p_c$.
For each gene pair, the SBX operation described above is applied
to generate new gene values $c_i$ and $d_i$, and otherwise
(with probability $1 - p_c$) the two offspring individuals
simply inherit the genes $a_i$ or $b_i$, respectively.
As for blend crossover,
if the above operation leads to gene values $c_i$
and/or $d_i$ that fall outside of the allowed range
of the genes of the individuals, their values are
clipped to the lower or upper limit of the allowed range.
In \evortran\ the probability $p_c$ has the default value
$p_c = 0.5$, but its value can be changed
by the user.
The parameter $\eta_c$ controls the distribution of offspring
genes. It has the default value $\eta_c = 1.0$ in
\evortran, but can also be modified by the user.
Smaller values of $\eta_c$ promote broader exploration,
allowing the genes of the offspring to deviate more
significantly from the genes of the parent individuals.
On the contrary, larger values of $\eta_c$ lead to
offspring genes that are centered more closely to the
genes of the parent individuals, which typically yields
faster convergence but less exploration of the solution space.
Compared to the blend crossover method discussed above,
SBX offers a more adjustable and often more efficient
trade-off between exploration (searching broadly
across the solution space) and exploitation (refining
and improving solutions near individuals with high fitness).
However, blend crossover requires fewer computation
steps and only has a single meta parameter $\alpha$.
Therefore, while SBX is often more effective for detailed
local optimization in promising regions of the
solution space, blend crossover is simpler and better suited
for broad, uniform searches across the entire
solution space where less precision is initially required.

\subsubsection{Mutation routines}
\label{sec:mutation}

We discuss here the mutation routines that
\evortran\ provides. Mutation introduces diversity
by randomly modifying gene values,
helping the GA escape local optima and explore
the search space more comprehensively.
The available mutation routines are summarized in
\cref{tab:mutation_methods}.
Each mutation routine operates in-place on a given
individual, i.e.~it directly modifies the genes of
the individual given as input.
After mutation, the fitness of the individual is
invalidated by a call to its internal
\texttt{reset\_fitness()} procedure.
This ensures that the next time the fitness is
accessed, it is correctly recalculated using the
mutated genes.

\begin{table}[t]
  \centering
  \setlength{\tabcolsep}{12pt}
  \renewcommand{\arraystretch}{1.4}
  \begin{tabular}{>{\bfseries}lcl}
     & Kinds & Optional arguments
      [default values] \\
    \hline
    uniform & I/F & \texttt{prob} [1/\texttt{ind\%length}] \\
    shuffle & I/F & \texttt{prob} [1/\texttt{ind\%length}] \\
    gaussian & F & \texttt{prob} [1/\texttt{ind\%length}],
      \texttt{sigma} [1.0]
  \end{tabular}
  \renewcommand{\arraystretch}{1.0}
  \caption{Mutation methods implemented in \evortran.
    The first column shows the names of the three
    mutation methods
    in \evortran.
    The respective functions are labeled in the second
    column as I, F, or I/F, indicating the type of
    genes they operate on: I for individuals with
    integer-valued genes, F for those with
    floating-point genes, and I/F for functions that
    can handle both types.
    The third column shows
    optional arguments of each routine and their default
    values.}
  \label{tab:mutation_methods}
\end{table}

The \textit{uniform} mutation routine can be applied to
both integer and float individuals.
Each gene has an independent probability \texttt{prob}
(defaulting to $1/\texttt{ind\%length}$, where
\texttt{ind\%length} is the length of the genes
of the individual) of being replaced
by a new value sampled uniformly from the full range of
allowable gene values.
A main advantage of uniform mutation is that it introduces
new gene values into the population over the whole possible
range. It is therefore particularly useful in early stages
of the GA, where the solution space should be covered
broadly without converging prematurely into a local minimum.
However, uniform mutation might be too disruptive at final
stages of the GA, when there are solutions that are already
close to the desired (global) optimum, since its coarse
nature may modify gene values far away from suitable values
instead of refining them.

Also compatible with both integer and float genes is
the \textit{shuffle} mutation routine.
With the specified probability \texttt{prob} per gene
(default $1/\texttt{ind\%length}$), each gene switches
the place in the array of genes with another randomly
selected gene. In contrast to the uniform mutation
discussed above, shuffle mutation only modifies
the order of the gene values, but it does not introduce
new gene values which were not present in the population
before. In a GA with a contineous solution space,
it should therefore be combined with
a crossover method that produces new gene values
(e.g.~blend or sbx crossover) instead
of only transferring gene values from parent to offspring
individuals.
Since shuffle mutation keeps the gene values,
only altering their positions,
the mutation is often less disruptive than uniform mutation
and can maintain some beneficial building blocks.
This makes it more suitable for permutation-based optimization
problems, such as scheduling, path finding or ordering tasks.
However, it might lack sufficient exploration power,
and it is usually not suitable for problems in which
the positions of the gene values within the sequence are
highly structured for good solutions to the problem at hand.

Available only for float individuals, \evortran\ offers a third
mutation routine called \textit{gaussian} mutation.
Here, each gene has a chance \texttt{prob} to be replaced by a
normally distributed random number with mean given by the
gene value to be replaced and standard deviation \texttt{sigma}.
This introduces smooth, local variations suitable for
continuous optimization problems.
The default value of \texttt{sigma} is set to 1.0 because the
default range of allowed gene values is from 0.0 to 1.0,
see the discussion in \cref{sec:floatpops}.
If the actual lower and upper limits of possible gene
values differ from this default, it is advisable to adjust
\texttt{sigma} accordingly to ensure that the
spread of mutation remains appropriate relative to the
full range of possible gene values.
Since the mean of the normal distribution is the original
gene value itself,
Gaussian mutation tends to produce values close to the original,
making this method less disruptive than uniform mutation,
where the gene value is replaced with a completely random
value from the entire allowed range.
The argument \texttt{sigma} controls the spread of
the distribution. A larger value increases variability and
allows for more exploratory mutations, whereas a smaller value
makes mutations more conservative, favoring fine-tuning.
Gaussian mutation is especially useful 
if the GA has already
determined solutions near a good solution in order
to explore the local neighborhood efficiently and fine-tune
the final solution.

\subsubsection{Elitism}
\label{sec:elitism}

Elitism is a common strategy in GAs that ensures the
preservation of the best-performing individuals across
generations. Its main purpose is to prevent the loss
of the best solutions at intermediate stages
of the GA due to the stochastic nature of selection,
crossover, and mutation. By retaining a subset of the
best individuals, elitism promotes convergence and often improves
the stability and performance of the GA. However, if the
number of elite individuals is too large, it might lead
to premature convergence and poor coverage of the solution space.

In \evortran, elitism is currently implemented in the most simple
form. Users can specify a number of elite individuals to
be preserved during reproduction. The individuals in the population
that have the lowest fitness values are directly appended
to the offspring.
It is worth noting that this elitism procedure can
become computationally significant for large population sizes
or when fitness evaluations are costly, as it requires computing
the fitness values of all individuals in the population and
sorting them based on those values.

\subsection{Evolution of a population}
\label{sec:evolution}

We have discussed above how to operate on individuals and
populations directly. In principle, a user can defined their
own GA using these functionalities to their specific needs.
However, with a certain optimization task at hand, it is usually
more practical to call a function which takes the fitness function
as input and performs a whole evolution of a population in order
to optimize the fitness function.
To this end, \evortran\ provides the function
\texttt{evolve\_population}, which corresponds to the next level of
abstraction and encapsulates the entire process of evolving a
population through the different stages of a GA.
This function iteratively applies the four fundamental operations
(selection, crossover, mutation, and elitism)
until either a maximum number of generations has been reached
or a predefined fitness target has been achieved.

This is one of the two interfaces that are most likely to be
used by users of \evortran\ (with the other one being
the \texttt{evolve\_migration} function discussed
in \cref{sec:migration}).
It allows users to easily apply GAs to their problems with
minimal boilerplate code. The only requirement is to
implement a user-defined fitness function conforming to the
abstract interface described in \cref{sec:intinds}.
Once the fitness function is defined, optimization is as
simple as calling \texttt{evolve\_population} with the
desired parameters.

The function \texttt{evolve\_population} is highly flexible,
offering a large number of optional arguments that enable
customization of the GA, such as gene initialization,
selection and mating strategies, mutation behavior, elitism,
and output tracking. The whole function declaration is as follow:
\begin{center}
\begin{tight}
\begin{lstlisting}[
  label={lst:evolve_population_decl},
  linewidth=\textwidth]
function evolve_population(  &
  pop_size,  &
  gene_length,  &
  fit_func,  &
  lower_lim,  &
  upper_lim,  &
  max_generations,  &
  fitness_target,  &
  verbose,  &
  gene_seed,  &
  add_ind,  &
  selection,  &
  selection_size,  &
  tourn_size,  &
  wheele_size,  &
  elitism,  &
  elite_size,  &
  mating,  &
  offspring_size,  &
  offspring_include_elite,  &
  mating_prob,  &
  blend_alpha,  &
  sbx_eta_c,  &
  sbx_p_c,  &
  uniform_mating_ratio,  &
  mutate,  &
  mutate_prob,  &
  mutate_gene_prob,  &
  mutate_gaussian_sigma,  &
  fittest_inds_from_gen,  &
  pops_from_gen,  &
  init_pop,  &
  final_pop) result(best_ind)

  integer, intent(in) :: pop_size
  integer, intent(in) :: gene_length
  procedure(func_abstract) :: fit_func
  real(wp), intent(in), optional :: lower_lim
  real(wp), intent(in), optional :: upper_lim
  integer, intent(in), optional :: max_generations
  real(wp), intent(in), optional :: fitness_target
  logical, intent(in), optional :: verbose
  real(wp), intent(in), optional :: gene_seed
  type(individual), intent(in), optional :: add_ind
  character(len=*), intent(in), optional :: selection
  integer, intent(in), optional :: selection_size
  integer, intent(in), optional :: tourn_size
  integer, intent(in), optional :: wheele_size
  character(len=*), intent(in), optional :: elitism
  integer, intent(in), optional :: elite_size
  character(len=*), intent(in), optional :: mating
  integer, intent(in), optional :: offspring_size
  logical, intent(in), optional :: offspring_include_elite
  real(wp), intent(in), optional :: mating_prob
  real(wp), intent(in), optional :: blend_alpha
  real(wp), intent(in), optional :: sbx_eta_c
  real(wp), intent(in), optional :: sbx_p_c
  real(wp), intent(in), optional :: uniform_mating_ratio
  character(len=*), intent(in), optional :: mutate
  real(wp), intent(in), optional :: mutate_prob
  real(wp), intent(in), optional :: mutate_gene_prob
  real(wp), intent(in), optional :: mutate_gaussian_sigma
  type(individual), intent(out), dimension(:),  &
    allocatable, optional :: fittest_inds_from_gen
  type(population), intent(out), dimension(:),  &
    allocatable, optional :: pops_from_gen
  type(population), intent(in), optional :: init_pop
  type(population), intent(out), optional :: final_pop
  type(individual) :: best_ind
\end{lstlisting}
\end{tight}
\end{center}
The arguments to this function are:
\begin{itemize}[label=--, topsep=2pt, itemsep=1pt, parsep=0pt]
  \item \texttt{pop\_size}: Number of individuals in the population.
  \item \texttt{gene\_length}: Number of gene values in each individual. Should be equal to the dimension of the fitness function.
  \item \texttt{fit\_func}: User-defined fitness function following
    the abstract interface \texttt{func\_abstract} given
    in \cref{sec:intinds}.
  \item \texttt{lower\_lim}, \texttt{upper\_lim}: Limits of the
    range of possible gene values (defaults are 0.0 and 1.0).
    These optional arguments can only be set together.
  \item \texttt{max\_generations}: Maximum number of generations to
    run (default is \texttt{pop\_size}).
  \item \texttt{fitness\_target}: Optimization stops early if this
    fitness value is reached.
  \item \texttt{verbose}: If true, prints evolving summary output
    during execution.
  \item \texttt{gene\_seed}: Seed used to initialize gene values
    of all individuals in the initial population.
  \item \texttt{add\_ind}: An individual to insert into the
    initial population.
  \item \texttt{selection}: The selection method, currently
    possible values are \texttt{tournament}, \texttt{rank}
    or \texttt{roulette}, see \cref{sec:selection}
    (default is \texttt{tournament}).
  \item \texttt{selection\_size}: Number of individuals to select
    (default is \texttt{pop\_size}).
  \item \texttt{tourn\_size}: Given as optional argument
    \texttt{tourn\_size} when tournament selection
    is used, see \cref{sec:selection}.
  \item \texttt{wheele\_size}: Given as optional argument
    \texttt{wheele\_size} if roulette wheele selection is
    used, see \cref{sec:selection}.
  \item \texttt{elitism}: Elitism mode, where
    currently only the default option \texttt{best\_fitness}
    is supported, see \cref{sec:elitism}.
  \item \texttt{elite\_size}: Number of elite individuals
    to keep (default is 1).
  \item \texttt{mating}: The crossover method, currently
    possible values are \texttt{one-point}, \texttt{two-point},
    \texttt{uniform} for both integer and float
    populations, and additionally \texttt{blend},
    and \texttt{sbx} only for float populations
    (default is \texttt{one-point}).
  \item \texttt{offspring\_size}: Number of offspring individuals
    to generate (default is \texttt{pop\_size}).
  \item \texttt{offspring\_include\_elite}: Whether to apply
    elitism, see \cref{sec:elitism} (default is \texttt{.true.}).
  \item \texttt{mating\_prob}: The mating probability,
    see \cref{sec:crossover} (default is 0.95).
  \item \texttt{blend\_alpha}: Parameter \texttt{alpha} for
    blend crossover, see \cref{sec:crossover} (default is 0.5).
  \item \texttt{sbx\_eta\_c}: Parameter \texttt{eta\_c} for
    simulated binary crossover, see \cref{sec:crossover}
    (default is 1.0).
  \item \texttt{sbx\_p\_c}: Parameter \texttt{p\_c} for
    simulated binary crossover, see \cref{sec:crossover}
    (default is 0.9).
  \item \texttt{uniform\_mating\_ratio}: Parameter \texttt{ratio}
    for uniform crossover, see \cref{sec:crossover}
    (default is 0.5).
  \item \texttt{mutate}: The mutation method, currently
    possible values are \texttt{uniform} and \texttt{shuffle}
    for both integer and float populations, and additionally
    \texttt{gaussian} only for float populations,
    see \cref{sec:mutation}.
    (default is \texttt{uniform}).
  \item \texttt{mutate\_prob}: Mutation probability
    (default is 0.1).
  \item \texttt{mutate\_gene\_prob}: Probability of mutating
    each gene if individual is mutated
    (default is 0.1).
  \item \texttt{mutate\_gaussian\_sigma}: Parameter
    \texttt{sigma} for \texttt{gaussian} mutation,
    see \cref{sec:mutation} (default is 1.0).
  \item \texttt{fittest\_inds\_from\_gen}: Stores the fittest
    individual from each generation in an array of
    length \texttt{max\_generations} if the final achieved
    fitness value is larger than \texttt{fitness\_target}.
    If the GA terminates because a fitness below
    \texttt{fitness\_target} has been achieved, the length
    of the array will be equal to the number of generations
    that were created up to this point.
  \item \texttt{pops\_from\_gen}: If present, stores the entire
    population at each generation. The user should ensure that
    sufficient memory is available.
  \item \texttt{init\_pop}: Initial population to start from
    instead of ranodmly generating one at the start of the GA.
  \item \texttt{final\_pop}: If present, stores the entire population
    after the GA has terminated.
\end{itemize}
The return object \texttt{best\_ind} of the function
is the individual with the best fitness
that was found during the process. A minimal call of
\texttt{evolve\_population} for a continuous
optimization problem (without specifying the
optional arguments) and printing out the minimal
value of the fitness function that was found, looks like this:
\begin{center}
\begin{tight}
\begin{lstlisting}[
  label={lst:evolve_population_decl},
  linewidth=\textwidth]
use evortran__individuals_float, only : individual
use evortran__evolutions_float, only : evolve_population

type(individual) :: best_ind

best_ind = evolve_population(1000, 20, func)
write(*,*) best_ind
\end{lstlisting}
\end{tight}
\end{center}
Here the function \texttt{func} with 20 arguments
is minimized using a GA with a population size of 1000.

In some applications, it may be beneficial to adapt
the behavior of the GA over time. For example,
starting with more exploratory operations such as
broad or disruptive crossover and mutation strategies,
and gradually transitioning to more fine-grained,
exploitative methods as the search progresses.
\evortran\ supports this type of staged evolution by
allowing multiple chained calls to
\texttt{evolve\_population}, each with different parameters.
Between calls, one can transfer either only the best individual
using the \texttt{add\_ind} argument or reuse the entire
final population from one stage as the initial population for
the next via the \texttt{final\_pop} and \texttt{init\_pop} arguments.
This modular design enables the construction of highly
flexible and dynamic GAs that evolve their strategies over
the course of the optimization process.

\subsection{Migration of populations}
\label{sec:migration}

In addition to the function \texttt{evolve\_population},
\evortran\ provides the function \texttt{evolve\_migration},
which operates on multiple populations simultaneously.
This represents the highest abstraction level in
the user-interface of \evortran.
Specifically, the function \texttt{evolve\_migration} evolves
multiple populations independently over a series of
\textit{epochs}, where each epoch consists of a number
of generations. After each epoch, individuals may migrate
between populations.
This form of co-evolution can help
to maintain genetic diversity by reducing premature convergence,
explore multiple areas of the search space concurrently,
and to increase robustness by yielding several good-fit
individuals (each corresponding to local minima of the
fitness function and potentially sufficiently good solutions
to the problem at hand).
Moreover,  this approach naturally allows for performant
and straightforward parallelization, as each population
can be evolved independently before migration steps are
applied, as is discussed in more detail in
\cref{sec:parallel}.

As in \texttt{evolve\_population}, users can configure the
GA, including selection, crossover, mutation, elitism, and
stopping criteria.
Additional parameters control the migration behavior.
The interface of the function is:
\begin{center}
\begin{tight}
\begin{lstlisting}[
  label={lst:evolve_population_decl},
  linewidth=\textwidth]
function evolve_migration(  &
  pop_number,  &
  epoches,  &
  pop_size,  &
  gene_length,  &
  fit_func,  &
  migration,  &
  migration_size,  &
  migration_order,  &
  lower_lim,  &
  upper_lim,  &
  max_generations,  &
  fitness_target,  &
  verbose,  &
  gene_seed,  &
  selection,  &
  selection_size,  &
  tourn_size,  &
  wheele_size,  &
  elitism,  &
  elite_size,  &
  mating,  &
  mating_prob,  &
  blend_alpha,  &
  sbx_eta_c,  &
  sbx_p_c,  &
  uniform_mating_ratio,  &
  offspring_size,  &
  offspring_include_elite,  &
  mutate,  &
  mutate_prob,  &
  mutate_gene_prob,  &
  mutate_gaussian_sigma,  &
  add_ind,  &
  fittest_inds_final_pops  &
  ) result(best_ind)

  integer, intent(in) :: pop_number
  integer, intent(in) :: epoches
  integer, intent(in) :: pop_size
  integer, intent(in) :: gene_length
  procedure(func_abstract) :: fit_func
  character(len=*), intent(in), optional :: migration
  integer, intent(in), optional :: migration_size
  character(len=*), intent(in), optional :: migration_order
  real(wp), intent(in), optional :: lower_lim
  real(wp), intent(in), optional :: upper_lim
  integer, intent(in), optional :: max_generations
  real(wp), intent(in), optional :: fitness_target
  logical, intent(in), optional :: verbose
  real(wp), intent(in), optional :: gene_seed
  character(len=*), intent(in), optional :: selection
  integer, intent(in), optional :: selection_size
  integer, intent(in), optional :: tourn_size
  integer, intent(in), optional :: wheele_size
  character(len=*), intent(in), optional :: elitism
  integer, intent(in), optional :: elite_size
  character(len=*), intent(in), optional :: mating
  real(wp), intent(in), optional :: mating_prob
  real(wp), intent(in), optional :: blend_alpha
  real(wp), intent(in), optional :: sbx_eta_c
  real(wp), intent(in), optional :: sbx_p_c
  real(wp), intent(in), optional :: uniform_mating_ratio
  integer, intent(in), optional :: offspring_size
  logical, intent(in), optional :: offspring_include_elite
  character(len=*), intent(in), optional :: mutate
  real(wp), intent(in), optional :: mutate_prob
  real(wp), intent(in), optional :: mutate_gene_prob
  real(wp), intent(in), optional :: mutate_gaussian_sigma
  type(individual), intent(in), optional :: add_ind
  type(individual), intent(out), dimension(:), allocatable,  &
      optional :: fittest_inds_final_pops
  type(individual) :: best_ind

end function evolve_migration
\end{lstlisting}
\end{tight}
\end{center}
This function contains the following arguments in addition
to the ones of the function \texttt{evolve\_population}
that were already discussed in \cref{sec:evolution}:
\begin{itemize}[label=--, topsep=2pt, itemsep=1pt, parsep=0pt]
  \item \texttt{pop\_number}: The number of populations evolved
    in parallel.
  \item \texttt{epoches}: The number of epochs.
  \item \texttt{migration}: The type of migration that is
    carried out after each epoche. The currently only
    implemented option is \texttt{rank}, where the best-fit
    individuals are selected to migrate to other populations.
  \item \texttt{migration\_size}: The number of individuals to
    migrate from each population per epoch.
  \item \texttt{migration\_order}: The migration order that
    determines the population to which individuals migrate
    from one to the other. Assuming that the populations
    are labeled by $i = 1, \dots, N$, with $N$ being equal
    to \texttt{migration\_size}, possible options are
    (default is \texttt{random}):
    \begin{itemize}[topsep=2pt, itemsep=1pt, parsep=0pt]
      \item[] \texttt{LR}: The individuals from population $i$
        migrate to population $i+1$, except the individuals
        from population $i = N$ migrate to population $i = 0$.
      \item[] \texttt{RL}: The individuals from population $i$
        migrate to population $i-1$, except the individuals
        from population $i = 1$ migrate to population $i = N$.
      \item[] \texttt{random}: For each population, after
        each epoch, the target
        population is randomly selected. In this case it is
        possible that individuals from different populations
        migrate to the same target population.
    \end{itemize}
  \item \texttt{fittest\_inds\_final\_pops}: Stores the best-fit
    individuals from each population upon completion.
\end{itemize}
The return value \texttt{best\_ind} is the best-fit individual
found across all populations after the GA completes,
either by reaching the fitness target or by reaching the
maximum number of epoches.

\subsection{Parallelization using OpenMP}
\label{sec:parallel}

GAs are particularly well-suited for parallelization due
to their inherently population-based structure.
Since individuals in a population evolve mostly independently
during fitness evaluation, selection, and crossover,
many of the operations in a GA can be efficiently distributed
across multiple CPU cores.\footnote{The suitability of
GAs for parallelization is reduced in special cases
where operations (such as the fitness function or the
selection procedure) depend not only on an individual’s genes,
but also collectively on the genes of multiple or all
individuals in the population.
Such global dependencies introduce synchronization constraints
that make parallel execution less efficient or even impossible.
These cases are not considered in this paper.
In \evortran, users can disable parallelization
if such dependencies are required for a specific application.}
This allows significant speed-up when tackling computationally
intensive optimization problems if a sufficient number of
CPU cores are available.

\evortran\ supports parallel execution on multi-core
CPUs through the \texttt{OpenMP} application programming
interface.
\texttt{OpenMP} is a widely adopted standard for
shared-memory parallel programming in Fortran, C, and C++,
which enables easy parallelization of loops and code
sections through compiler directives that are supported
by many Fortran compilers.
\evortran\ uses \texttt{fpm} as build system, where
\texttt{OpenMP} support is activated by adding it
as a meta-package in the \texttt{fpm.toml} configuration
file, as discussed in more detail in \cref{sec:build}.

The way parallelization is applied in \evortran\
depends on the abstraction level used by the user.
When operating at the level of individual populations
(e.g.~via \texttt{evolve\_population}), \evortran\
parallelizes internal loops over individuals.
For instance, fitness evaluation, selection, and
offspring generation can each be executed in parallel
across individuals within a population.
The user does not need to modify the runtime call to
benefit from this.
At the higher level of abstraction using
\texttt{evolve\_migration}, \evortran\ evolves
multiple populations in parallel. In this case,
the outer loop over populations is parallelized,
which is often more efficient than parallelizing individual
operations, because fewer threads need to be launched
and managed, thus reducing overhead.
When selecting how many CPU cores to use, it can be
beneficial to consider these differences:
With \texttt{evolve\_population}, performance may improve
if the number of individuals in the population is a
multiple of the available CPU cores.
With \texttt{evolve\_migration}, best efficiency is
typically achieved when the number of populations matches
or is a multiple of the number of available cores.

To ensure correctness during parallel execution,
the user must make sure that the custom fitness function
is thread-safe, i.e.~it should not depend on or modify
shared state in a non-synchronized manner.
All internal components of \evortran\ that are involved
in parallel execution, such as random number generation
and sorting routines, are implemented to be thread-safe,
as will be discussed in more detail in
\cref{sec:cores}.
The number of threads used during execution can
be controlled by the user via the \texttt{OMP\_NUM\_THREADS}
environment variable at compile time
(see also the discussion in \cref{sec:installation}).
While this must be set before compilation, this is not a
major limitation in practice, since \evortran\ can be
compiled in just a few seconds.\footnote{\TB{The number
of threads can be modified at runtime using
the \texttt{omp\_lib} module, e.g.~by calling the
subroutine \texttt{omp\_set\_num\_threads}. However, it cannot
be set to a value that exceeds the maximum number of threads
specified at compile time by the \texttt{OMP\_NUM\_THREADS}
environment variable.}}

\TB{The current implementation of \evortran\ focuses
on shared-memory parallelization via \text{OpenMP} directives.
This prioritises portability and simplicity for users working
on typical desktop and single-node workstations, where many
scientific users conduct their computations and optimization tasks.
Here, \texttt{OpenMP} provides efficient scaling across multi-core
CPUs, while being widely supported by modern Fortran compilers.
It should be noted, however, that this approach is inherently limited
to single-node systems.
For large-scale applications (for instance, the reconstruction of
gravitational wave spectra from LISA data, see \cref{sec:lisa}),
future extensions of \evortran\ could benefit from distributed-memory
parallelization using \texttt{MPI}, or from offloading computationally
intensive task to GPUs through \texttt{OpenMP} target direcives
or the \texttt{CUDA} framework developed by NVIDIA, which provides
the \texttt{nvfortran} compiler.
The efficiency of GPU acceleration depends strongly on the form of the
fitness function, where only computationally intensive and side-effect-free
fitness functions are likely to benefit significantly from GPU offloading.
In some applications, the evaluation of the fitnesses of all individuals
contained in a population may be suitable for offloading as a whole
to the GPU, for instance, if the computation across the individuals can
be expressed in terms of matrix operations.
Support for GPU-accelerated GAs in \evortran\ is planned for
future releases.}

\subsection{Core utilities and numerical tools}
\label{sec:cores}

This section discusses essential building blocks that
underpin the operation of GAs, and their implementation
in \evortran. These core utilities include the
pseudo-random number generators~(PRNGs),
which are fundamental to introduce randomness
into the algorithm, and sorting routines, which are
crucial for implementing selection mechanisms based
on fitness values of the individuals.
In addition to these core components,
\evortran\ also includes a numerical interpolation module
that can be valuable for regression or surrogate modeling
tasks.

\subsubsection{Pseudo-random number generation}
\label{sec:prng}

Pseudorandom number generation is a central component in
GAs, as it governs stochastic processes such as initialization,
mutation, selection, and crossover.
In parallelized GAs, the PRNG desirably
is thread-safe to avoid race
conditions and ensure reproducibility.
\evortran\ provides two options for generating random numbers.

The default and recommended PRNG in
\evortran\ is a thread-safe implementation of the
\textit{Mersenne Twister algorithm}~\cite{10.1145/369534.369540,
10.1145/272991.272995}
(specifically, the 64-bit variant MT19937-64).
The implementation contained in \evortran\
is adapted from Ref.~\cite{twistermodern}.
The adaptation integrates it with
\evortran's internal real kind working precision (\texttt{wp})
and modifies it for parallel execution using \texttt{OpenMP}.
Each \texttt{OpenMP} thread is assigned its own instance of
the PRNG and initialized with a unique seed at the start
of the program. This design ensures that
random number generation is thread-safe, and that
the results of the GA are deterministic and reproducible
in parallel execution.
Moreover, the
performance is competitive compared to Fortran
intrinsic PRNGs, making it suitable for
large-scale stochastic sampling.\footnote{The Merseene
Twister algorithm is also used in the modernizded \texttt{fpm}
version of \texttt{Pikaia}~\cite{pikaiamodern}, whereas
the original version uses the ``Minimal standard''
PRNG~\cite{10.1145/63039.63042}.}

Alternatively, \evortran\ offers the option to
generate random integer and float numbers based on the
Fortran intrinsic function \texttt{random\_number()}.
While slightly faster in sequential execution, this option
is not thread-safe, since concurrent calls by multiple
threads result in race conditions.
As a consequence, results are no longer deterministic when
executed in parallel.
This option is useful when strict reproducibility is not
required and performance is paramount.
However, typically the thread-save implementation
of the Mersenne Twister is strongly recommended.

Before using any functionality of the \evortran\ library,
the PRNG must be explicitly initialized.
This is done by calling the subroutine
\texttt{initialize\_rands}. For instance, to use
the Mersenne Twister PRNG, one has to call:
\begin{center}
\begin{tight}
\begin{lstlisting}[
  label={lst:evolve_population_decl},
  linewidth=\textwidth]
use evortran__prng_rand, only : initialize_rands

call initialize_rands(mode='twister', seed=0)
\end{lstlisting}
\end{tight}
\end{center}
The second argument \texttt{seed} is an optional integer argument
that sets an initial seed value for the Mersenne Twister PRNG
(the default value is 0).
Alternatively, to use the Fortran intrinsic PRNG function,
one has to call:
\begin{center}
\begin{tight}
\begin{lstlisting}[
  label={lst:evolve_population_decl},
  linewidth=\textwidth]
call initialize_rands(mode='intrinsic')
\end{lstlisting}
\end{tight}
\end{center}
Using this mode, the optional \texttt{seed} argument
is ignored.
If this routine is not called, the program will result
in a runtime error if \evortran\ was compiled in
\texttt{DEBUG} mode, or in undefined behavior if
compiled without it.
The compilation flags, including enabling or disabling
\texttt{DEBUG} mode, are discussed in detail
in \cref{sec:installation}.

\subsubsection{Sorting methods}
\label{sec:sort}

Sorting individuals based on their fitness values is a
central operation in GAs, especially for ranking-based
selection methods and elitism.
In \evortran, sorting is used internally to determine the
ordering of individuals according to their fitness,
and the library provides two efficient and thread-safe
sorting algorithms that are designed to be called in
parallel using \texttt{OpenMP}.

The first sorting option is a parallel \textit{merge sort} algorithm,
which is the default sorting method in \evortran.
It is based on the merge sort implementation from the
\texttt{orderpack} library~\cite{orderpack}.
The version used in \evortran\ has been adapted from a
parallel implementation provided in a public
github repository~\cite{mergepara}, and modified for
compatibility with the real working precision kind
and data structures used in the library.
The second available option is \textit{quick sort}, implemented using
a recursive algorithm adapted from a publicly available Fortran
quicksort implementation~\cite{quickpara}.
Like the merge sort routine, it is thread-safe, allowing
concurrent execution in \texttt{OpenMP}-parallelized regions
of user code.

Users can explicitly select the sorting algorithm used by
\evortran\ by calling the subroutine \texttt{set\_rank\_method}
with the desired mode:
\begin{center}
\begin{tight}
\begin{lstlisting}[
  label={lst:evolve_population_decl},
  linewidth=\textwidth]
use evortran__sorting_ranking, only : set_rank_method

! Select merge sort
call set_rank_method(mode='merge')

! Or select quicksort
call set_rank_method(mode='quick')
\end{lstlisting}
\end{tight}
\end{center}
In typical use cases within \evortran, both sorting methods
were found to exhibit similar runtime performance.
However, for specific use cases of the GA the user may
wish to specifically chose one of the two options.
Merge sort guarantees $\mathcal{O}(n \log n)$ time
complexity in all cases and is stable (preserves relative ordering
of equal elements), which can be advantageous in some applications.
Quick sort may be faster in practice due to lower
constant factors, but its performance can degrade to
$\mathcal{O}(n^2)$ in the worst case, although this is rare
with randomized pivoting.
Because of its more predictable performance and stability,
merge sort is the default sorting method in \evortran{}.

\subsubsection{Function interpolation}
\label{sec:interp}

While not a core component of GAs, \evortran\ includes a
utility for performing cubic spline interpolation,
which can be valuable in a range of optimization contexts
where functions are approximated via discretization
and a finite number of points.
This functionality is implemented through pure Fortran procedures,
making the routines thread-safe and thus suited for
parallel execution.
In GA applications, spline interpolation can be useful when
the fitness function is based on experimental or computational
data sampled at discrete values. Instead of restricting the
optimization to those discrete points, a spline-based
interpolant can provide a smooth and continuous approximation
of the function, allowing for evaluations at arbitrary points
in the parameter space. Additionally, spline interpolation may
be helpful in hybrid optimization strategies where
gradient-free methods like GA are combined with smooth
approximations to facilitate local refinement or sensitivity
analysis, by enabling smooth representations of otherwise
discrete data.

The following code snippet demonstrates how to use the cubic spline
interpolation feature to approximate the value of a
function between a set of known data points:
\begin{center}
\begin{tight}
\begin{lstlisting}[
  label={lst:evolve_population_decl},
  linewidth=\textwidth]
use evortran__util_kinds, only : wp
use evortran__util_interp_spline, only : spline_construct
use evortran__util_interp_spline, only : spline_getval

integer, paramter :: n = 4 ! Number of known points
real(wp) :: x(n)           ! Array of x-values
real(wp) :: y(n)           ! Corresponding y-values = f(x)
real(wp) :: b(n)           ! Spline coefficient array
real(wp) :: c(n)           ! Spline coefficient array
real(wp) :: d(n)           ! Spline coefficient array
real(wp) :: xi             ! Point at which to evaluate the spline
real(wp) :: yi             ! Interpolated value at xi

! Define interpolation points: f(x) = x^2
x = [1.0e0_wp, 2.0e0_wp, 3.0e0_wp, 4.0e0_wp]
y = [1.0e0_wp, 4.0e0_wp, 9.0e0_wp, 16.0e0_wp]

xi = 2.5e0_wp ! Target x-value for interpolation

! Compute spline coefficients based on input data
call spline_construct(x, y, b, c, d, n)

! Evaluate the spline at xi = 2.5
yi = spline_getval(xi, x, y, b, c, d, n)
\end{lstlisting}
\end{tight}
\end{center}
Specifically, this example interpolates the function
$f(x) = x^2$ using four data points.
After constructing the spline coefficients using
\texttt{spline\_construct}, it evaluates the interpolated value
at a midpoint $x_i = 2.5$ using \texttt{spline\_getval}.

\section{User instructions}
\label{sec:instructs}

In this section we give practical guidance on how to get
started with \evortran. In \cref{sec:installation},
we explain how to install the library and its dependencies.
In \cref{sec:usage} we introduce the key functionalities of
\evortran\ and demonstrates how to develop a custom GA.

\subsection{Installation}
\label{sec:installation}

To use \evortran, a few prerequisites must be installed
on the system. These are standard tools in modern
Fortran development and are typically easy
to set up on most platforms.

\subsubsection{Prerequisities}
\label{sec:prereqs}

\evortran\ is written in modern Fortran using an
object-oriented user interface.
The recommended compiler to build \evortran\
is the GNU Fortran compiler \texttt{gfortran}.\footnote{While other
modern Fortran compilers exist, they are either not yet
compatible with key features used in \evortran{} or lack
integration with the Fortran Package Manager.
\texttt{lfortran}, currenly in alpha stage and
under active development, still lacks full support for
\texttt{OpenMP}. Moreover,
LLVM \texttt{flang} is not yet widely
supported by \texttt{fpm} (as of version 0.12.0).}
On Ubuntu or Debian-based systems \texttt{gfortran}
can be installed with:
\begin{center}
\begin{tight}
\begin{lstlisting}[
  language=bash,
  linewidth=\textwidth]
sudo apt update
sudo apt install gfortran
\end{lstlisting}
\end{tight}
\end{center}
\evortran\ was developed and tested using the
versions 11, 12 and 13 of \texttt{gfortran}.

For the build process and for managing dependencies,
\evortran\ uses the Fortran Package Manager
\texttt{fpm}~\cite{ehlert:hal-03355768}.
There are various ways to install \texttt{fpm},
for instance, using package managers like \texttt{pip},
\begin{center}
\begin{tight}
\begin{lstlisting}[
  language=bash,
  linewidth=\textwidth]
pip install fpm
\end{lstlisting}
\end{tight}
\end{center}
or conda,
\begin{center}
\begin{tight}
\begin{lstlisting}[
  language=bash,
  linewidth=\textwidth]
conda config --add channels conda-forge
conda create -n fpm fpm
conda activate fpm
\end{lstlisting}
\end{tight}
\end{center}
One can also install \texttt{fpm} by downloading
a binary for the latest stable release which
are available for Windows, MacOS, and
Linux, or build fpm from source, see Ref.~\cite{fpm}.

\subsubsection{Building evortran}
\label{sec:build}

With \texttt{gfortran} and \texttt{fpm} installed,
one can build \evortran.
The first step is to clone the repository and to
navitage to the \evortran\ directory:
\begin{center}
\begin{tight}
\begin{lstlisting}[
language=bash,
linewidth=\textwidth]
git clone https://gitlab.com/thomas.biekoetter/evortran
cd evortran
\end{lstlisting}
\end{tight}
\end{center}
Then one has to
source \texttt{fpm} environment
variables which define compiler flags and the number
of \texttt{OpenMP} threads that should be used.
This can be done by sourcing one of two provided scripts.
Running
\begin{center}
\begin{tight}
\begin{lstlisting}[
  language=bash,
  linewidth=\textwidth]
source exports_debug.sh
\end{lstlisting}
\end{tight}
\end{center}
sets compiler flags appropriate for debugging,
including options that enable stricter compile-time checks.
More importantly, it activates a wide range of runtime argument
checks specific to \evortran\ that help ensure correct
usage of the library.
These checks are only available in debug mode and are
strongly recommended while setting up or developing the GA.
For performance runs, one should instead run
\begin{center}
\begin{tight}
\begin{lstlisting}[
  language=bash,
  linewidth=\textwidth]
source exports_run.sh
\end{lstlisting}
\end{tight}
\end{center}
which disables these runtime checks through preprocessor
directives, activates compiler optimization
(\texttt{-O3 -march=native}) and omits
array out-of-bounds checking.
This results in significantly faster execution but
should only be used once the implementation is
known to be correct.
In both cases, the number of threads used for parallel
regions is controlled by the \texttt{OMP\_NUM\_THREADS}
environment variable, which is set to 8 by default in
both files but can be modified as needed.

As was already mentioned in \cref{sec:floatinds},
by default the real kind working precision
\texttt{wp} corresponds to double precision,
as defined in the module
\texttt{evortran\_util\_kinds} using
\texttt{selected\_real\_kind(15, 307)}.
This corresponds to a precision of at least 15
significant digits.
The precision can be upgraded to quadruple precision,
using \texttt{selected\_real\_kind(30, 4931)},
by setting the \texttt{QUAD} preprocessor flag.
In this case, \evortran\ operates with at least 30
significant digits.
The real kind precision can be changed at compile
time by appending \texttt{-DQUAD} to the
\texttt{FPM\_FFLAGS} environment variable, for example:
\begin{center}
\begin{tight}
\begin{lstlisting}[
language=bash,
linewidth=\textwidth]
export FPM_FFLAGS="$FPM_FFLAGS -DQUAD"
\end{lstlisting}
\end{tight}
\end{center}
Alternatively, the user can add the \texttt{-DQUAD}
flag directly to the \texttt{exports\_debug.sh} or
\texttt{exports\_run.sh} files,
where \texttt{FPM\_FFLAGS} is defined.
These files should be sourced before compilation to
ensure the correct build configuration is used,
as discussed above.

Once the environment variables are sourced,
one can compile the project with:
\begin{center}
\begin{tight}
\begin{lstlisting}[
language=bash,
linewidth=\textwidth]
fpm build
\end{lstlisting}
\end{tight}
\end{center}
The build process is handled by \texttt{fpm},
which automatically downloads and compiles all required
dependencies. These include the meta-dependency
\texttt{openmp}, which enables multithreaded parallelization,
and a Fortran error-handling module:
\begin{center}
\begin{tight}
\begin{lstlisting}[
language=bash,
linewidth=\linewidth]
[dependencies]
openmp = "\*"
error-handling = { git = "https://github.com/SINTEF/fortran-error-handling.git", tag = "v0.2.0" }
\end{lstlisting}
\end{tight}
\end{center}
In addition, \texttt{dev-dependencies} are specified in the
\texttt{fpm.toml} file which are used in example applications and
test programs:
\begin{center}
\begin{tight}
\begin{lstlisting}[
language=bash,
linewidth=\linewidth]
[dev-dependencies]
csv-fortran = { git = "https://github.com/jacobwilliams/csv-fortran.git" }
pikaia = { git = "https://github.com/jacobwilliams/pikaia.git", tag = "2.0.0" }
\end{lstlisting}
\end{tight}
\end{center}
The package \texttt{pikaia} is another GA framework implemented
in Fortran~\cite{pikaia1,pikaiamodern}.
It is included as a \texttt{dev-dependency} because
it is used for comparisons with \evortran\ in example applications,
see \cref{sec:Michalewicz} and \cref{sec:dropwave}.
The package \texttt{csv-fortran} is used for exporting data
to csv-files.

To use \evortran\ as a dependency in another \texttt{fpm}
project, one can simply add the following entry to the
\texttt{fpm.toml} file of the project:
\begin{center}
\begin{tight}
\begin{lstlisting}[
language=bash,
linewidth=\textwidth]
[dependencies]
evortran = { git = "https://gitlab.com/thomas.biekoetter/evortran" }
\end{lstlisting}
\end{tight}
\end{center}
This makes it straightforward to integrate \evortran\
into existing programs and libraries.

After building \evortran\, one can install the compiled
executables that are contained in the \texttt{app} folder
locally by running:
\begin{center}
\begin{tight}
\begin{lstlisting}[
language=bash,
linewidth=\textwidth]
fpm install
\end{lstlisting}
\end{tight}
\end{center}
This command copies the built executables to a default
location, typically \texttt{\textasciitilde/.local/bin}, and the compiled module
files are copied to \texttt{\textasciitilde/.local/include}.
If these paths are included in the \texttt{PATH}
environment variable, one can run the executable from
any directory without having to reference the full build path.
If needed, you can change the installation prefix using
the \texttt{--prefix} flag.

\subsubsection{Running tests and applications}

evortran includes a set of test programs to verify
that the library is working correctly.
Tu run a specific test program, use the fpm test
command followed by the name of the test program:
\begin{center}
\begin{tight}
\begin{lstlisting}[
language=bash,
linewidth=\textwidth]
fpm test <name_of_test_program>
\end{lstlisting}
\end{tight}
\end{center}
This will execute the selected test and output the
results to the terminal. To view the available
test programs, navigate to the \texttt{test} directory
and its sub-directories.

Similarly, evortran provides example application programs
located in the \texttt{app} directory, which demonstrate how
to use \evortran\ in practical scenarios.
To run a specific application, one can use the \texttt{fpm run}
command followed by the name of the application program:
\begin{center}
\begin{tight}
\begin{lstlisting}[
language=bash,
linewidth=\textwidth]
fpm run <name_of_app_program>
\end{lstlisting}
\end{tight}
\end{center}
This will compile and execute the chosen application,
displaying its output in the terminal. One can explore
the available application programs by browsing the
\texttt{app} directory and its sub-folders.
These applications include several of the examples
discussed in \cref{sec:apps}. These are intended to
be a convenient starting point for understanding the library
and for developing new programs that use \evortran.

\subsection{Quick start: basic usage and main features}
\label{sec:usage}

To help new users quickly get started with \evortran,
this section presents a minimal yet complete example that
demonstrates how to use the library to solve an
optimization problem. The example program \texttt{quick\_start}
is located in the \texttt{app} directory and can be
executed using the following command:
\begin{center}
\begin{tight}
\begin{lstlisting}[
language=bash,
linewidth=\textwidth]
fpm run quick_start
\end{lstlisting}
\end{tight}
\end{center}
This program uses a GA to find the global minimum of the
well-known \textit{Rosenbrock function}~\cite{rosenbrock},
which is commonly used as a benchmark for optimization algorithms.
The Rosenbrock function is defined as
\begin{equation}
f(x, y) = (a - x)^2 + b(y - x^2)^2,
\label{eq:rosenbrock}
\end{equation}
where $a = 1$ and $b = 100$ are parameters that define the
shape of the function.
The function has a global minimum at $(x, y) = (a,a^2) = (1, 1)$,
where $f(a, a^2) = 0$.
The Rosenbrock function is a suitable test case for optimzation
algorithms, as it poses challenges related to convergence to
the global minimum since the minimum is located inside a long, narrow,
parabolic-shaped flat valley.
The source code for this example is the following:
\begin{center}
\begin{tight}
\begin{lstlisting}[
linewidth=\textwidth]
program quick_start

  use evortran__util_kinds, only : wp
  use evortran__individuals_float, only : individual
  use evortran__evolutions_float, only : evolve_population
  use evortran__prng_rand, only : initialize_rands

  implicit none

  type(individual) :: best_ind
  real(wp) :: x
  real(wp) :: y

  real(wp), parameter :: a = 1.0e0_wp
  real(wp), parameter :: b = 1.0e2_wp
  real(wp), parameter :: xmin = -2.0e0_wp
  real(wp), parameter :: xmax = 2.0e0_wp
  real(wp), parameter :: ymin = -1.0e0_wp
  real(wp), parameter :: ymax = 3.0e0_wp

  call initialize_rands(mode='twister')

  best_ind = evolve_population(  &
    100, 2, rosenbrock,  &
    mating='blend',  &
    elite_size=1,  &
    fitness_target=1.0e-10_wp,  &
    mutate='gaussian',  &
    mutate_prob=0.5e0_wp,  &
    mutate_gene_prob=0.5e0_wp,  &
    mutate_gaussian_sigma=1.0e-3_wp)

  call get_xy(best_ind

  write(*,*) 'Rosenbrock function:'
  write(*,*) '  Minimum at x, y =', x, y
  write(*,*) '  f(x,y) =', best_ind

contains

  pure subroutine rosenbrock(ind, f)

    class(individual), intent(in) :: ind
    real(wp), intent(out) :: f

    real(wp) :: x
    real(wp) :: y

    call get_xy(ind

    f = (a - x)**2 + b * (y - x**2)**2

  end subroutine rosenbrock

  pure subroutine get_xy(genes, x, y)

    real(wp), intent(in) :: genes(2)
    real(wp), intent(out) :: x
    real(wp), intent(out) :: y

    x = xmin + genes(1) * (xmax - xmin)
    y = ymin + genes(2) * (ymax - ymin)

  end subroutine get_xy

end program quick_start
\end{lstlisting}
\end{tight}
\end{center}
The program illustrates the key components needed to
define and solve an optimization problem using \evortran:
\begin{itemize}[label=--, topsep=2pt, itemsep=1pt, parsep=0pt]
\item The program starts by importing the relevant modules and
  defining parameters for the Rosenbrock function and the domain
  of the variables $x$ and $y$.
\item The pseudo-random number generator is initialized with
  the Mersenne Twister algorithm.
\item The call to \texttt{evolve\_population} is the main entry
  point for the optimization. Here, we define:
  \begin{itemize}[label=--, topsep=2pt, itemsep=1pt, parsep=0pt]
  \item A population size of 100 and gene length 2
    (for variables $x$ and $y$),
  \item The objective function \texttt{rosenbrock} as the
    fitness function,
  \item Blend crossover as mating method, and Gaussian mutation,
  \item Elitism with one elite individual per generation,
  \item A termination criterion based on reaching a fitness value
    below $10^{-10}$.
  \end{itemize}
  \item The fitness function is defined as a pure subroutine
    taking an individual and returning the value of the
    Rosenbrock function.\footnote{Recall that, unlike many
    other GA frameworks which maximize the fitness function,
    \evortran\ minimizes it, such that the Rosenbrock function
    is implemented as shown in \cref{eq:rosenbrock}.}
  \item The helper subroutine \texttt{get\_xy} rescales the genes
    values from the normalized interval $[0,1]$ to the
    domain $-2 \leq x \leq 2$ and $-1 \leq y \leq 3$ in which
    the global minimum should be determined by the algorithm.
  \item Finally, the location of the found minimum and the
    corresponding value of the Rosenbrock function is printed.
\end{itemize}
Executing this example yields the output:
\begin{center}
\begin{tight}
\begin{lstlisting}[
language=bash,
deletekeywords={function},
linewidth=\textwidth]
 Rosenbrock function:
   Minimum at x, y =  0.99996858723522219       0.99993706186036069
   f(x,y) =   9.8805221531614218E-010
\end{lstlisting}
\end{tight}
\end{center}
This demonstrates that \texttt{evortran} successfully locates
the global minimum of the Rosenbrock function with high accuracy.

Users can adapt this minimal example to define and optimize
their own objective functions by simply modifying the fitness
function and adjusting the meta-parameters of the GA accordingly.
Instead of defining the fitness function inside of the main
program after the \texttt{contains} statement, one can also
define the fitness function in a separate module and import
the function at the beginning of the program.

\subsection{Using evortran from Python}
\label{sec:python}

To increase accessibility and broaden the potential user base,
a Python interface to the core optimization routines of \evortran\
is provided through a separate Fortran \texttt{fpm} package called
\texttt{pyevortran}. This interface allows users to take advantage
of the performance of Fortran-based GAs while working
within Python environments.

The \texttt{pyevortran} package is a minimal Fortran project that
uses \evortran\ as a dependency. It contains only two modules:
the module \texttt{pyevortran\_\_evolutions\_float}
provides a C-binding wrapper for the
\texttt{evolve\_population} routine, and
the module \texttt{pyevortran\_\_migrations\_float}
provides a C-binding wrapper for the
\texttt{evolve\_migration} routine.
So far only the routines acting on float populations are implemented.
The two routines are compiled into a shared library which is then used
by the Python package \texttt{pyevortran} that is installed using \texttt{pip}.
Within this Python package, the shared library is accessed using
Python’s \texttt{ctypes} module to expose the two functions
\texttt{evolve\_population} and \texttt{evolve\_migration}.
These functions can be called from Python to optimize arbitrary Python
functions of the form \texttt{f(x)}, where \texttt{x}
is a \texttt{NumPy} array.

The Python interface currently supports only the two
mentioned high-level optimization routines. Nevertheless,
these routines allow the use of a variety of GA configurations
by specifying optional arguments, just as in the native
Fortran implementation.
Another limitation to note is performance: when using the
Python interface, the fitness function is typically written
in Python and may run significantly slower than a compiled
Fortran equivalent. If the evaluation of the fitness function
dominates runtime of the GA, it may be worthwhile to
translate the function into Fortran and use the
native \evortran\ package for optimal performance.

For installation, the \texttt{pyevortran}
interface requires a Python v.3 installation,
the \texttt{gfortran} Fortran compiler,
and \texttt{fpm} version~$\geq$~0.12.0,
as earlier versions do not support building shared libraries.
The version of \texttt{fpm} currently distributed via PyPI is
outdated (currently version 0.10.0) and will not work.
To install a suitable version of \texttt{fpm},
one can download a recent
binary from the GitHub releases page~\cite{fpm},
make it executable,
and move it to a directory in the system’s \texttt{PATH}, e.g.:
\begin{center}
\begin{tight}
\begin{lstlisting}[
language=bash,
deletekeywords={local},
linewidth=\textwidth]
chmod +x fpm
mv fpm ~/.local/bin/
\end{lstlisting}
\end{tight}
\end{center}
In addition, \texttt{patchelf} must be installed to embed
runtime library paths (RPATH) into the shared libraries,
ensuring that Python can find them at runtime without needing
to manually set \texttt{LD\_LIBRARY\_PATH}.
On Debian-based systems, \texttt{patchelf} can be installed with:
\begin{center}
\begin{tight}
\begin{lstlisting}[
language=bash,
linewidth=\textwidth]
sudo apt install patchelf
\end{lstlisting}
\end{tight}
\end{center}

With the above mentioned prerequisites installed, one
can proceed to install \texttt{pyevortran}.
The first step is to clone the repository and
navigate to its main folder:
\begin{center}
\begin{tight}
\begin{lstlisting}[
language=bash,
linewidth=\textwidth]
git clone https://gitlab.com/thomas.biekoetter/pyevortran
cd pyevortran
\end{lstlisting}
\end{tight}
\end{center}
Then one must source
one of two provided environment setup scripts.
As for \evortran, see the discussion in \cref{sec:build},
the script \texttt{exports\_debug.sh} is intended for
development and enables runtime checks and assertions,
while \texttt{exports.sh} is optimized for production use
by enabling compiler optimizations and disabling checks.
These scripts also set up the number of \texttt{OpenMP} threads.
To use the debug version, run:
\begin{center}
\begin{tight}
\begin{lstlisting}[
language=bash,
linewidth=\textwidth]
source exports_debug.sh
\end{lstlisting}
\end{tight}
\end{center}
Alternatively, to use the optimized configuration, run:
\begin{center}
\begin{tight}
\begin{lstlisting}[
language=bash,
linewidth=\textwidth]
source exports.sh
\end{lstlisting}
\end{tight}
\end{center}
Once the environment is configured,
\texttt{pyevortran} can be built and installed with:\footnote{\TB{Updating
Python packaging tools may help
prevent potential installation issues: \texttt{python -m pip install --upgrade
pip setuptools wheel}.}}
\begin{center}
\begin{tight}
\begin{lstlisting}[
language=bash,
linewidth=\textwidth]
make pyevortran
\end{lstlisting}
\end{tight}
\end{center}
The makefile invokes \texttt{fpm} to build the shared library,
uses \texttt{pip} to install the Python wrapper into the
current environment, and finally applies \texttt{patchelf}
to embed the necessary RPATHs into the shared libraries.
After successful installation, users can import
\texttt{pyevortran} in Python and use the
\texttt{evolve\_population} and \texttt{evolve\_migration}
functions to apply GAs to their own objective functions.

To test the installation, and to demonstrate the
usage of \texttt{pyevortran}, a simple test program
\texttt{test\_rastrigin.py} is provided in the
\texttt{python/test} folder:
\begin{center}
\begin{tight}
\begin{lstlisting}[language=Python, linewidth=\textwidth]
import numpy as np
from pyevortran.evolutions import evolve_population

A = 10 
def rastrigin(x):
    return A * len(x) + np.sum(x**2 - A * np.cos(2 * np.pi * x))

print()
print("Minimizing Rastrigin function:")
print()
for dim in range(2, 11):
    print("  Number of dimensions =", dim)
    xmin = evolve_population(
        rastrigin, dim,
        pop_size=1000,
        lower_lim=-5.12, upper_lim=5.12,
        max_generations=1000,
        fitness_target=1e-9, verbose=False,
        selection="rank", selection_size=100,
        elite_size=100)
    print("    Minimum at xmin =", xmin)
    print("            f(xmin) =", rastrigin(xmin))
    print()
\end{lstlisting}
\end{tight}
\end{center}
This script demonstrates the
use of the \texttt{evolve\_population} function from
the Python interface to minimize the Rastrigin function,
see \cref{sec:rastrigin} for details. The script performs
minimization for dimensions ranging from two to ten
and prints both the location of the minima and
the corresponding function values.
We note that only the first two arguments,
the fitness function (\texttt{rastrigin}) and the
dimensionality of the problem (\texttt{dim}), are required.
All other arguments are optional and allow the user to
specify GA parameters such as population size,
selection strategy, elite size, and termination criteria.

\section{Example applications}
\label{sec:apps}

To demonstrate the flexibility and performance of
\evortran, this section presents several example applications
ranging from standard mathematical benchmarks to
physics-motivated use cases. In \cref{sec:benchfuncs}
we discuss the minimization of common multimodal test functions
that are frequently used
to evaluate optimization algorithms.
Next, in \cref{sec:physapps} we showcase two realistic
applications from theoretical physics. First, we present
a fit of a beyond-the-Standard-Model~(BSM)
theory with an extended Higgs sector
against data from the Large Hadron Collder~(LHC)
in \cref{sec:lhc}.
Afterwords, we show how \evortran\ can be used
to reconstruct a stochastic gravitational
wave background predicted by a cosmological
phase transition in the early Universe
from mock data of the Laser Interferometer Space
Antenne~(LISA) in \cref{sec:lisa}.

\subsection{Minimizing multimodal benchmark functions}
\label{sec:benchfuncs}

Multimodal test functions are commonly used in the
evaluation and benchmarking of optimization algorithms,
particularly GAs, due to their complex landscapes characterized
by multiple local minima. Successfully identifying the
global minimum of such functions is a core strengths of GAs.
In this subsection, we use \evortran\ to
minimize several standard
multimodal functions to demonstrate its performance and to
serve as templates that can be adapted by the user to
new optimization problems.

\subsubsection{Rastrigin function}
\label{sec:rastrigin}

\begin{figure}[t]
\centering
\includegraphics[height=6.4cm]{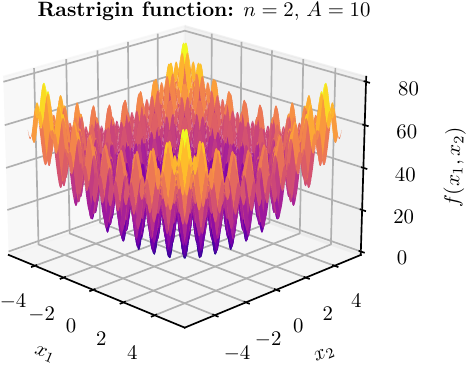}~~
\includegraphics[height=6.6cm]{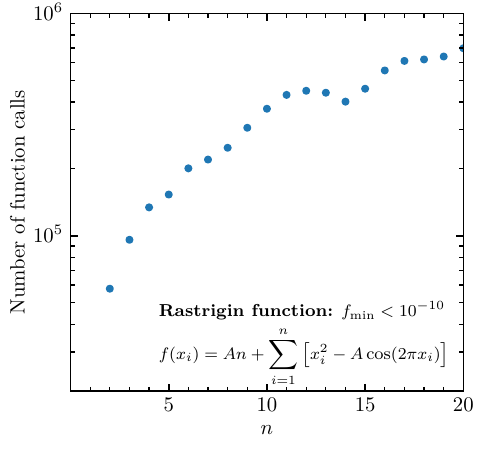}
\caption{Left: Surface plot of the $n=2$ dimensional
Rastrigin function 
with $A = 10$ in the domain $x_{1,2} \in [-5.12, 5.12]$.
Right: Number of calls to the Rastrigin function
against the number of dimensions $n$ until the global
minimum has been determined with a function value
$f_{\rm min} < 10^{-10}$ using the example application
\texttt{benchmark\_rastrigin}.}
\label{fig:rastrigin}
\end{figure}

The Rastrigin function~\cite{1570854175361395200,
10.1007/BFb0029787,MUHLENBEIN1991619} is a well-known and widely used
multimodal benchmark function in global optimization.
It is particularly challenging for optimization algorithms
due to its highly repetitive landscape filled with many
local minima. This makes it an excellent test function for
GAs to locate the globally best solution under the presence of
many sub-optimal candidate solutions.
The Rastrigin function is defined as
\begin{equation}
f(x_i) = A \cdot n + \sum_{i=1}^{n}
  \left[x_i^2 - A \cdot \cos(2 \pi x_i)\right] \, ,
\end{equation}
where $A = 10$ and $x_i = x_1, \dots, x_n$,
with $n$ being the number of dimensions.
The global minimum is located at $x_i = 0$ for all $i$,
where the function reaches the minimum value
$f(x_i = 0) = 0$.
The Rastrigin function is typically minimized in the
domain $x_i \in [-5.12,5.12]$.
The left plot of \cref{fig:rastrigin}
shows a surface plot of the Rastrigin function
in two dimensions.

To demonstrate the capabilities of \evortran,
we performed a series of optimization runs using the
\texttt{benchmark\_rastrigin} example program included in
the \texttt{app} folder of the repository.
In this benchmark, the goal is to find the global minimum with
a precision of $f_{\rm min} < 10^{-10}$ for dimensions
ranging from $n = 2$ to $n = 20$.
The GA is executed for each dimensionality separately, and the
number of fitness function evaluations until convergence,
as well as the final minimum value of the function found,
are stored in a CSV file.
The core of the \texttt{benchmark\_rastrigin} program
consists of a loop over the number of dimensions
(\texttt{currdim}) in which the \texttt{evolve\_population}
function is called:
\begin{center}
\begin{tight}
\begin{lstlisting}[
language=Fortran,
linewidth=\textwidth]
best_ind = evolve_population(  &
  10000, currdim, rastrigin,  &
  mating='blend',  &
  elite_size=100,  &
  lower_lim=-5.12e0_wp,  &
  upper_lim=5.12e0_wp,  &
  selection='rank',  &
  selection_size=100,  &
  fitness_target=1.0e-10_wp,  &
  mutate_prob=0.1e0_wp,  &
  mutate_gene_prob=0.1e0_wp)
\end{lstlisting}
\end{tight}
\end{center}
Here, a large population size of 10,000 individuals and
a relatively high number of elite individuals (100) are used to
ensure robust exploration across all dimensions.
The GA uses blend crossover, rank-based selection,
and gaussian mutation, see \cref{sec:operations} for details.
The GA terminates once a fitness value of
$10^{-10}$ has been found, which is set
via the argument \texttt{fitness\_target}.\footnote{Using
a fitness target value to control the desired precision
of the solution is only appropriate in this case because the
global minimum of the Rastrigin function is known to be exactly zero.
For functions where the global minimum is not known,
convergence criteria must be defined differently.}

Despite using the same GA settings for all dimensions,
this basic setup is sufficient to reliably identify the
global minimum of the Rastrigin function in all tested dimensions.
The performance could be further optimized by tuning the
GA parameters individually for each dimension (for example,
using smaller population sizes in lower-dimensional cases)
but the aim here is to demonstrate the generality and
robustness of the algorithm rather than maximizing efficiency.
Even with this general setup, the full benchmark test program
completes in approximately 7 seconds on a standard laptop
equipped with 8 CPU threads.
In the right plot of \cref{fig:rastrigin} we show the number of
fitness function evaluations required to reach the desired
precision of $f < 10^{-10}$ as a function of the number of
dimensions $n$. The plot reveals that approximately 60,000
evaluations are needed for the two-dimensional test case,
while the number increases gradually with dimensionality,
reaching about 700,000 evaluations at $n = 20$.
This growth is expected due to the exponential increase
in the search space volume with dimension.

\begin{figure}[t]
\centering
\includegraphics[height=6.6cm]{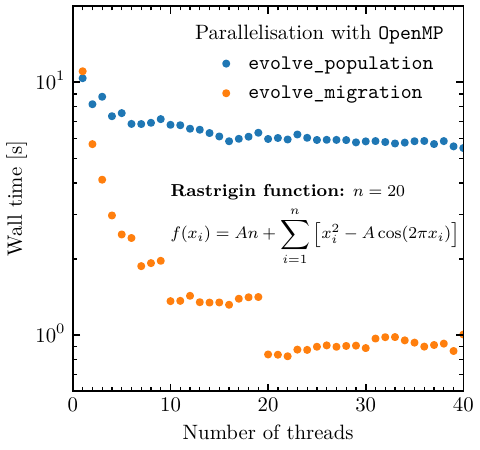}
\caption{\TB{Wall time as a function of the number of threads for the
example program \texttt{benchmark\_rastrigin\_nthreads}. For each
number of threads, the 20-dimensional Rastrigin function was minized
ten times, and the plot shows the average wall time per minimzation.
The blue points show the wall time using the \texttt{evolve\_population}
routine, and the orang points show the wall time using the
\texttt{evolve\_migration} routine.}}
\label{fig:rastrigin_nthreads}
\end{figure}

\TB{To illustrate the effects of parallelization in a
case where fitness evaluations are computationally inexpensive,
we now consider the case of minimizing the 20-dimensional
Rastrigin function. For an increasing number of threads,
up to a maximum of 40 threads, the minimization was
repeated ten times each, and the average wall time per
minimization is shown in \cref{fig:rastrigin_nthreads}.
The blue points correspond to results obtained with
the \texttt{evolve\_population} routine, and the orange
points correspond to the \texttt{evolve\_migration} routine.
Minimization with \texttt{evolve\_population} was performed
as discussed above, except that the \texttt{fitness\_target}
argument was left unset, such that the GA runs through all
generations without early convergence.
For the minimization with the \texttt{evolve\_migration}
routine, a population size of 500 and a total number
of 20 populations was used, giving a total of 10,000 individuals,
the same as for the minimization using the
\texttt{evolve\_population}. As a result, the wall
time for both routines is comparable.
As one can see in \cref{fig:rastrigin_nthreads},
the \texttt{evolve\_population} routine shows only modest speedup
in this case, with the wall time decreasing from
roughly eleven seconds using a single thread to just
below six seconds using more than ten threads.
The \texttt{evolve\_migration} routine benefits more
significantly from parallelization here, with the
wall time dropping to
about five seconds with only two threads,
to roughly one second with ten threads, and to around
0.8 seconds for 20 threads or more.
In both cases, no further improvement is observed beyond
approximately 20 threads, as the runtime is then
dominated by the constant overhead of the GA itself.
The difference in runtime improvement between both
routines can be attributed to the larger
parallelization overhead of \texttt{evolve\_population},
which is more advantageous when fitness evaluations
are costly, whereas \texttt{evolve\_migration} incurs
smaller overhead and scales more efficiently with thread
count in this example.
The runtime improvements with parallelization become more
significant when the evaluation of the fitness function
is computationally expensive and dominating the overall
runtime, as will be demonstrated in a more realistic
application in \cref{sec:lisa}.}

\subsubsection{Michalewicz function}
\label{sec:Michalewicz}

\begin{figure}[t]
\centering
\includegraphics[height=5.5cm]{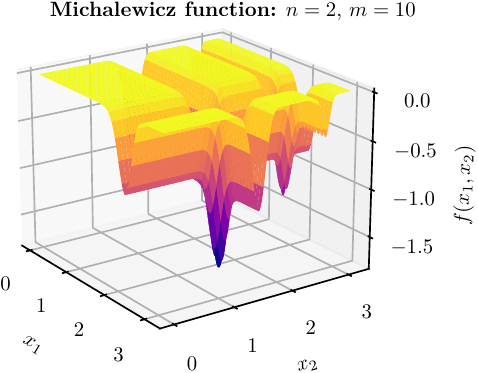}~~~
\includegraphics[height=6.0cm]{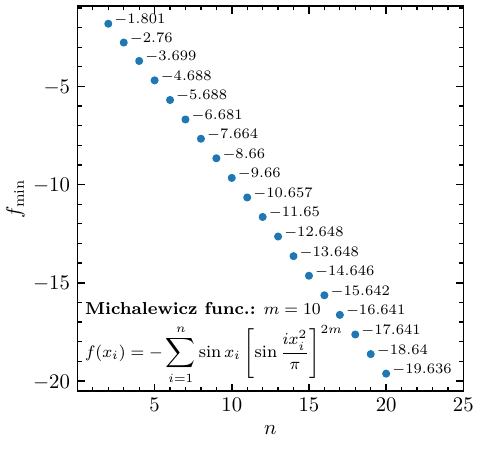}
\caption{
Left: Surface plot of the $n=2$ dimensional
Michalewicz function
with $m = 10$ in the domain $x_{1,2} \in [0, \pi]$.
Right: Minimal function values $f_{\rm min}$
of the Michalewicz function for $m = 10$
against the number of dimensions $n = 2,\dots 25$
that were found using the example application
\texttt{benchmark\_michalewicz}.}
\label{fig:micha1}
\end{figure}

Another widely used multimodal test function for evaluating
the performance of global optimization algorithms is
the Michalewicz function~\cite{mischa}.
The function is defined in $n$ dimensions as
\begin{equation}
f(x_i) = -\sum_{i=1}^{n} \sin(x_i) \left[
  \sin\left( \frac{i x_i^2}{\pi} \right) \right]^{2m}
\end{equation}
where the search domain is
$x_i \in [0, \pi]$, and the parameter $m$ controls the
sharpness of the valleys and ridges that structure the function.
We show a surface plot of the two-dimensional Michalewicz function
in the left plot of \cref{fig:micha1}.
A common and recommended value is $m = 10$, which is also
adopted in our analysis.
As $m$ increases, the local minima become steeper and narrower,
making the optimization landscape significantly more
difficult to navigate. In total, the Michalewicz function has
$n!$ local minima, and only one of them corresponds to
the global minimum.

A key distinctions between the Michalewicz function
and the Rastrigin function discussed in \cref{sec:rastrigin}
lies in the nature of their local minima.
The local minima of the Rastrigin function are located
in a regular pattern, whereas the Michalewicz function features
broad, almost flat valleys that can mislead an optimizer
into converging prematurely. These valleys exhibit very small
gradients, which can render gradient-based methods
ineffective. At the same time, the global minimum is
confined to a narrow, sharply defined region of the parameter
space, further increasing the difficulty of the search.

To demonstrate how \evortran\ can be used to tackle
this problem, we include a test program
\texttt{benchmark\_michalewicz} in the \texttt{app}
folder of the repository. This program runs a GA on the
Michalewicz function for dimensions ranging from
$n = 2$ to $n = 100$.
To search for the global minimum,
we use the following call to \texttt{evolve\_population}
from evortran:
\begin{center}
\begin{tight}
\begin{lstlisting}[
language=Fortran,
linewidth=\textwidth]
best_ind = evolve_population( &
  1000, n, michalewicz, &
  lower_lim=0.0e0_wp, &
  upper_lim=pi, &
  selection='rank', &
  mating='blend', &
  mutate='gaussian', &
  selection_size=100, &
  elite_size=10, &
  mutate_prob=0.4e0_wp, &
  mutate_gene_prob=0.1e0_wp)
\end{lstlisting}
\end{tight}
\end{center}
This configuration uses a population size of 1,000,
applies rank-based selection, blend crossover and
gaussian mutation, and it includes both
elitism with 10 individuals per generation.
In the right plot of \cref{fig:micha1} we show the
minimum value of the test function achieved at the
end of the GA as a function of the dimension $n$
up to $n = 25$. For $n=2,5,10$ the global minimum
of the Michalewicz function is known, and we find
good agreement with the minimum found using \evortran.

\begin{figure}[t]
\centering
\includegraphics[height=7.2cm]{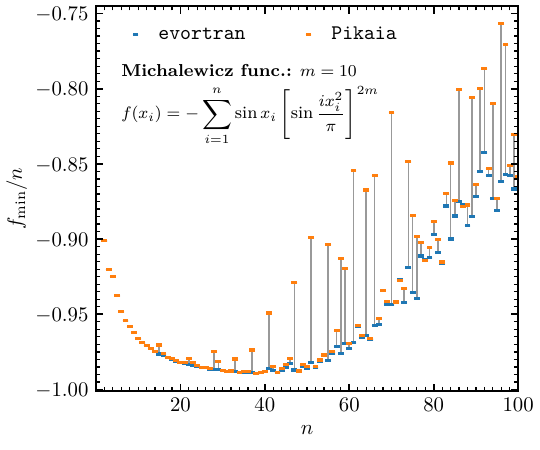}
\caption{
Minimal function values $f_{\rm min}$
of the Michalewicz function normalized to
the dimensionality $n$ for $m = 10$
as a function of the number of dimensions $n = 2,\dots 100$
that were found using the example application
\texttt{benchmark\_michalewicz}. Blue and orange points
indicate the values obtained with \evortran\ and \texttt{Pikaia},
respectively, see text for details.}
\label{fig:micha2}
\end{figure}

For larger number of dimensions, the global minium
of the Michalewicz function is
in general not known.
To assess the quality of the solutions found by \evortran,
we therefore apply also the \texttt{Pikaia} GA framework to
the same problem, see also the discussion in \cref{sec:introduction}.
For the comparison, we use the default settings of
\texttt{Pikaia}, except we increase the population size to
1,000 and the maximum number of generations to 10,000.
The latter is ten times higher than the corresponding setting
in the \evortran\ run. We found that in this way both algorithms
use roughly the same runtime.
It is important to emphasize that our goal is not a
performance comparison between \texttt{Pikaia} and
\evortran. The goal of this comparison is simply to verify
that \evortran\ reliably identifies minima that are at
least as deep as those found by \texttt{Pikaia},
and to therefore be able to validate the \evortran\ result
in the absence of known solutions to the optimization problem.

In \cref{fig:micha2} we show the minimum values of the
Michalewicz function $f_{\rm min}$ devided by the number
of dimensions $n$ as a function of $n$ that
were found using \evortran\
(blue) and \texttt{Pikaia} (orange). One can observe that
for each value of $n$ the values of $f_{\rm min} / n$ determined
with \evortran\ are in agreement
or slightly smaller than the values obtained with \texttt{Pikaia}.
For $n \gtrsim 50$, we observe that for both \evortran\
and \texttt{Pikaia} the values of $f_{\rm min} / n$ start to grow with
increasing value of $n$. This potentially indicates that
both codes struggle to determine the exact global minima
in this regime given the settings used
in the example program \texttt{benchmark\_michalewicz}.

\subsubsection{Himmelblau function}
\label{sec:himmelblau}

\begin{figure}[t]
\centering
\includegraphics[height=6cm]{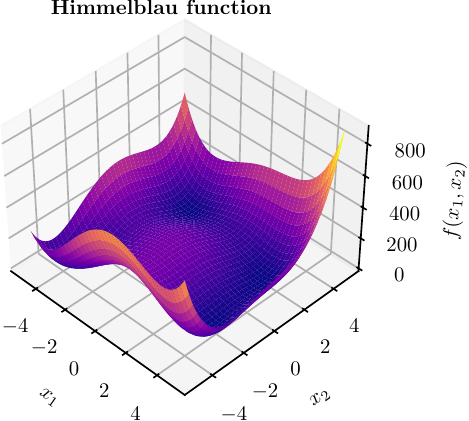}
\caption{
Surface plot of the $n = 2$ dimensional
Himmelblau function in the domain $x,y \in [-5, 5]$.}
\label{fig:himmel}
\end{figure}

The Himmelblau function~\cite{himmelblau}
serves as a classic benchmark in the study of multimodal
optimization problems. Unlike the Rastrigin and Michalewicz functions
discussed above, which feature a single global minimum
(albeit in a complex landscape), the Himmelblau function presents a
qualitatively different challenge because it possesses multiple global
minima of equal depth. The four minima are separated by relatively shallow
barriers. This makes it an ideal benchmark function for evaluating the
ability of an optimization framework not just to find a global minimum,
but to recover multiple degenerate global optima.
The Himmelblau function is defined as
\begin{equation}
f(x, y) = (x^2 + y - 11)^2 + (x + y^2 - 7)^2 \, .
\end{equation}
It has four known global minima in the search
domain $x,y \in [-5, 5]$, located approximately at
\begin{align}
(x, y) &\approx (3.0,\ 2.0), \\
(x, y) &\approx (-2.805,\ 3.131), \\
(x, y) &\approx (-3.779,\ -3.283), \\
(x, y) &\approx (3.584,\ -1.848),
\end{align}
all with a function value of $f(x,y) = 0$.
In \cref{fig:himmel} we
show a surface plot of the Himmelblau function
in the search domain.

To demonstrate how \evortran\ can successfully identify all of
these minima, we make use of the \texttt{evolve\_migration} routine,
which is designed to evolve multiple populations in parallel,
see the discussion in \cref{sec:migration}.
In this example, we disable migration between the evolving
populations (by chosing a single epoche, see the discussion below),
such that each population evolves independently.
Due to the stochastic nature of GAs, different populations initialized
with random gene values can then converge to different minima.
Using a sufficient number of evolving populations enhances
the likelihood of recovering all four global minima
of the Himmelblau function  within a single execution.
The corresponding example program \texttt{benchmar\_himmelblau}
can be found in the \texttt{app} folder of the repository.
In this program,
the call to the \texttt{evolve\_migration} function is as follows:
\begin{center}
\begin{tight}
\begin{lstlisting}[
language=Fortran,
linewidth=\textwidth]
best_ind = evolve_migration( &
  pop_number=20, &
  epoches=1, &
  pop_size=50, &
  gene_length=2, &
  fit_func=himmelblau, &
  mating='sbx', &
  elite_size=1, &
  lower_lim=-5.0e0_wp, &
  upper_lim=5.0e0_wp, &
  max_generations=100, &
  fittest_inds_final_pops=fittest_inds_final_pops)
\end{lstlisting}
\end{tight}
\end{center}
20 independent populations of size 50 are evolved in parallel
for 100 generations. Each individual encodes two genes,
corresponding to the two-dimensional input of the Himmelblau function,
with gene values bounded between -5 and 5. Simulated binary crossover
is used for mating, and one elite individual is retained in each generation.
Only a single epoch is used, which means no migration of individuals
occurs between populations since migration in \texttt{evolve\_migration}
is only applied across multiple epochs.
This setup allows each population to converge independently.
The function returns the best-fit individual across all populations
(\texttt{best\_ind}). Since we are interested in finding all minima,
we additionally store the best-fit individual of each population
after the completion of the GA
using the optional argument \texttt{fittest\_inds\_final\_pops}.

Upon completion, the \texttt{benchmark\_himmelblau} program
prints the best solution found across all populations,
followed by a detailed list of the best soloutions from within
each of the 20 independent populations:
\begin{center}
\begin{tight}
\begin{lstlisting}[
language=bash,
deletekeywords={function,in},
linewidth=\textwidth]
 Fittest overall individual:
  Genes:   3.5844E+00 -1.8481E+00
  Fitness:   0.0000E+00

 Fittest individuals in each population:
     i           x           y        fmin
     1  3.0000E+00  2.0000E+00  4.4727E-21
     2  3.5844E+00 -1.8481E+00  0.0000E+00
     3  3.5844E+00 -1.8481E+00  0.0000E+00
     4  3.0000E+00  2.0000E+00  9.6754E-19
     5  3.0000E+00  2.0000E+00  4.2069E-26
     6 -2.8051E+00  3.1313E+00  7.8886E-31
     7 -2.8051E+00  3.1313E+00  7.8886E-31
     8 -2.8051E+00  3.1313E+00  1.3647E-28
     9  3.5844E+00 -1.8481E+00  1.1177E-22
    10  3.0000E+00  2.0000E+00  1.1292E-25
    11 -3.7793E+00 -3.2832E+00  2.9196E-22
    12  3.5844E+00 -1.8481E+00  0.0000E+00
    13  3.0000E+00  2.0000E+00  0.0000E+00
    14  3.5844E+00 -1.8481E+00  0.0000E+00
    15  3.0000E+00  2.0000E+00  0.0000E+00
    16 -2.8051E+00  3.1313E+00  1.3996E-24
    17  3.5844E+00 -1.8481E+00  0.0000E+00
    18 -3.7793E+00 -3.2832E+00  7.1510E-27
    19  3.0000E+00  2.0000E+00  5.8376E-28
    20 -3.7793E+00 -3.2832E+00  3.1554E-30
\end{lstlisting}
\end{tight}
\end{center}
The table lists the population
index, the gene values, i.e.~the $x$ and $y$ coordinates, of
the fittest individual, and the achieved minimum of the Himmelblau
function. In the example shown, all four known global minima
of the Himmelblau function are discovered by different populations.
This demonstrates the effectiveness of evolving multiple
independent populations in parallel for identifying multiple
(approximately) degenerate
solutions of multimodal objective functions.

\subsubsection{Drop-Wave function}
\label{sec:dropwave}

\begin{figure}[t]
\centering
\includegraphics[height=6cm]{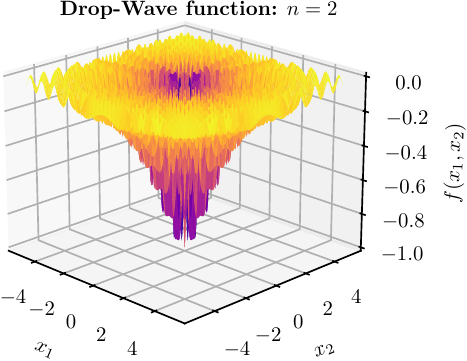}~~~
\includegraphics[height=6.2cm]{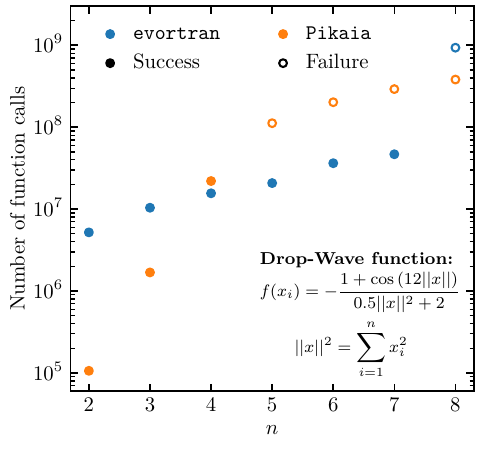}
\caption{Left: Surface plot of the $n=2$ dimensional
Drop-Wave function in the domain $x_{1,2} \in [-5.12, 5.12]$.
Right: Number of calls to the Drop-Wave function
against the number of dimensions $n$ until the GA
completes by either finding the global minimum
with a precision of $10^{-4}$ (filled points) or
by reaching the maximum number of epoches/generations
(empty points) using the example application
\texttt{benchmark\_dropwave\_pikaia}.
Blue and orange points correspond to using
\evortran\ and \texttt{Pikaia}, respectively,
with the GAs configured as described in the text.}
\label{fig:drop}
\end{figure}

As a final benchmark, we consider the Drop-Wave
function, a challenging multimodal test
function often used to evaluate the robustness of
global optimization methods.
The definition of the Drop-Wave function is
\begin{equation}
f(x_i) = -\dfrac{1 + \cos\left(12 ||x||\right)}
  {0.5 ||x||^2 + 2} \quad \textrm{with }
||x||^2 = \sum_{i=1}^n x_i^2 \, .
\end{equation}
Its global minimum with $f(x_i) = -1$ is
located at the center $x_i = 0$ of the search
domain $x_i \in[-5.12, 5.12]$.
In \cref{fig:drop} we show a surface plot of the Drop-Wave
function in two dimensions over the search domain.
Due to the presence of steep central basins
surrounded by increasingly shallow, oscillating ripples
that form many local minima,
the landscape of the function misleads optimization algorithms
away from the global minimum.
This structure makes it extremely difficult for optimization
methods to locate the global minimum, even for a relatively
small number of dimensions.
The Drop-Wave function is particularly well-suited to test
the ability of GAs to escape local minima.
In this benchmark, we again compare the results obtained
with \evortran\ with the ones obtained using the
\texttt{Pikaia} library.

We apply both frameworks to search for the global minimum of
the Drop-Wave function in dimensionalities ranging from
$n = 2$ to $n = 8$. The corresponding example program
\texttt{benchmark\_dropwave\_pikaia} is located in the
\texttt{app} folder of the \evortran\ repository.
For \evortran, we use the \texttt{evolve\_migration} routine
with the following call:
\begin{center}
\begin{tight}
\begin{lstlisting}[
language=Fortran,
linewidth=\textwidth]
best_ind = evolve_migration(  &
  pop_number=90,  &
  epoches=1000,  &
  pop_size=200,  &
  max_generations=300,  &
  gene_length=i,  &
  fit_func=dropwave,  &
  migration_size=10,  &
  elite_size=10,  &
  selection_size=100,  &
  mating='blend',  &
  fitness_target=-1.0e0_wp + 1.0e-4_wp,  &
  selection='roulette',  &
  wheele_size=7,  &
  mutate='uniform',  &
  mutate_prob=0.2e0_wp)
\end{lstlisting}
\end{tight}
\end{center}
Here the argument \texttt{i} for the \texttt{gene\_size}
corresponds to the number of dimensions which
are iterated over.
This setup results in a total of 18,000 individuals evolving
in parallel in 90 different populations of size 200 each,
over a maximum number of 300 generations and
1000 epoches, see the discussion in \cref{sec:migration}
for details. We make use
of elitism (10 individuals in each population),
roulette wheel selection (with \texttt{wheele\_size}=7),
blend crossover, and uniform mutation
(with a mutation probability of 20\%).\footnote{A total
number of 90 populations is chosen because the program
was executed on a cluster with 90 CPU cores available.}
Migration exchanges 10 individuals between populations after
each epoch. The algorithm halts when the fitness falls
below -0.9999, which is within $10^{-4}$ of the known
global minimum of the Drop-Wave function, or when
the maximum number of epochs is reached.

For comparison, \texttt{Pikaia} is configured with similar
computational resources by assigning it the same total
number of individuals and a maximum of 5000 generations.
The initialization is performed as follows:
\begin{center}
\begin{tight}
\begin{lstlisting}[
language=Fortran,
linewidth=\textwidth]
call pikaia
  i, xl, xu,  &
  dropwave_pikaia,  &
  status_pikaia,  &
  ngen=5000,  &
  np=200*90)
\end{lstlisting}
\end{tight}
\end{center}
The large number of maximum generations allows
\texttt{Pikaia} to perform up to approximately 100
million fitness evaluations, which is a number
comparable to what is typically required by
\evortran\ to converge to the global minimum
in more than 5 dimensions.
The arguments \texttt{xl} and \texttt{xu} are arrays that determine
the search domain in each $x_i$-direction.
Since \texttt{Pikaia} maximizes the fitness function,
whereas \evortran\ minimizes, we define a separate
function \texttt{dropwave\_pikaia} here, which differs
by an overall minus sign from the fitness function
\texttt{dropwave} used in the \texttt{evolve\_migration}
call of \evortran.
We again would like to emphasize
that this setup is not intended to directly compare
runtime performance between \evortran\ and \texttt{Pikaia},
as the frameworks differ in design and parallelization,
but only to validate the \evortran\ results against
a similar available GA implementation.

In the right plot of \cref{fig:drop} we show the
number of times the Drop-Wave function was evaluated
for the different values of dimensions $n$ when
using the example program \texttt{benchmark\_dropwave\_pikaia},
with the orange points indicating the calls from
\evortran\ and the blue points indicating the calls
from \texttt{Pikaia}.\footnote{We slightly modified
the \texttt{Pikaia} algorithm for this example by
changing the convergence condition. We implemented
that the \texttt{Pikaia} algorithm only halts if
either the global minimum was found with a precision
of $10^{-4}$, or if the maximum number of generations
was reached. In this way the \texttt{Pikaia} algorithm
allows for a larger number of function calls and behaves
more similarily to the \evortran\ algorithm.}
The filled points indicate if the GA converged successfully
by converging to the global minimum, whereas the empty
points indicate if the GA failed by converging to
a local minimum. We observe that the GA applied
with \evortran\ is able to find the global minimum
of the Drop-Wave function up to $n = 7$ dimensions,
requiring a total number of about $5 \cdot 10^7$ function calls
for $n=2$ dimensions and of about $5 \cdot 10^8$ function calls
for $n=7$ dimensions.
The GA applied with \texttt{Pikaia} is able to find the
global minimum of the Drop-Wave function with a substantially
smaller number of function evaluations for $n=2$ and
$n=3$ dimensions, and with a similar number of functions
calls than the GA applied with \evortran\ for $n=4$ dimensions.
For $n>4$ the GA applied with \texttt{Pikaia} is not
able to locate the global minimum in the give maximum
number of generations, and thus shows a larger number of
function calls for $n=4,5,6$ than the GA applied with
\evortran\, because the latter halts once the global
minimum is found within the required precision.

To demonstrate the flexibility of the \evortran\ library,
and to compare the performance of different crossover methods,
we conducted a comprehensive set of runs using the
\texttt{evolve\_migration} function to minimze the Drop-Wave
function using different crossover routines.
For each dimensionality from $n = 2$ to $n = 10$, we tested
each crossover operator (\texttt{blend}, \texttt{sbx}, \texttt{one-point},
\texttt{two-point}, \texttt{uniform}) in combination with all
possible permutations of the three available selection
routines (\texttt{tournament}, \texttt{rank}, \texttt{roulette})
and the three mutation strategies (\texttt{uniform}, \texttt{shuffle},
\texttt{gaussian}). Each unique combination was run 10 times,
resulting in a total of $3 \cdot 3 \cdot 10 = 90$ runs per crossover
method and dimension.
The GA was configured via the following call:
\begin{center}
\begin{tight}
\begin{lstlisting}[
language=Fortran,
linewidth=\textwidth]
best_ind = evolve_migration(  &
  pop_number=90,  &
  epoches=1000,  &
  pop_size=200,  &
  max_generations=300,  &
  gene_length=number_dimensions,  &
  fit_func=dropwave,  &
  migration_size=2,  &
  elite_size=2,  &
  selection_size=100,  &
  selection=selection_opts(i_selection),  &
  mating=mating_opts(i_mating),  &
  mutate=mutate_opts(i_mutate),  &
  fitness_target=-0.999,  &
  wheele_size=7,  &
  verbose=verbose)
\end{lstlisting}
\end{tight}
\end{center}
Here, \texttt{evolve\_migration} was used to evolve 90 populations
in parallel with light migration (\texttt{migration\_size=2})
between them. The number of epochs was set to 1000, and each
population consisted of 200 individuals, evolved over up to
300 generations per epoch. The \texttt{fitness\_target} value
was set to -0.999, which lies below the value of any local
minimum of the Drop-Wave function, thereby serving as an
effective criterion to identify convergence to the
true global minimum.
Default values were used for all optional arguments of the
selection, crossover and mutation routines, except for
a \texttt{wheele\_size} of 7 for \texttt{roulette} selection.

\begin{figure}[t]
\centering
\includegraphics[width=\textwidth]{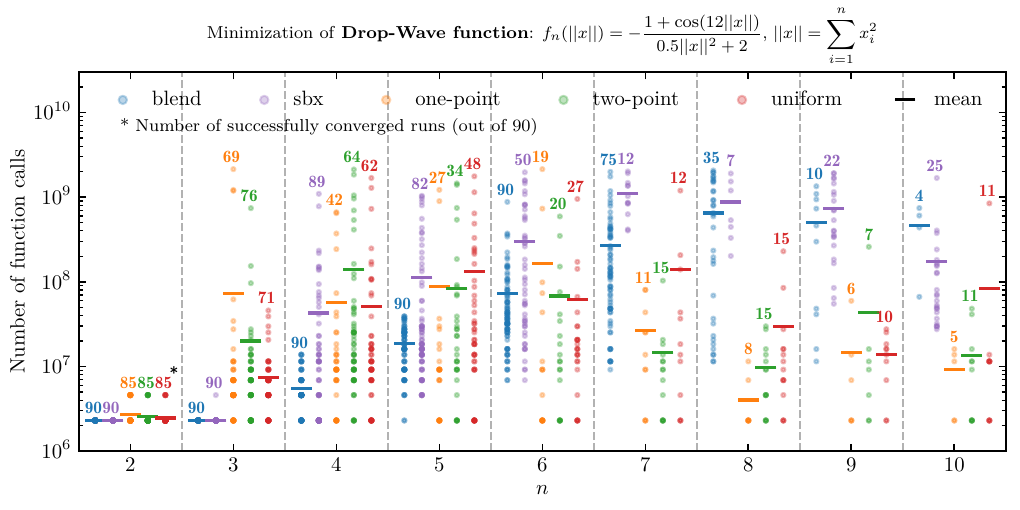}
\caption{Number of calls to the Drop-Wave function against the
number of dimensions $n$ until the GA successfully converged
to the global minimum for the different crossover methods
implemented in \evortran. The horizontal bars indicate the mean
values per dimension for each crossover method, and the colored
numbers show the total number of successfully converged runs
out of a total number of 90 runs (see the discussion in the
text for details).}
\label{fig:drop_cxs}
\end{figure}

The results are summarized in \cref{fig:drop_cxs}, which shows
the number of function evaluations required for successful
convergence versus dimensionality $n$. The total number of successful
runs is indicated with numbers
per method and dimension to quantify robustness.
Also shown with horizontal bars are the mean values of the function calls
per dimension for each crossover method, including only the
successfully converged runs.
The plot reveals several key features regarding the robustness
of the different crossover strategies.
The crossover methods specifically designed for continuous
optimization problems, \texttt{blend} and \texttt{sbx},
outperform the more generic binary-inspired crossovers,
\texttt{one-point}, \texttt{two-point}, and \texttt{uniform},
particularly in lower and intermediate dimensions $n \leq 6$.
This suggests that using continuous crossover operators
can significantly enhance the reliability of a GA in continuous
solution spaces.
Up to dimension $n=7$, the \texttt{blend} crossover shows the
highest robustness, achieving successful convergence in at
least 75 out of 90 runs, highlighting its effectiveness across
a wide range of settings.
For higher dimensionalities ($n=8,\dots,10$), the \texttt{sbx}
crossover becomes more robust, maintaining a convergence
success rate of over 20\%.
Interestingly, the \texttt{uniform} crossover, while
generally less robust at lower dimensionality, shows a
certain robustness in performance at high dimensionality,
matching the robustness of \texttt{blend} and \texttt{sbx}
for $n=8$ and beyond.
It maintains a minimum success rate of roughly
10\% across all dimensions.

These results underscore the importance of carefully
selecting the components of a GA to match the characteristics
of the specific optimization problem.
The flexible design of \evortran\ allows users to easily
explore and combine different evolutionary operators.
This modularity is a key advantage of the library,
making it a powerful tool in wide areas
of in scientific computing.

\subsection{Physics applications}
\label{sec:physapps}

To demonstrate the capabilities of
\evortran\ in real-world research settings,
we present two physics-motivated applications that go
beyond the mathematical benchmark problems discussed above.
The first application focuses on a global fit of an
extended Higgs sector in a beyond-the-Standard-Model~(BSM)
scenario to existing data from the Large Hadron Collider~(LHC).
This includes fitting the observed signal strengths
of the Higgs boson at 125~GeV as well as
incorporating cross section
limits from searches for additional Higgs bosons that
have (so far) not given rise to a discovery.
The second application addresses a problem in cosmology:
reconstructing the frequency spectrum of a stochastic
gravitational wave background produced by a cosmological
first-order phase transition, such as the ones expected
from an electroweak phase transition in the early universe.
Here, the goal is to reconstruct a consistent signal
from LISA mock data, which includes both an injected gravitational
wave signal and realistic instrumental noise.

\subsubsection{Confronting extended Higgs sector with LHC data}
\label{sec:lhc}

A common challenge in studies of BSM physics is the efficient
exploration of high-dimensional parameter spaces subject
to a complex set of constraints. In particular, models with
extended scalar/Higgs sectors often feature many free
parameters (eleven in the first scenario studied below,
and 14 in a second more general scan). These parameters are
subject to both theoretical consistency requirements
(such as vacuum stability or perturbativity) and a wide
range of experimental constraints. Among the latter are
95\% confidence level exclusion bounds on production cross
sections from direct searches for additional Higgs
bosons at the LHC, as well as precision measurements of the
observed 125~GeV Higgs boson. Many of these constraints are
either non-differentiable (e.g., hard cutoffs on excluded
cross sections) or involve discrete features in the
parameter space, which makes gradient-based
optimization methods less effective.

GAs offer a natural solution to this type of problem.
Their population-based and non-gradient nature allows them
to effectively navigate large, non-linear, and non-smooth
landscapes, making them well-suited for parameter scans that
need to identify viable regions consistent with
a diverse set of constraints.
An additional advantage of GAs is their ability to
uncover multiple, qualitatively distinct regions of
parameter space that are all consistent with theoretical and
experimental constraints within their respective uncertainties.
This is particularly important when different parameter
regions compatible with the constraints yield different
phenomenology. Since these parameter regions may correspond
to local minima of the $\chi^2$-function or
likelihood that are only marginally suboptimal compared
to the global minimum within the margin of uncertainty,
identifying (ideally) all of these regions may be crucial.
GAs are well-suited for this task because their stochastic and
population-based search can naturally discover such local
optima, rather than converging exclusively to a
single best-fit solution.

As an example, we focus here on the singlet-extended two Higgs
doublet model~(S2HDM)~\cite{Biekotter:2021ovi},
a well-motivated BSM scenario that augments
the Standard Model by a second Higgs doublet and a
complex scalar singlet.
We utilized a GA to explore the phenomenology
of the S2HDM in previous studies in Refs.~\cite{Biekotter:2021ovi,
Biekotter:2023jld,Biekotter:2023oen} using the Python
GA framework \texttt{DEAP}~\cite{DEAP_JMLR2012}.
In the S2HDM,
the presence of the second Higgs doublet
opens the possibility of a
strong first-order electroweak phase transition, which is
an essential requirement for electroweak baryogenesis
to explain the baryon asymmetry of the universe, and which
cannot be realized in the Standard Model. Additionally,
the complex singlet respects a softly-broken global U(1)
symmetry, which is spontaneously broken when the real
component of the singlet acquires a vacuum expectation value.
This results in a pseudo-Nambu-Goldstone boson from
the imaginary part of the singlet field, which is stable and
provides a viable Higgs-portal dark matter candidate.
Notably, a pseudo-Nambu-Goldstone
dark matter particle can achieve the observed relic
abundance through the standard thermal freeze-out mechanism,
while evading the stringent limits from dark matter
direct detection experiments due
to momentum-suppressed scattering cross sections at
leading order~\cite{Gross:2017dan,Biekotter:2022bxp}.

A comprehensive discussion of the S2HDM can
be found in Ref.~\cite{Biekotter:2021ovi}. Here,
we briefly summarize the key features relevant
for the following analysis. The model contains the
same fermionic and gauge field content as the
Standard Model, but due to the extended scalar sector
the physical spectrum includes a total of three
CP-even neutral scalar Higgs bosons $h_i$
($i = 1,2,3$) which
mix with each other, a CP-odd pseudoscalar Higgs boson $A$,
a pair of singly charged Higgs bosons $H^\pm$, and
a stable spin-0 dark matter candidate $\chi$.\footnote{The
Standard Model of particle physics predicts a single
CP-even Higgs boson.}
We consider two Yukawa structures: first, a so-called
Type~I structure where only one Higgs doublet is coupled
to fermions, and second, a flavour-aligned setup
which can be parameterized by three so-called
flavour alignment parameters $\xi_{u,d,\ell}$~\cite{Pich:2009sp}.
The other free parameters used in the analysis below
include the masses $m_{h_{1,2,3}}$, $m_A$, $m_{H^\pm}$,
$m_\chi$, the three CP-even scalar mixing angles
$\alpha_{1,2,3}$, the ratio of the Higgs doublet vacuum
expectation values $\tan\beta$, the singlet vacuum
expectation value $v_S$, and an additional mass scale
parameter $M$, which controls the allowed mass
range of the BSM Higgs states.

To identify phenomenologically viable points in the
parameter space of the S2HDM, we construct a global
$\chi^2$-function that incorporates a range of
theoretical and experimental constraints.
This function is then minimized using a GA
implemented in \texttt{evortran}.
Since evaluating the fitness function requires interfacing
with external codes, namely \texttt{HDECAY}~\cite{Djouadi:1997yw,
Muhlleitner:2016mzt}
for computing Higgs boson decay properties and
\texttt{HiggsTools}~\cite{Bechtle:2008jh,
Bechtle:2013xfa,Bahl:2022igd,Bahl:2025}
for checking consistency against
LHC data (see below), the evaluation is computationally
expensive. For this reason, we employ a relatively
economical GA with a population size of only
40 individuals, evolved using \evortran’s
\texttt{evolve\_population} routine over a maximum
of 1000 generations,
applying tournament selection,
blend crossover and uniform mutation.
\TB{All source code used to obtain the results
presented in this section is publicly available
in a dedicated Git repository~\cite{gitrepos2hdm}.}

The theory constraints included in the analysis are
the following:
\begin{itemize}[label=-, topsep=2pt, itemsep=1pt, parsep=0pt]
  \item \textbf{Vacuum stability:} We impose that the
    scalar potential is bounded from below,
    ensuring that no field direction leads to the
    potential tending to $-\infty$ for large field values.
    At tree-level,
    this leads to analytic conditions requiring certain
    combinations of quartic couplings appearing in the
    potential to be positive.
    The corresponding expressions can be found in
    Refs.~\cite{Klimenko:1984qx,Muhlleitner:2016mzt}.
  \item \textbf{Perturbative unitarity:} We verify that
    the leading-order
    $2 \to 2$ scalar scattering amplitudes remain
    below the threshold $8\pi$ in the high-energy limit.
    This results in a set of inequalities on combinations of
    the scalar quartic couplings, whose precise form
    can be found in Ref.~\cite{Biekotter:2021ovi}.
\end{itemize}
Both theory constraints are implemented as hard cuts and introduce non-differentiable features into the $\chi^2$ function.
The experimental constraints that we consider here are
the following:
\begin{itemize}[label=-, topsep=2pt, itemsep=1pt, parsep=0pt]
  \item \textbf{125 GeV Higgs cross sections:} We include
    cross section measurements of the observed Higgs boson
    in various production and decay channels,
    including also the measurements presented in the
    form of the Simplified Template Cross Section~(STXS)
    framework.
  \item \textbf{Limits from BSM spin-0 resonance searches:}
    We apply 95\% confidence level exclusion limits from a
    large set of direct LHC searches for additional neutral
    and charged Higgs bosons.
    These constraints are implemented as hard cuts.
  \item \textbf{Electroweak precision observables:} We
    require compatibility with the measured electroweak
    parameter $\Delta\rho$ within its experimental
    uncertainty~\cite{ParticleDataGroup:2024cfk},
    based on its one-loop prediction
    in the S2HDM~\cite{Biekotter:2021ovi}.
    This constraint is included as a hard cut, rejecting
    parameter points predicting a value of $\Delta\rho$
    deviating by more than two standard deviations from
    the experimental central value, corresponding to
    a confidence level of about 95\%.
\end{itemize}
The check against the LHC data is carried
out in our analysis with the help
of the public C++ package \texttt{HiggsTools},
which performs a global $\chi^2$ fit to the
cross section measurements of the 125~GeV Higgs bosons
and checks if all existing cross section limits
from BSM scalar searches are satisfied.
The \texttt{HiggsTools} library
can be exposed to Fortran via the
\texttt{iso\_c\_binding} module. We created
a minimal working example
for using \texttt{HiggsTools} within an \texttt{fpm}
project that is available in a dedicated
GitLab repository~\cite{htfpmcaller}.

The global $\chi^2$ function minimized by
\evortran\ includes the contributions from
\texttt{HiggsTools} for the 125~GeV Higgs measurements,
denoted $\chi^2(h_{125})$ in the following,
while all other constraints are incorporated via
large fixed penalty values if they are not satisfied.
This setup effectively translates the task
of finding allowed parameter points into
minimizing $\chi^2(h_{125})$
under the condition that all other theoretical and
experimental constraints are satisfied, for
which derivative-free optimization strategies
like GAs are especially useful.
To define a meaningful condition at which
the GA shall halt, we subtract the Standard Model
prediction $\chi^2_\mathrm{SM}(h_{125})$ from the
total $\chi^2$, such that $\chi^2 = 0$ corresponds
to the S2HDM fitting the LHC data as well as
the Standard Model (while satisfying all other constraints),
while $\chi^2 < 0$ indicates a better fit
than the Standard Model.
The GA halts if a parameter point predicting $\chi^2 < 0$
was found.
During the minimization process,
we not only record the best-fit point but also store
all parameter points generated during the
evolution process that predict
$\chi^2 = \chi^2(h_{125}) - \chi^2_{\rm SM}(h_{125})
\leq 6.18$,
which approximately corresponds to a 95\%
confidence region assuming two degrees of
freedom and assuming that the Standard Model
prediction is a good approximation of the
best fit to the LHC data
(see Ref.~\cite{Bahl:2025} for details).
To obtain a diverse sample of viable
parameter points across the scanned regions
of parameter space, we run the GA multiple
times with different random seeds.

\begin{table}[t]
\centering
\renewcommand{\arraystretch}{1.2}
\begin{tabular}{lr}
\hline
\textbf{Parameter} & \textbf{Range} \\
\hline
$m_{h_1}$ & 125.1 GeV (fixed) \\
$m_{h_2}, m_{h_3}, m_A$, $m_{H^\pm}$,
  $m_\chi$ & [30, 1000] GeV \\
$\alpha_{1,2,3}$ & [$-\frac{\pi}{2}$, $\frac{\pi}{2}$] \\
$\tan\beta$ & [1, 20] \\
$M (= m_{12}^2 / (\sin\beta \cos\beta))$ & [30, 1000] GeV \\
$v_S$ & [10, 3000] GeV \\
\hline
Scenario 1: & Yukawa Type I \\
Scenario 2: & Flavour alignment: $\xi_{u,d,\ell}
  \in [10^{-3}, 10^3]$ (log prior) \\
\hline
\end{tabular}
\renewcommand{\arraystretch}{1.0}
\caption{The top rows show the
S2HDM parameter ranges used in
the scans. The last two rows indicate
the Yukawa structure chosen for
the scenario~1 and the scenario~2.}
\label{tab:scan_ranges}
\end{table}

\medskip

\noindent{\textbf{Scenario 1: Scanning the S2HDM Type~I}
-- 11 free parameters: $\{ m_{h_2}, m_{h_3}, m_A, m_{H^\pm},
  m_\chi,$ $\tan\beta, \alpha_{1,2,3}, M, v_S \}$\\
In the first scenario, we use \evortran\ to perform a
global scan of the S2HDM with Yukawa Type~I,
using the full parameter space summarized in
\cref{tab:scan_ranges}. This constitutes a
very generic scan where the goal is to test
whether the GA is capable of identifying regions in
parameter space that yield a 125~GeV Higgs boson
(here $h_1$) with properties resembling those
predicted by the Standard Model within current
experimental precision, while also satisfying all
other theoretical and experimental constraints
discussed above. In the case where the additional
Higgs bosons are substantially heavier, the
properties of the 125~GeV Higgs boson are mainly
governed by the mixing angles $\alpha_{1,2,3}$
and $\tan\beta$,
which must be tuned with some precision in
order to reproduce the observed Higgs
boson signal strengths.
The other parameters are relevant for satisfying
the remaining constraints. For instance, the
masses of the BSM states have to be chosen in
a range that is not excluded by LHC searches.
Their allowed values are also strongly
correlated with the parameter $\tan\beta$,
the mass parameters $M$ and $v_S$,
and the mass splitting among the BSM states
is constrained by the $\Delta \rho$ parameter.

Technically, this parameter scan is a
minimization of a $\chi^2$ function defined over
eleven free parameters. The complexity of the
task is significant since the $\chi^2$ function
encodes roughly 150 independent LHC measurements of
the 125~GeV Higgs boson cross sections,
in addition to a vast number of cross section limits
from direct searches for BSM Higgs bosons.
This results in a highly structured and non-trivial
eleven-dimensional optimization landscape.

In \cref{fig:lhc_1_1} we show results from scenario
1, where each point corresponds to a valid parameter
point identified by the GA. The left plot shows the
distribution of $\chi^2(h_{125})$ values obtained with
\texttt{HiggsTools} against the coupling coefficient
$\kappa_V$ of the CP-even Higgs boson $h_1$ to vector bosons
$V = W,Z$, defined as
$\kappa_V = g_{h_1 VV} / g_{h_{\rm SM} VV}$,
where $g_{h_1 VV}$ is the coupling predicted in the
S2HDM and $g_{h_{\rm SM} VV}$ is the coupling predicted
for a Standard Model Higgs boson of the same mass.
This plot demonstrates that the GA successfully identified
parameter points consistent with the 125~GeV Higgs
boson data at the level of the Standard Model
($\chi^2 \approx 0$) and even slightly better
($\chi^2 < 0$), though the latter is not a statistically
significant improvement considering the larger number
of additional parameters of the S2HDM compared to
the Standard Model.

\begin{figure}
\centering
\includegraphics[height=6cm]{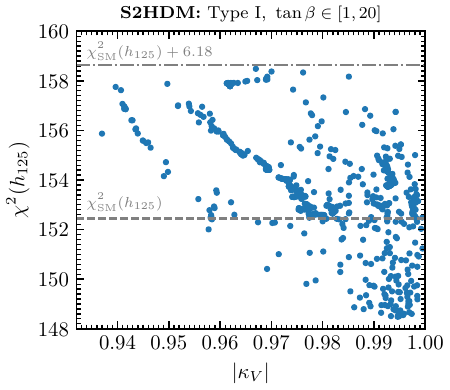}~~
\includegraphics[height=6cm]{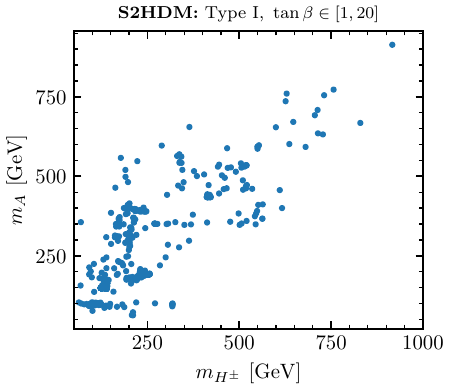}
\caption{Each point in the two plots corresponds to
a valid parameter point found with the GA
in the scenario 1 in the S2HDM with
type~I Yukawa structure.
Left: $\chi^2(h_{125})$ obtained with
\texttt{HiggsTools} against the coupling coefficient
$|\kappa_V|$ of the 125~GeV Higgs boson $h_1$.
Right: mass of the CP-odd Higgs boson
$m_A$ against the mass of the charged
Higgs bosons $m_{H^\pm}$.}
\label{fig:lhc_1_1}
\end{figure}

The right plot of \cref{fig:lhc_1_1} shows
the mass of the CP-odd Higgs boson $m_A$
against the mass of the charged Higgs boson
$m_{H^\pm}$. The GA identified valid parameter
points over a wide mass range up to the upper limit 1~TeV
of the scan for the masses of the BSM states.
The lightest masses found
for $A$ and $H^\pm$ are around 200~GeV.
Below this, experimental searches from the LHC
increasingly constrain the parameter space,
and a light $H^\pm$ also introduces substantial
loop-level corrections to the di-photon decay
rate of the 125~GeV Higgs boson $h_1$,
making it harder to achieve agreement with the data.
However, this does not imply that smaller
values of $m_A$ or $m_{H^\pm}$ are excluded
in the S2HDM. Uncovering valid parameter space
in this regime would require a dedicated scan
that biases the GA toward lower mass ranges
(or higher values of $\tan\beta$), instead of
the setup applied here that constitutes a broad
and generic exploration of parameter space.
Biasing the GA towards smaller masses could, for instance,
be achieved using a logarithmic prior instead
of linear prior for the masses of the BSM Higgs bosons
when the GA is initialized, see also the discussion
in \cref{sec:lisa}.

In \cref{fig:lhc_1_2} we show the couplings and masses
of the three CP-even Higgs bosons $h_1$, $h_2$, and
$h_3$ in the S2HDM with type I Yukawa structure for
the valid parameter points found in scenario 1.
In each plot, for each valid parameter point three
points are displayed with the following color-coding:
blue for $h_1$, orange for $h_2$, and green for $h_3$.
The left plot shows the vector boson coupling coefficient
$\kappa_V(h_i)$ versus the Higgs boson mass
$m_{h_i}$.
Here it should be noted that due to unitarty there
is a sum rule for these coupling coefficients:
$\sum_i \kappa_V(h_i)^2 = 1$. Since LHC measurements
require $\kappa_V(h_1) \gtrsim 0.93$,
see left plot of \cref{fig:lhc_1_1}, this forces the
couplings of $h_2$ and $h_3$ to electroweak gauge
bosons to be small. The GA identified parameter points
with this coupling configuration by accurately sampling
the mixing angles $\alpha_{1,2,3}$, as is
visible in the plot where both $\kappa_V(h_2)$ and
$\kappa_V(h_3)$ lie below about 0.3.
Accordingly,
for $h_1$ the GA efficiently converges
to $\kappa_V(h_1) \approx 1$, in agreement with
LHC data requiring the 125~GeV Higgs boson to
resemble the Standard Model prediction within
an experimental precision at the level of~10\%.

\begin{figure}
\centering
\includegraphics[height=6cm]{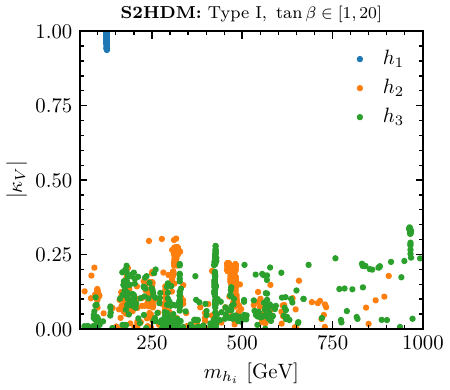}~~
\includegraphics[height=6cm]{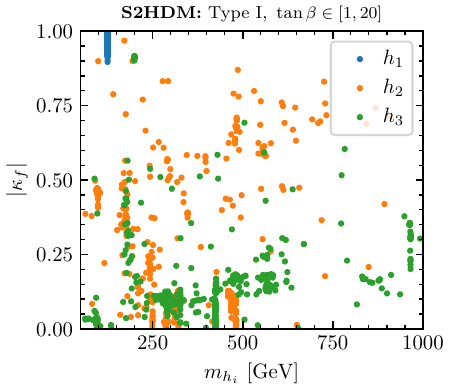}
\caption{For each valid parameter point found with the GA
in the scenario 1 in the S2HDM with
type~I Yukawa structure,
the plots show three points with properties related
to the Higgs bosons $h_1$, $h_2$ and $h_3$ in blue,
orange and green, respectively.
Left: coupling coefficient $\kappa_V$
of the respective state $h_i$ against
the mass of the Higgs boson $m_{h_i}$.
Left: coupling coefficient $\kappa_f$
of the respective state $h_i$ against
the mass of the Higgs boson $m_{h_i}$.}
\label{fig:lhc_1_2}
\end{figure}

Using the Type~I Yukawa structure, the modifications of
the Higgs boson couplings to fermions can be expressed
by means of a single coupling coefficient $\kappa_f(h_i)$
for each Higgs boson, independently of the fermion kind.
The right plot shows the fermion coupling coefficients
defined as $\kappa_f(h_i) = g_{h_i f\bar{f}} /
g_{h_\textrm{SM} f\bar{f}}$, where $g_{h_i f\bar{f}}$ is
the coupling predicted for the state $h_i$ in the S2HDM, and
$g_{h_\textrm{SM} f\bar{f}}$ is the coupling predicted in
the Standard Model for a Higgs boson of the same mass.
In the type I Yukawa structure,
there is an upper bound of $|\kappa_f(h_i)| \leq 1$.
Unlike the gauge couplings, $|\kappa_f(h_2)|$ and
$|\kappa_f(h_3)|$ can be sizable, i.e.~as large as
$|\kappa_f(h_1)|$, without violating current LHC data.
This behavior is visible in the right plot.
Again, as before for $|\kappa_V(h_1)|$ shown in
the left plot of \cref{fig:lhc_1_2},
the GA converges to $\kappa_f(h_1) \approx 1$ for
the 125~GeV Higgs boson, such that the state $h_1$
resembles a SM Higgs boson.

Since the scan covers all physically distinct combinations
of the three scalar mixing angles $\alpha_{1,2,3}$,
see \cref{tab:scan_ranges}, no distinction between
$h_2$ and $h_3$ is imposed on their
possible masses and couplings.
As a result, the orange points corresponding to the
state $h_2$ and the green points corresponding
to the state $h_3$
show a similar distribution in the two plots
of \cref{fig:lhc_1_2}.
Furthermore,
the mass ranges for $h_2$ and $h_3$ span from
about 63~GeV to 1~TeV. The lower bound arises because
masses below $62.5$~GeV would open the decay
channels $h_1 \to h_2 h_2/h_3 h_3$, which tends to
spoil the Standard-Model-like nature of $h_1$
unless the corresponding couplings governing
these decays are fine-tuned to suppress the decays.
The GA does not converge to such tuned regions of
parameter space in this broad scan.
On the upper end, the scan reaches the maximum mass
range of 1~TeV set in the input.

We finally note that some points shown in the
plots in \cref{fig:lhc_1_1} and \cref{fig:lhc_1_2}
are clustered along a line in the plot,
which reflects the fact that during the GA evolution,
all parameter points satisfying $\chi^2 \leq 6.18$ are stored,
and not just the final best-fit point.
As a result, parameter points originating from gradual
changes to a successful candidate solutions during the
optimization appears as an approximately continuous
trajectory in parameter space.

\medskip

\noindent{\textbf{Scenario 2: Scanning the flavor-aligned
S2HDM}
-- 14 free parameters: $\{ m_{h_2}, m_{h_3}, m_A, m_{H^\pm},$ $
  m_\chi,\tan\beta, \alpha_{1,2,3}, M, v_S, \xi_{u,d,\ell}
\}$\\
In this second scenario, we consider a more general
Yukawa sector for the S2HDM, namely the
flavor alignment mentioned above.
In contrast to the Type~I Yukawa structure where
only one Higgs doublet couples to all fermions,
the flavor-aligned setup allows both Higgs
doublets to couple to fermions, controlled via three
additional parameters $\xi_u$, $\xi_d$, and $\xi_\ell$.
These alignment parameters act as proportionality factors
between the Yukawa couplings of the two Higgs doublets
and determine which doublet dominantly couples
to a given fermion type: up-type quarks $u$,
down-type quarks $d$, and charged leptons $\ell$.
Specifically, values of $\xi_{u,d,\ell} \ll 1$
correspond to dominant couplings to the first Higgs
doublet, while $\xi_{u,d,\ell} \gg 1$ correspond to
dominant couplings to the second doublet.

An important consequence is that compared to the
Type~I Yukawa structure, where
the couplings of the Higgs bosons to fermions are
governed by a universal modifier $\kappa_f$
for all fermion types, the flavor alignment leads to
an independent coupling modifier for each fermion kind:
$\kappa_u$, $\kappa_d$, and $\kappa_\ell$,
defined analogously to $\kappa_f$ as the ratio
of the coupling of a state $h_i$ in the S2HDM to the
corresponding coupling of a Higgs boson in the Standard
Model with the same mass as the state $h_i$.
This results in additional freedom to accommodate the
LHC Higgs boson measurements.
The coupling coefficients $\kappa_{u,d,\ell}$ depend
on the parameters $\xi_{u,d,\ell}$, $\tan\beta$ and
the mixing angles $\alpha_{1,2,3}$, such that the
couplings of the 125~GeV Higgs bosons depends on various
independent parameters, and where various entirely different
regions of the complex parameter space can predict
a Higgs boson $h_1$ in agreement with the LHC measurements.

\begin{figure}
\centering
\includegraphics[height=6cm]{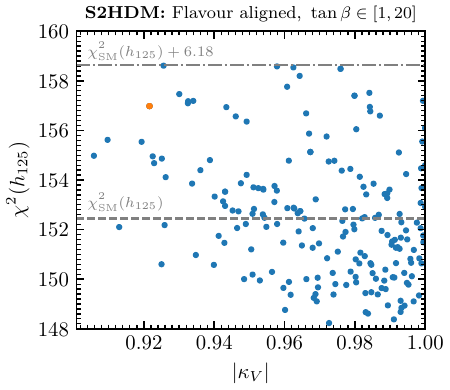}~~
\includegraphics[height=6cm]{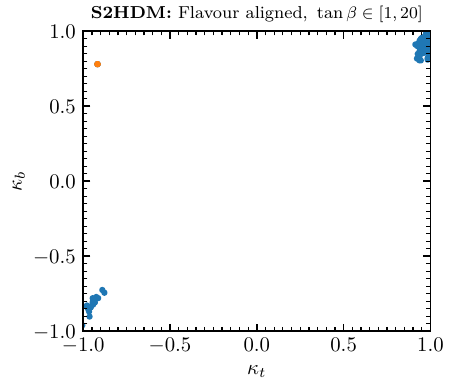}
\caption{For each valid parameter point found with the GA
in the scenario 2 in the flavor-aligned S2HDM,
the left plot shows the the values of $\chi^2(h_{125})$
obtained with \texttt{HiggsTools} against the
coupling coefficient $|\kappa_V|$
of the 125~GeV Higgs boson $h_1$.
The right plot shows the $h_1$ coupling coefficient
$\kappa_t$ against the coupling coefficient $\kappa_b$.
In both plots, the orange points belong to the
same parameter point which stands out as the only
parameter point for which $\kappa_t$ and $\kappa_b$
have the opposite sign.}
\label{fig:eta}
\end{figure}

In \cref{fig:eta} we show the results obtained for
scenario 2, where the GA was used to scan the
flavor-aligned S2HDM, see \cref{tab:scan_ranges}.
Each point in both plots
corresponds to a valid parameter point that
satisfies all theoretical and experimental constraints.
In the left plot, we show the values of the
global $\chi^2(h_{125})$ as computed with
\texttt{HiggsTools} plotted against the
absolute value of the coupling coefficient
$|\kappa_V|$ of the 125~GeV Higgs boson $h_1$.
The plot demonstrates that the GA successfully identifies
parameter points where $|\kappa_V| \approx 1$
with $\chi^2(h_{125})$ close to or below the Standard
Model value $\chi^2_{\rm SM}(h_{125})$,
indicating good agreement with the LHC measurements
of the Higgs boson. This confirms that even in
the more general flavor-aligned setup with a total
of 14 free parameters the GA is able to determine
parameter space regions with the desired features.
In this scenario, we find that smaller values of
the coupling coefficient $|\kappa_V|$
are still compatible with current LHC data.
Values down to $|\kappa_V| \approx 0.912$ are allowed
while still yielding valid parameter points.
This is in contrast to the more restrictive scenario~1
with Yukawa Type~I, see the left plot of \cref{fig:lhc_1_1},
where viable parameter points were only found
for $|\kappa_V| \gtrsim 0.935$. Notably, even some
parameter points with $|\kappa_V| \lesssim 0.92$
result in a total $\chi^2(h_{125})$ below the
Standard Model value $\chi^2_{\rm SM}(h_{125})$.
This indicates that, compared to the Yukawa Type~I,
the added freedom of the flavor-aligned Yukawa structure
allows the model to better accommodate the LHC
measurements in case of sizable deviations in $\kappa_V$
from the SM prediction $\kappa_V = 1$.

In the right plot of \cref{fig:eta}, we show the
values of the coupling modifiers $\kappa_u$ vs.\
$\kappa_d$, which specifically correspond to the ratios of
the $h_1$ couplings to top and bottom quarks, respectively,
compared to the Standard Model.
Unlike in the Yukawa Type~I, these two couplings
are now controlled independently via the
alignment parameters $\xi_u$ and $\xi_d$.
The distribution shows that most viable parameter points
cluster in the region where both $\kappa_u$ and
$\kappa_d$ are close to either $+1$ or $-1$,
reflecting Standard-Model-like behavior.
However, one point
clearly stands out as it corresponds to a
parameter point where $\kappa_u$ and $\kappa_d$ have
opposite signs.
The two points in the left and the right plot
of \cref{fig:eta} corresponding to this parameter
point are highlighted in orange.
This sign configuration between $\kappa_u$ and $\kappa_d$
(also referred to as wrong-sign Yukawa coupling regime)
is a distinctive feature of the flavor-aligned S2HDM
compared to the S2HDM Type~I and can, as observed here,
still be consistent with current data, though it
is associated with
modified interference effects in loop-induced
Higgs processes (such as gluon fusion
production or the $h \to \gamma\gamma$ decay).

In summary, the two example scenarios demonstrate
that GAs are effective tools for scanning the
complex parameter space of BSM theories.
Despite the high dimensionality of the search
(eleven and fourteen free parameters) and the
large amount of experimental constraints
(including around 150 measurements of the 125~GeV
Higgs boson and a plethora of exclusion limits from
searches for additional scalars), the GA reliably
identifies viable parameter regions. Notably,
one of the key strengths of GAs is their ability
to uncover distinct and potentially isolated
solutions in parameter space that are consistent
with all constraints and may even yield a better
fit to the data. Such solutions might easily be
missed by more local or gradient-based methods.
This is exemplified in scenario~2 by the
identification of a parameter point with
opposite signs in the top and bottom Yukawa coupling
modifiers, a feature that
nonetheless yields an acceptable fit to the LHC data.
The S2HDM serves here as one specific but
sufficiently complex example to showcase the power
of GAs, demonstrating their potential for
parameter scans in BSM theories.
More broadly, GAs hold great promise for
exploring a wide range of
particle physics models with a substantial
number of unknown parameters and
with similar or even greater complexity.

\subsubsection{Reconstructing gravitational wave spectrum from
LISA mock data}
\label{sec:lisa}

To explore the utility of GAs, and in particular the
\evortran\ library, in a cosmological context, we consider
the problem of reconstructing a stochastic gravitational wave
signal generated by a first-order phase transition in
the early universe.
The analysis begins with the construction of mock observational
data for the upcoming LISA experiment, into which a synthetic
gravitational wave signal is injected based on a physically
motivated template 
predicted for such a cosmological phase transition.
To simulate realistic observational conditions,
Gaussian noise simulating the LISA instrument sensitivity is
added to the signal. This introduces a stochastic component to
the data, which complicates the parameter inference and
makes traditional gradient-based minimization techniques less
effective, whereas a GA is well suited for this type of problem.
For the fitting process we use a GA implemented with
\evortran\ to minimize an overall $\chi^2$-function that
quantifies the deviation between the theoretical signal
(including LISA’s sensitivity curve) and the noisy mock data.
This procedure is repeated multiple times to generate
a set of reconstructed signals.
Each reconstructed signal corresponds to a set of cosmological
parameters that govern the spectral shape and amplitude of the
gravitational wave signal,
such as the strength, duration, and energy release of the
phase transition. The resulting distribution of these
parameter sets reveals the regions of parameter space
compatible with the data, and allows for a comparison with the
values originally used for signal injection.

The gravitational wave spectrum produced during a
cosmological phase transition carries information about
the underlying physics of the early universe.
The shape and amplitude of the signal can be
parametrized in terms of the strength of the phase
transition $\alpha$, defined as the ratio of the
vacuum energy released to the radiation energy density,
the inverse duration of the transition normalized to
the Hubble rate, $\beta/H$, the transition temperature
$T_*$ at which the signal is generated, the number of
relativistic degrees of freedom in the plasma at
that temperature, $g_*$, and the terminal velocity
$v_w$ of the bubble walls expanding through the plasma.
In this study, we use signal templates derived
from numerical simulations that model the generation
of gravitational waves produced during
a first-order phase transitions.
The total gravitational wave spectral power
density $\Omega_{\rm GW} h^2$,
with $h = 0.68$ being the dimensionless Hubble
constant, is composed of three main contributions,
\begin{equation}
  \Omega_{\rm GW} h^2 = \Omega_{\rm sw} h^2 +
    \Omega_{\rm turb} h^2 + \Omega_{\rm coll} h^2 \, .
  \label{eq:gwtemplate}
\end{equation}
Each of these components contributes with a characteristic
spectral shape and peak frequency, dependent on the physical
parameters mentioned above.
The sound wave contribution $\Omega_{\rm sw} h^2$
results from gravitational wave production from
the acoustic oscillations in the plasma after bubble
collisions, which dominates the signal in
scenarios in which the bubble walls reach
a terminal velocity before colliding~\cite{Hindmarsh:2013xza}.
The turbulence contribution $\Omega_{\rm turb} h^2$ arises
from magnetohydrodynamic turbulence,
i.e.~nonlinear plasma motion generated
after the transition. The turbulence contribution are
not yet well understood, but numerical simulations
indicate that it peaks at
slightly higher frequencies than the sound wave contribution
and with peak amplitudes that are substantially smaller
than the one of the sound wave contribution~\cite{Caprini:2009yp}.
Finally, $\Omega_{\rm coll} h^2$ contains
the contributions from
bubble wall collisions, which account for the
direct collisions of expanding bubbles during the
transition~\cite{Kosowsky:1992vn}.
Specifically, we model the three contributions
using the following templates available in the literature:
the sound wave contribution follows the power-law
parametrization given in Ref.~\cite{Caprini:2019egz},
the turbulence spectrum is implemented using
the results of Ref.~\cite{Ellis:2020awk},
and the bubble collision part
is described by the broken power-law given
in Ref.~\cite{Caprini:2024hue}.

To simulate realistic LISA mock data and
model the detector sensitivity, we follow the
methodology outlined in Ref.~\cite{Caprini:2019pxz}.
The total power spectral density includes contributions
from instrumental noise due to the optical metrology system
and mass acceleration, assuming a standard LISA
configuration with an arm length of $L=2.5 \cdot 10^6$~km.
We incorporate the full LISA response function as
detailed in the same reference.
For simplicity, we do not include uncertainties on the
effective functions parametrizing the two noise components.
In a more realistic scenario, these could  be constrained
by measuring in the low- and high-frequency regions where
no gravitational wave signal is expected.
\TB{Following the Welch method~\cite{1161901},}
the data are simulated over a frequency range from
$f_{\rm min} = 3 \cdot 10^{-5}$~Hz to
$f_{\rm max} = 0.5$~Hz, with a resolution of
$\Delta f = 10^{-6}$~Hz, determined by the length of
the time stream. To inject a signal, we generate
at each frequency 94 individual signal power values with
Gaussian noise and compute their average to obtain the
final mock data. This mimics an expected 4-year
observational run of LISA with approximately 75\% observing
efficiency, which results in 94 statistically independent
data chunks.

To reconstruct the gravitational wave signal from
the simulated LISA data, we use \evortran\ to minimize
a $\chi^2$-function that quantifies the difference
between the model and the data, with the data including
the injected signal. Specifically, we minimize
\begin{equation}
\chi^2(\alpha, \beta / H, T_*, g_*, v_w) =
  N_{\rm chunks} \sum_i \frac{1}{2}
  \left[
    \frac{\bar{D}_i - \Omega_{\rm GW} h^2 - \Omega_s h^2}
    {\sigma_i}
  \right]^2 \, ,
\label{eq:chisqlisa}
\end{equation}
where $\Omega_{\rm GW} h^2$ is the template for the
gravitational wave spectrum from a cosmological phase
transition, see \cref{eq:gwtemplate}, depending on the
five physical parameters $\alpha$, $\beta / H$,
$T_*$, $G_*$, and $v_w$, as discussed above.
$\Omega_s h^2$ is the LISA face sensitivity curve,
$\bar{D}_i$ are the averaged simulated signal powers
at frequency bin $i$, and $\sigma_i$ are the variances
over the $N_{\rm chunks} = 94$ individual realizations
that were averaged to produce the mock data
(see Ref.~\cite{Caprini:2019pxz} for details).
The sum in \cref{eq:chisqlisa} runs over the
frequency bins from the minimum
frequency $3 \cdot 10^{-5}$~Hz up to a frequency
of $10^{-2}$~Hz, and the bins at higher frequency
are discarded since the gravitational wave signals
are far below the LISA sensitivity there.

We employ the \texttt{evolve\_population}
function of \evortran\ to perform the optimization and locate
parameter values that minimize $\chi^2$.
The GA is configured with tournament
selection, sbx crossover and
uniform mutation. The populations size is
200, the selection size is 100, and the elite size is four.
The population is evolved over a maximum number
of 500 generations, or until a minimal $\chi^2$ threshold
value of $1.11 \cdot 10^{-2}$ was achieved. This
value was determined heuristically because we observed
that below this value the instrumental noise prevents
improving the signal reconstruction to a higher
level of precision.
For each example presented below,
which differ by the injected signal or the
parameter set to be reconstructed from the mock data,
the minimization is repeated 200 times to explore
potential variations and degeneracies in the fit.
The resulting distribution of reconstructed parameter
sets is then compared to the true parameters used
to inject the signal, providing insight into the
precision with which LISA may constrain phase transition
parameters if a stochastic gravitational wave
background is observed.
\TB{All source code used to obtain the results
presented in this section is publicly available
in a dedicated Git repository~\cite{gitrepolisa}.}

Before turning to the discussion of specific example
scenarios, we briefly comment on the simplifications made
in this analysis. First, we do not include uncertainties in
the LISA sensitivity curve, which in a full analysis would
arise from limited knowledge of the noise power spectral
densities and calibration uncertainties~\cite{Caprini:2019pxz}.
Second, we ignore
stochastic astrophysical foregrounds, such as the unresolved
foregrounds from stellar binary systems like
white dwarfs, neutron stars or black holes~\cite{Farmer:2003pa,
Rosado:2011kv},
which are expected
to contribute in the relevant frequency range and may
complicate signal extraction. Third, we assume that the
shape of the gravitational wave spectrum follows fixed template
forms based on hydrodynamic simulations, rather than
reconstructing the spectrum in a binned, model-independent
way from the data, as considered in various LISA
forecasting studies.
Furthermore, we do not perform a statistically comprehensive
likelihood analysis to determine allowed parameter
ranges at a given confidence level. Instead, we
effectively over-fit the data to provide signal
reconstructions that agree well with the observations.
Finally, we note that future improvements
in theoretical modeling might reveal additional features
in the signal templates, which could allow more precise
parameter extraction if present in real data.
These simplifications are justified here since our
goal is to illustrate how \evortran\ can be used for
gravitational wave signal reconstruction as a
general-purpose, customizable tool. A comprehensive treatment
that includes these more realistic aspects
and a more sound statistical interpretation is
left for future work. For studies that address these
issues in detail, see e.g.~Refs.~\cite{Caprini:2019pxz,Caprini:2024hue}.

\begin{figure}
\centering
\includegraphics[height=6cm]{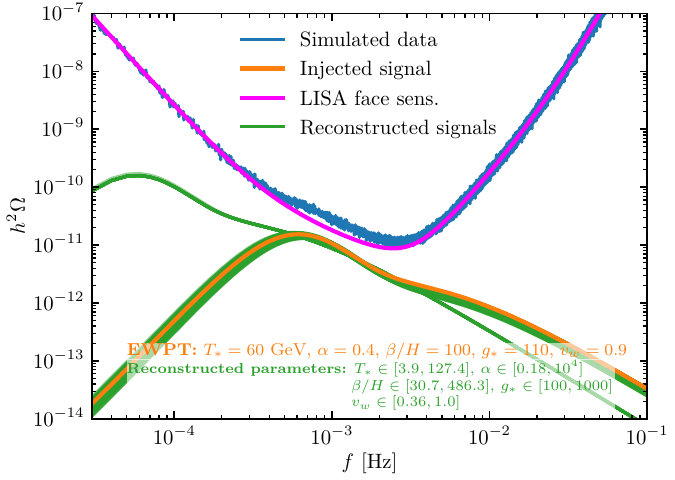}\\[1em]
\includegraphics[width=13cm]{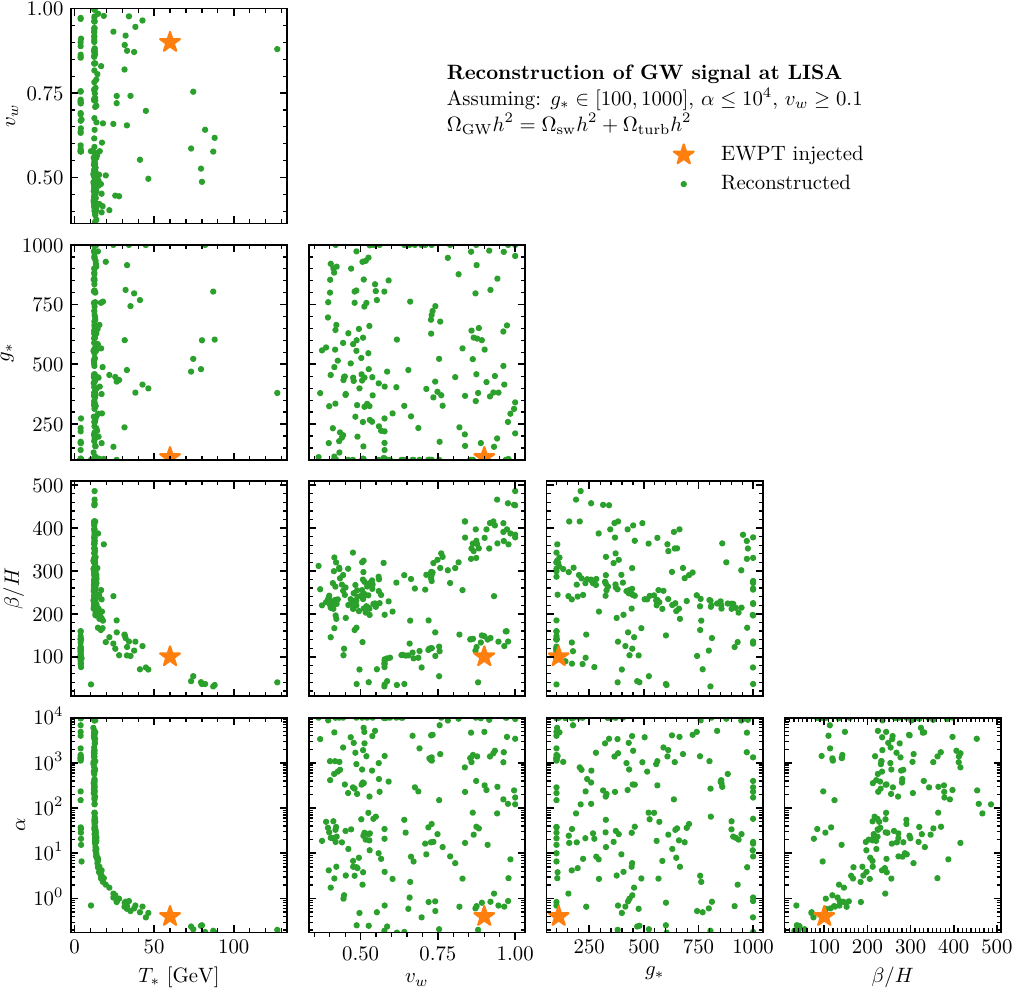}
\caption{Top: Gravitational wave power spectrum as
a function of frequency for the scenario~1.
The magenta line shows the LISA power-law face
sensitivity curve, the orange line is the injected
gravitational wave signal composed of sound wave and
turbulence contributions, and the blue line shows the
mock data including Gaussian noise.
The green lines represent 200 reconstructed gravitational wave
signals obtained using the \evortran\ by minimizing the
$\chi^2$ function given in \cref{eq:chisqlisa}.
Bottom: Corner plot showing the distributions of
the phase transition parameters corresponding to
the reconstructed signals. Each green point marks
a parameter set that produced a signal consistent with
the mock data, while the orange star indicates the
parameters used to generate the injected signal.}
\label{fig:gw_general}
\end{figure}

\medskip

\noindent{\textbf{Scenario 1: Reconstructing EWPT signal}
-- Free parameters: $\{ \alpha,\beta/H,T_*,g_*,v_w \}$\\
As a first example, we consider a gravitational wave signal
injected into the LISA mock data that is consistent with
an electroweak phase transition in the early universe.
The underlying parameters for this signal reflect a realistic
scenario in BSM physics. We assume a transition temperature
$T_* = 100$~GeV, corresponding to the order of
the electroweak scale. For the transition strength we assume a value of
$\alpha = 0.4$, which lies at the upper end of what can be
achieved in simple scalar extensions of the Standard
Model such as models with an additional singlet~\cite{Ellis:2022lft,
andrii} and/or
a second Higgs doublet~\cite{Biekotter:2022kgf,Bittar:2025lcr,
Biekotter:2025fjx}, without invoking significant
supercooling. The terminal velocity of the expanding
bubbles is taken to be $v_w = 0.9$, close to the speed of
light, since in strong transitions with $\alpha \gtrsim 0.1$
the bubble expansion is expected to proceed as
relativistic detonations~\cite{Krajewski:2024gma}.
The number of effective relativistic degrees of freedom
is set to $g_* = 110$, accounting for the ones
of the Standard Model plus a modest number of BSM states
that facilitate a strong electroweak phase transition.
Finally, we use $\beta / H = 100$ as a representative inverse
duration of the transition, which is typical for such scenarios.

The upper plot of \cref{fig:gw_general} shows the LISA face
sensitivity curve (magenta), the injected gravitational wave
signal (orange), and the mock data after adding Gaussian
detector noise (blue). As described above, we use \evortran\
to minimize the $\chi^2$ function shown in \cref{eq:chisqlisa}
in order to reconstruct the signal from this noisy data.
This reconstruction is performed 200 times to explore
degeneracies and noise-induced variance, where we include
all five parameters $\{ \alpha,\beta/H,T_*,g_*,v_w \}$ as
free parameters to be fitted to the data, under the conditions
that $g_* \leq 10^3$, $\alpha \leq 10^4$ and $v_w \geq 0.1$.
The resulting reconstructed spectra are shown in green.
For this specific scenario, the contribution from bubble
collisions is omitted, as the expanding bubbles are not expected to
runaway in a typical electroweak phase transition,
and the resulting collision signal is expected
to be subdominant compared to sound waves and turbulence in
the LISA frequency band.

The plot reveals an important degeneracy in the signal
reconstruction. \evortran\ consistently identifies two
qualitatively distinct classes of solutions that fit
the data taking into account instrumental noise.
The first class closely resembles the
injected signal, with the peak from the sound wave contribution
lying within the most sensitive frequency range of LISA.
In this case, the turbulence contribution remains largely
irrelevant, as its amplitude is suppressed, and its
peak falls at higher frequencies where the experimental
sensitivity is strongly reduced. The second class of
reconstructed signals, however, fits the data using the
turbulence peak instead. Here, the entire signal is
shifted to lower frequencies, and although the corresponding
sound wave component is much stronger, it is shifted to
the left of the LISA sensitivity curve and thus
effectively undetected. Despite arising from entirely different
physical parameters and microphysics, these two
alternative reconstructions produce a gravitational
wave signal that would be indistinguishable from the
injected signal within the noise level of the simulated data.
This example highlights the practical challenges and
degeneracies involved in interpreting gravitational wave
observations from cosmological phase transitions.

In the same plot, we also show the distributions of the
reconstructed parameter values corresponding to the
signals that are consistent with the mock data.
One can see that none of the parameters can be extracted
given the available data. In particular, the parameters
$\alpha$ and $g_*$ remain effectively unconstrained.
Importantly, this imprecision is not solely due to
the existence of the two different classes of viable signals,
but also persists when considering only the subset of
reconstructed signals that closely resemble the
injected one. This residual uncertainty reflects inherent
degeneracies in the dependence of the gravitational wave
spectrum on the underlying parameters.
In the future, if LISA will detect a signal consistent
with an electroweak phase transition,
these degeneracies might severely
limit the possibility of distinguishing
between different BSM theories
that might have given rise to the electroweak phase transition.

The lower plot in \cref{fig:gw_general} is a corner plot
showing the parameter distributions of the 200
reconstructed signals. Each green point corresponds to
a parameter combination that yields a signal
compatible with the mock data
(green lines in the top plot), while the orange star
indicates the true values of the injected signal
(orange line in the top plot).
The corner plot shows the presence of strong degeneracies,
as the green points are broadly scattered, filling
large portions of the allowed parameter space across
all pairwise projections.
\TB{This behavior does not indicate a shortcoming of
the applied GA or a lack of convergence.
The reconstruction problem in the chosen parameter basis
(called ``thermodynamical'' parameters) is known to
exhibit strong parameter degeneracies~\cite{Caprini:2024hue},
which prevent precise constraints regardless of the sampling technique employed.}
In particular, a prominent
degeneracy is visible between the parameters $\alpha$,
$\beta / H$ and $T_*$. This degeneracy will
be further analyzed and discussed in the next example.

\begin{figure}
\centering
\includegraphics[height=7cm]{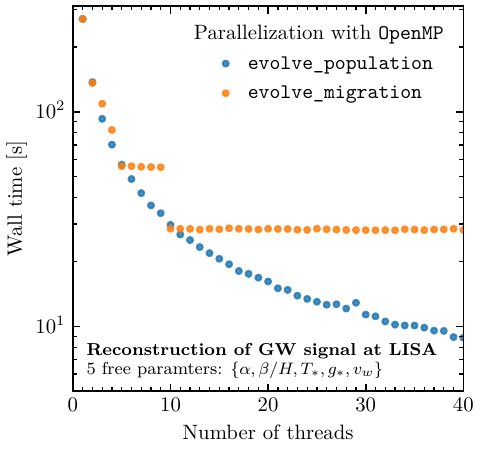}
\caption{\TB{Wall time as a function of the number of threads for
the scenario 1, using the \texttt{evolve\_population}
(blue points) and \texttt{evolve\_migration}
(orange points) routines.}}
\label{fig:gw_general_nthreads}
\end{figure}

\TB{To study the performance of the parallel implementation
of \evortran\ in a realistic setting (the scaling was studied
for the minimization of the Rastrigin benchmark function
in \cref{sec:rastrigin}) with a computationally expensive
fitness function, we analyzed the scaling of wall time with
the number of threads in this LISA signal reconstruction example.
The results are shown in \cref{fig:gw_general_nthreads}.
For both the \texttt{evolve\_population} (blue points) and
the \texttt{evolve\_migration} (orange points) routines,
the minimization of the $\chi^2$ function was performed here without
setting a \texttt{fitness\_target} to ensure that all generations
were executed. In the case of \texttt{evolve\_population},
a population size of 1000, a maximum of 250 generations,
and a selection size of 1000 were used.
For \texttt{evolve\_migration}, ten populations with 100 individuals
each were evolved over five epochs of up to 50 generations,
resulting in the same total number of individuals and
total number of generations.
Moreover, the selection size was set equal to the population size.
This setup yields similar wall times for both routines when
using a single thread, although it leads to some overfitting
in this scenario. The configuration was chosen deliberately to
increase the total runtime so that the scaling with thread
count remains visible before reaching the regime where
algorithmic overhead dominates.

One can see in \cref{fig:gw_general_nthreads} that the
\texttt{evolve\_population} routine exhibits a significant
reduction of the wall time with thread count.
Starting from almost 300~seconds for a single thread,
the wall time roughly halves when using two threads,
decreases to about 5~seconds with five threads,
and continues to improve gradually with even larger
numbers of threads, dropping below one second at 30~threads.
With the maximum of 40 threads, the wall time reaches about
0.9~seconds, corresponding to an overall speedup by a factor of
more than~300. The \texttt{evolve\_migration} shows similar improvement
up to about five threads, where the wall time decreases to
roughly 5~seconds. Then the wall time stagnates until ten threads,
where the wall time improves the last time, reaching
approximately 3~seconds, corresponding to a speedup of
about two orders of magnitude relative to a
run with only one thread.
For even larger number of threads, no further improvement
of the runtime is achieved.
The better scaling of
the \texttt{evolve\_population} routine in this example
compared to \texttt{evolve\_migration} is the opposite of
what was observed for the optimization of the Rastrigin function discussed
in \cref{sec:rastrigin}, see \cref{fig:rastrigin_nthreads},
where the \texttt{evolve\_migration} routine performed significantly
better using parallel execution.
The important difference here is
the higher computational cost of the fitness function which
makes the parallelization over individuals within a population,
as implemented in the \texttt{evolve\_population} routine,
highly effective. In contrast, the parallelization implemented in
the \texttt{evolve\_migration} over the different populations
is limited here by the small number of only five simultaneously
evolving populations, such that using significantly more than
about five threads provides no additional improvement in runtime.}

\begin{figure}
\centering
\includegraphics[height=6cm]{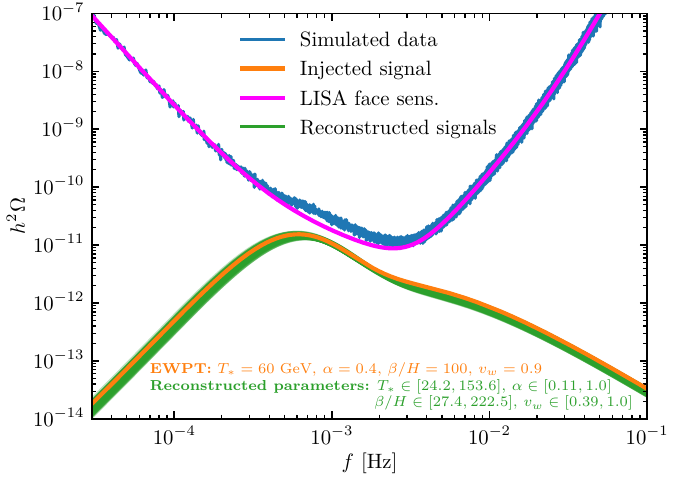}\\[1em]
\includegraphics[width=14cm]{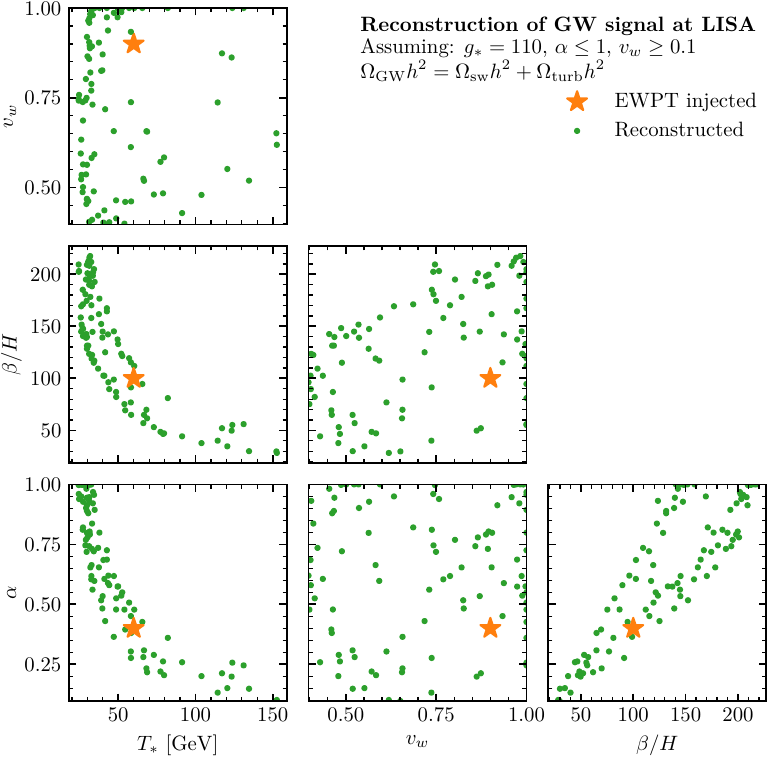}
\caption{As in \cref{fig:gw_general} for the scenario~2.}
\label{fig:gw_nocol_gxfix}
\end{figure}

\medskip

\noindent{\textbf{Scenario 2: Reconstructing EWPT signal
assuming $\alpha \leq 1$ and $g_* = 110$}
-- Free parameters: $\{ \alpha,\beta/H,T_*,v_w \}$\\
In the second example, we build upon the previous analysis
by introducing additional constraints that reflect a
more model-dependent interpretation of the gravitational wave
signal. Specifically, we restrict the strength of the phase
transition to values $\alpha \leq 1$, and we fix the
effective number of relativistic degrees of freedom
to $g_* = 110$. These choices are motivated by the
expectation that an electroweak phase transition occurring
in minimal extensions of the Standard Model, such as
those with a singlet scalar, a second Higgs doublet,
or a Higgs triplet, will involve only a modest increase
in the particle content and will not exhibit significant
supercooling. As a result, this setup provides a more
focused reconstruction of the signal under the
assumption that the underlying physics corresponds to
a specific electroweak-scale model.
In contrast, the broader parameter space explored in
the previous example remains more agnostic to the
origin of the signal and is also compatible with other
cosmological phase transitions that might
have occurred in the early universe.
We again take into account only the sound wave and
turbulence contribution to the gravitational wave signal,
which is consistent with the considered range of $\alpha$.

The results of this second example are shown in
\cref{fig:gw_nocol_gxfix}, with the top plot displaying
the power spectra and the bottom plot presenting
the corner plot of reconstructed parameter values.
Due to the additional assumptions on $\alpha$ and $g_*$,
only reconstructed signals that closely resemble the
injected one are found. The second class of solutions
observed in the previous example, characterized by a
peak at lower frequencies dominated by the turbulence
contribution as visible in the upper plot of \cref{fig:gw_general},
disappears. This is a consequence of requiring $\alpha \leq 1$ here.
As a result of the additional assumptions, the parameter reconstruction
becomes more useful. From the distribution of reconstructed
signals, we infer that the transition temperature $T_*$
is constrained between about \TB{24~GeV and 154~GeV}, $\alpha$
 must be larger than about 0.1, $\beta/H$ lies between
27 and 223, and the bubble wall velocity $v_w$
is reconstructed to be above about 0.4. Here one should
note that the lower limit on $T_*$ and the upper bound
on $\beta/H$ are both consequences of the upper limit
assumed on $\alpha$, and thus not a direct consequence of
the fitting procedure.

The corner plot at the bottom of
\cref{fig:gw_nocol_gxfix} further reveals that the bubble wall
velocity $v_w$ remains practically unconstrained.
However, fixing $g_* = 110$ gives rise to a clearer
correlation between the other three parameters
$T_*$, $\alpha$ and $\beta/H$,
with $\alpha$ and $\beta/H$ decreasing with
increasing values of $T_*$, see the two bottom
panes in the left column
of the corner plot. The size of the bands
in which the green points are concentrated in these
plots results from the unknown bubble wall velocity
$v_w$ which is also fitted to the data in this example.
In the following example we will investigate the
correlations between the parameters by further assuming
that a prediction for $v_w$ is available for the
electroweak phase transition, in which case $v_w$ does
not have to be reconstructed from the LISA data but
can be set to the predicted value.

\begin{figure}
\centering
\includegraphics[height=6cm]{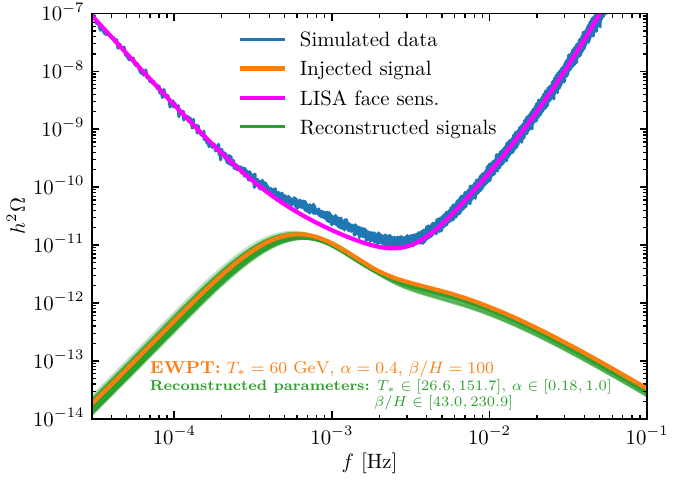}\\[1em]
\includegraphics[width=14cm]{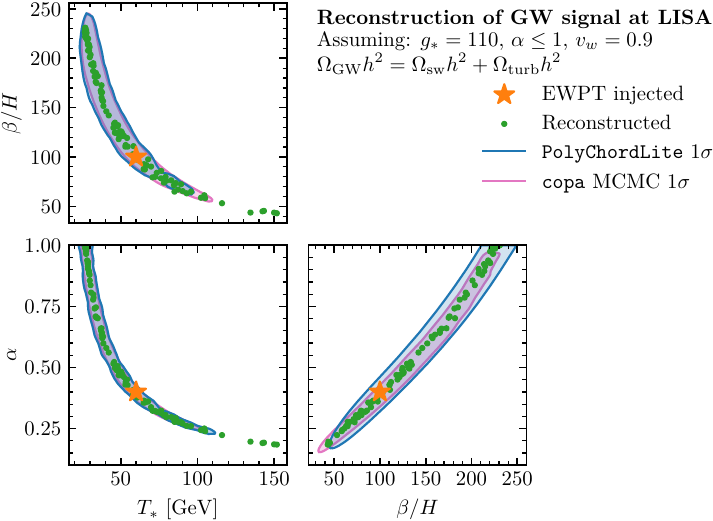}
\caption{As in \cref{fig:gw_general} for the
scenarion 3.
\TB{The corner plot additionally shows the $1\sigma$ confidence level
credible regions obtained with two independent Bayesian inference
methods: nested sampling using \texttt{PolyChordLite}
(blue shaded regions) and an ensemble Markov Chain Monte Carlo
sampler implemented using \texttt{copa} (purple shaded regions).}}
\label{fig:gw_nocol_gxvwfix}
\end{figure}

\medskip

\noindent{\textbf{Scenario 3: Reconstructing EWPT signal
assuming $\alpha \leq 1$, $g_* = 110$ and $v_w = 0.9$}
-- Free parameters: $\{ \alpha,\beta/H,T_* \}$\\
The third example builds upon the second scenario with
an even more constrained fit. In addition to assuming
$\alpha \leq 1$ and fixing the effective number of degrees
of freedom to $g_* = 110$, we now also fix the bubble wall
velocity to $v_w = 0.9$, motivated by the expectation of
relativistic expansion for phase transitions with
$\mathcal{O}(0.1$--$1)$ values of $\alpha$.
This setup reflects a more model-dependent interpretation
in which we assume that in the future when LISA is in
operation it may be possible to reliably compute
the bubble wall velocity from first principles
in a given BSM scenario. Consequently, the reconstruction
is now performed only for the three parameters
$\alpha$, $\beta/H$ and $T_*$. While these additional
assumptions reduce generality, they significantly enhance
the precision of the parameter reconstruction,
as we will demonstrate.
The injected signal parameters remain the same as in the
previous two examples discussed above, and we again only
consider the sound wave and turbulence contribution
to the gravitational wave signal.

In \cref{fig:gw_nocol_gxvwfix} we show the results for this
third example, again showing the reconstructed power spectra
in the top plot and the reconstructed values of the fitted parameters
in the corner plot at the bottom.
The upper plot illustrates that the signal is now
reconstructed with high precision across the frequency range
relevant for LISA. However, despite the overall good match
between the reconstructed signals and the injected one,
the underlying parameter values corresponding to the
injected signal are still not recovered. This is a result
of the degeneracy between the three free parameters
$\alpha$, $\beta/H$ and $T_*$. As shown in the corner plot,
the viable parameter values consistent with the data
lie along thin lines in the parameter space.
This indicates that a detection of a stochastic
signal with LISA could effectively be used to
express two of the parameters as functions of the third.

While the observed degeneracy between the parameters
still prevents precise extraction of
the parameter values of the injected signal,
the reconstruction becomes significantly
more informative. In particular, it enables meaningful model
discrimination power. If a given BSM theory can predict a
set of values for $\alpha$, $\beta/H$ and $T_*$ that fall
on top of the reconstructed lines in the corner plot,
that model remains viable within experimental uncertainty
and could provide an explanation for the detected signal.
Conversely, if the predicted parameter correlations in
a specific BSM theory fall outside the reconstructed
viable region, not overlapping with the lines in
the corner plot, the model could potentially be ruled out.
It is important to note, however, that this type of
model testing hinges on the assumption that the
bubble wall velocity $v_w$ is known and fixed.
If a precise prediction for $v_w$ is not available
in the future when LISA is in operation,
the discrimination power between different models is
significantly worse, see the scenarios discussed above.

\TB{In addition to the reconstructed points obtained
with \texttt{evortran}, the corner plot in \cref{fig:gw_nocol_gxvwfix}
also shows the $1\sigma$ credible regions derived from
Bayesian parameter inference, obtained using nested sampling
with the Fortran library
\texttt{PolyChordLite}~\cite{Handley:2015fda,2015MNRAS.453.4384H}
(blue shaded regions)
and an ensemble Markov Chain Monte Carlo sampler implemented in
the \texttt{fpm} project \texttt{copa}~\cite{gitrepocopa}
(purple shaded regions).
The chains produced by both the ensemble MCMC sampler
and the nested sampler were processed using the
Python package \texttt{GetDist}~\cite{Lewis:2019xzd}
to determine the $1\sigma$ credible regions,
making use of kernel density estimation to construct
smooth posterior distributions.
The comparison of the signal reconstruction using
\evortran\ with with Bayesian samplers is carried out only for
this example scenario, since the previous two scenarios exhibit
multiple degeneracies in the parameter space, which make a
meaningful construction of credible regions difficult.
In contrast, in the present scenario, the degeneracy between the
three free parameters $\alpha$, $\beta/H$, and $T_*$ is confined
to a single, relatively well-defined direction in parameter space,
allowing for a more direct comparison between the results obtained
with \evortran\ and the statistically inferred credible regions
from \texttt{PolyChordLite} and \texttt{copa}, respectively.

In the comparison between the results from \evortran\
against the ones from \texttt{PolyChord}
and \texttt{copa}, it is important to keep in mind that
GAs and  Bayesian
sampling methods serve complementary but distinct purposes.
GAs are designed to efficiently identify the best-fit solutions
that minimize the $\chi^2$ function, whereas Bayesian sampling
explores the posterior probability distribution around these solutions,
providing statistically meaningful confidence regions that
quantify parameter uncertainties. However, sampling methods can
struggle to locate the global best-fit solutions if the likelihood
landscape is sufficiently complex or if some optima appear isolated
from the others in the analyized parameter space.
In such cases, GAs provide an ideal tool to verify whether the
samplers have identified all viable regions of parameter space.

One can see that most of the reconstructed parameter points obtained
with \evortran\ lie within the $1\sigma$ credible regions,
and both the \texttt{PolyChordLite} and \texttt{copa} results
agree very well with each other. The strong degeneracy among
the parameters $\alpha$, $\beta/H$, and $T_*$ is visible
across all methods, confirming that it is an intrinsic feature
of the reconstruction problem rather than a limitation of
the GA optimization or the sampling methods.
A few of the \evortran\ points at low transition temperatures
and transition strengths appear outside of the
$1\sigma$ credible regions. This behavior arises because
the credible regions from the Bayesian analyses were
constructed using uniform sampling, which, combined with
finite resolution, can artificially truncate the regions
at the lowest parameter values along the flat direction
in the $\chi^2$ function. The fact that \evortran, where
also uniform initialization of the initial population
was employed, identifies viable solutions in this
region highlights this limitation and demonstrates the
advantage of combining GA-based reconstructions with
Bayesian sampling. A GA efficiently explores the global
parameter space more exploratory and can expose paramter regions
that may be undersampled in a statistical analysis.
However, also the results of GAs depend on the priors
that are used to initialize the population.
The impact of different prior choices on the inferred parameter
regions is discussed in more detail in the following example.}

\begin{figure}
\centering
\includegraphics[height=6cm]{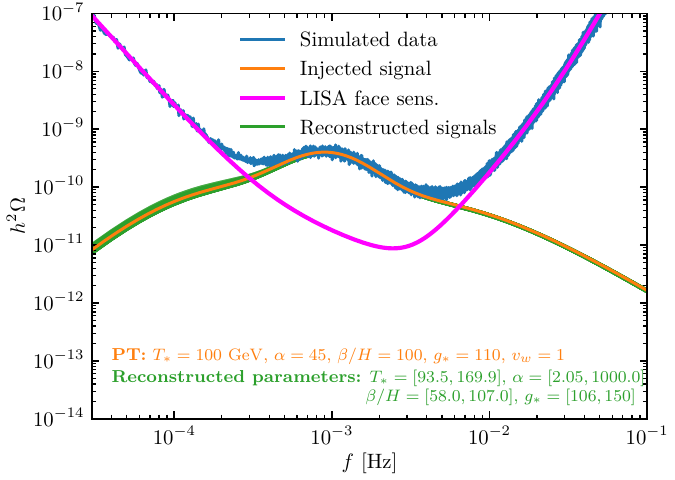}\\[1em]
\includegraphics[width=14cm]{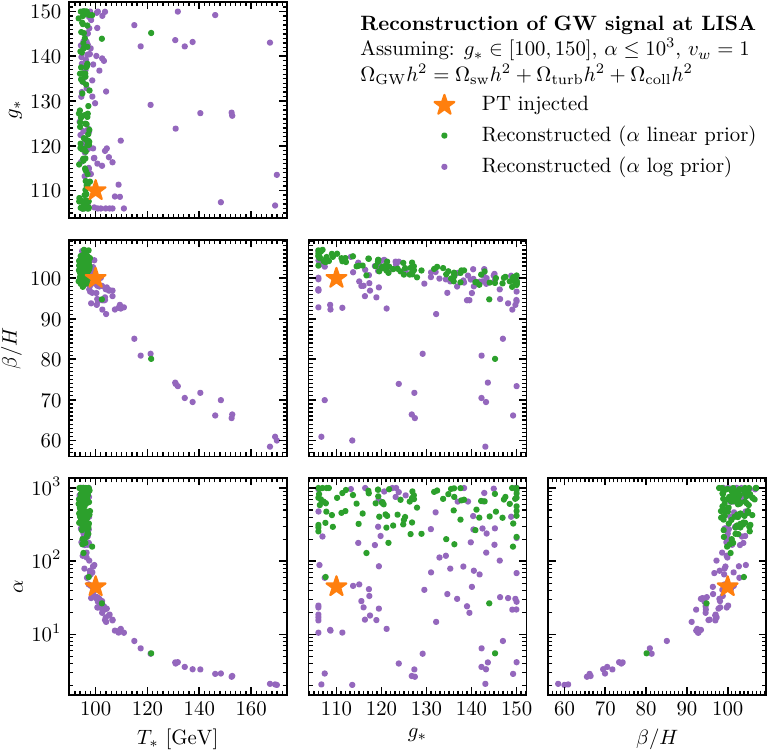}
\caption{Same as in \cref{fig:gw_general} for the
scenario~4. The corner plot additionally shows reconstructed
parameter values using a logarithmic prior on the
parameter $\alpha$ at initialization of the GA.}
\label{fig:gw_yescol}
\end{figure}

\medskip

\noindent{\textbf{Scenario 4: Reconstructing stronger
signal assuming $\alpha \leq 10^3$,
$g_* \leq 150$ and $v_w = 1$}
-- Free parameters: $\{ \alpha,\beta/H,T_*,g_* \}$\\
In this fourth and final scenario, we return to a
more general setup and consider a substantially stronger
gravitational wave signal originating from a
cosmological phase transition with a large strength
parameter of $\alpha = 45$. The motivation behind this
choice is to investigate how well the signal reconstruction
and parameter inference is improved when the injected signal
has a significantly higher signal-to-noise ratio,
thereby enhancing its detectability across a wider range
in the LISA frequency band. In this example,
the contribution from bubble collisions might not
be negligible, and we therefore include this third
source component in the signal template.
The bubble collision peak appears at slightly lower
frequencies than the sound wave and turbulence contributions,
providing additional structure in the spectrum that
could potentially help break some of the degeneracies
observed in earlier examples.

A key focus of this example is to explore the
influence of the prior distribution used for the
reconstructed parameters.
In particular the sampled range of the
strength parameter $\alpha$ span several orders
of magnitude. To investigate the impact of different
initialization of gene values in a GA, we perform
two separate reconstructions: one using a
linear prior on $\alpha$ and the other using a
logarithmic prior. The comparison between the resulting
distributions of reconstructed parameters highlights
that the performance and coverage of the solution space
of GAs can depend sensitively on the initial seeding of
gene values, i.e.~on how the sampling space
is explored from the start. As this example will
demonstrate, the choice of prior can have significant
consequences for the robustness and reliability
of the inferred parameter ranges in
multi-scale parameter spaces.

In \cref{fig:gw_yescol} we show the results for the
fourth scenario, where a strong gravitational wave signal
is injected with parameters
$T_* = 100$~GeV, $\alpha = 45$, $\beta/H = 100$
$g_* = 110$, and $v_w = 1$.
The large signal-to-noise ratio allows for a very precise
reconstruction of the spectral shape, as visible in
the top plot, with the sound wave peak placed near
the maximum sensitivity of LISA and the turbulence and
collision peaks located to either side but still
within the sensitive band.
In the corner plot below, we show two sets of
reconstructed parameter values. The green points
correspond to a linear prior on $\alpha$, while purple
points come from using a log prior. The linear prior
leads to a more concentrated reconstruction,
especially for $g_*$, as visible in the upper pane of
the corner plot. This might misleadingly suggest that
this parameter is well constrained by the data.
However, the results using the log prior reveal that a wide range of
values, including the whole sampled range of $g_*$,
are actually compatible with the injected signal.
This discrepancy arises because the linear prior
oversamples values of $\alpha$ at the upper end
of the sampled interval,
whereas the log prior allows the GA to explore
several orders of magnitude more evenly
as visible in the bottom row of the corner plot.
Notably, only the log prior reconstruction
recovers the injected parameter values (orange stars),
while the linear prior biases the fit
toward larger $\alpha$ and $\beta/H$ values.
This example illustrates the importance of prior
choice in GA-based inference, especially when
parameters span several orders of magnitude.

\section{Conclusions}
\label{sec:conclus}

We have introduced \evortran, a lightweight, flexible,
and efficient genetic algorithm (GA) library written in
modern Fortran.
The \texttt{evortran} package is available at:
\begin{center}
\texttt{\href{https://gitlab.com/thomas.biekoetter/evortran}
  {https://gitlab.com/thomas.biekoetter/evortran}}.
\end{center}
With its modular design and simple
user interface, \evortran\ enables users to easily
apply evolutionary strategies to complex optimization problems,
including those with non-differentiable, discontinuous,
or noisy fitness functions. The library supports
customization of GA components, native real and integer
encodings, and parallel execution via \texttt{OpenMP}.
\evortran\ is installed with the Fortran package manager
\texttt{fpm}, ensuring a straightforward
dependency management, compilation and
installation process, as well as a seamless integration into
both simple scripts and larger code bases.
To further enhance accessibility, \evortran\ provides
Python bindings available at:
\begin{center}
\texttt{\href{https://gitlab.com/thomas.biekoetter/pyevortran}
  {https://gitlab.com/thomas.biekoetter/pyevortran}}.
\end{center}
This interface allows users to run the core
optimization routines of \evortran\ directly from Python.

To demonstrate its robustness and versatility, we
first validated \evortran\ on a set of well-known
multi-modal benchmark functions commonly used in
global optimization, showing reliable convergence
and the ability to locate global optima even in
rugged fitness landscapes.
As a complex, real-world application from particle physics,
we used \evortran\ to perform high-dimensional
parameter scans of the Singlet-extended Two-Higgs-Doublet
Model~(S2HDM), involving eleven to 14 free parameters,
a combination of theoretical constraints,
and an extensive set of LHC data.
As a second physics application from cosmology,
we then applied the library to the reconstruction of primordial
gravitational wave signals and their underlying parameters
from mock data of the upcoming LISA space observatory.
In both cases, \evortran\ performed successfully,
identifying viable solutions efficiently in
challenging search spaces.

While our focus here has been on specific scientific use
cases, the design of \evortran\ makes it broadly applicable
to a wide range of optimization problems in physics,
engineering, and other fields requiring global search strategies.
We hope that \evortran\ becomes a useful tool for
GAs in the Fortran ecosystem and scientific computing.

\section*{Acknowledgements}

I gratefully acknowledge the open-source Fortran
community for their continued efforts in developing
and modernizing the Fortran ecosystem, with special
thanks to the contributors of the Fortran Package
Manager~(FPM), whose work was beneficial for the
development of \texttt{evortran}.
The project that gave rise to these
results received the support of a
fellowship from the ``la Caixa''
Foundation (ID 100010434). The
fellowship code is  LCF/BQ/PI24/12040018.
We acknowledge the support of the Spanish Agencia
Estatal de Investigaci\'on through the grant
``IFT Centro de Excelencia Severo Ochoa CEX2020-001007-S''.

\appendix

\bibliographystyle{JHEP}
\bibliography{lit}

\end{document}